\documentclass[apj,iop]{emulateapj}
\usepackage{times,mathptmx,color}

\shorttitle{{\sc Disk-planets interactions and Kepler diversity}}
\shortauthors{{\sc Clement Baruteau and John C. B. Papaloizou}}

\begin{document}
\title{Disk-planets interactions and the diversity of period ratios in Kepler's
  multi-planetary systems}

\author{Clement Baruteau and John C. B. Papaloizou}
\affil{DAMTP, University of Cambridge, Wilberforce Road, Cambridge CB3 0WA, U.K.}
\email{C.Baruteau@damtp.cam.ac.uk; J.C.B.Papaloizou@damtp.cam.ac.uk}


\begin{abstract}
  The Kepler mission is dramatically increasing the number of planets
  known in multi-planetary systems.  Many adjacent planets have
  orbital period ratios near resonant values, with a tendency to be
  larger than required for exact first-order mean-motion resonances.
  This intriguing feature has been shown to be a natural outcome of
  orbital circularization of resonant planetary pairs due to
  star-planet tidal interactions.  However, this feature holds in
  multi-planetary systems with periods longer than ten days, for which
  tidal circularization is unlikely to provide efficient divergent
  evolution of the planets orbits.  Gravitational interactions between
  planets and their parent protoplanetary disk may instead provide
  efficient divergent evolution. For a planet pair embedded in a disk,
  we show that interactions between a planet and the wake of its
  companion can reverse convergent migration, and significantly
  increase the period ratio from a near-resonant value.  Divergent
  evolution due to wake-planet interactions is particularly efficient
  when at least one of the planets opens a partial gap around its
  orbit.  This mechanism could help account for the diversity of
  period ratios in Kepler's multiple systems comprising super-Earth to
  sub-jovian planets with periods greater than about ten days.
  Diversity is also expected for pairs of planets massive enough to
  merge their gap. The efficiency of wake-planet interactions is then
  much reduced, but convergent migration may stall with a variety of
  period ratios depending on the density structure in the common
  gap. This is illustrated for the Kepler-46 system, for which we
  reproduce the period ratio of Kepler-46b and c.
\end{abstract}

\section{Introduction}
\label{sec:intro}
Planetary astrophysics is undergoing an epoch of explosive growth
driven by the discovery of about 850 exoplanets in two decades.
Continuous improvement in detection techniques will uncover many more
planets in the near future which are too small or too far from their
host star to be detected at present. Observations suggest that planet
formation is ubiquitous, with more than 40\% of nearby Sun-like stars
harboring at least one planet less massive than Saturn
\citep{Mayor09}.  These prolific discoveries have revealed the amazing
diversity of exoplanetary systems and indicated that planets may form
under a wide range of conditions and have considerable mobility.

The diversity of exoplanets has been particularly highlighted by the
Kepler's space mission. To date, Kepler has detected over 2300 planet
candidates \citep{Batalha13}, from which more than a hundred are
confirmed planets (http://kepler.nasa.gov). About one third of
Kepler's candidates are associated with multiple transiting systems
and the vast majority of them are expected to be real multi-planetary
systems \citep{Lissauer12}. Some of these systems show remarkable
properties, like the five coplanar, alternating rocky and icy planets
orbiting Kepler-20 in less than 80 days \citep{Kepler20}. Interactions
between planets and their parent protoplanetary disk are likely to
have played a major role in shaping such compact system.

\begin{figure}
\centering
\includegraphics[width=\hsize]{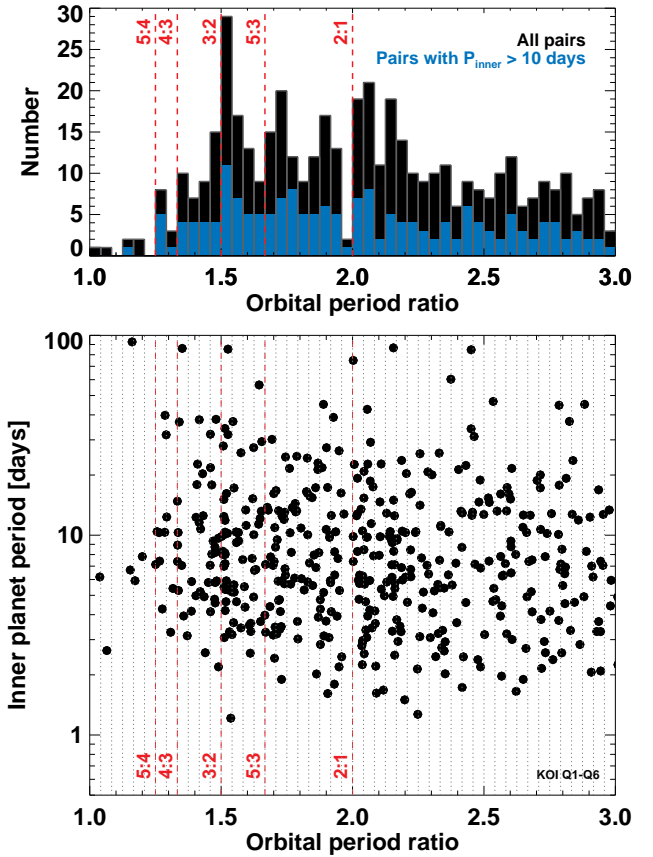}
\caption{\label{fig:Kepler}Bottom: ratio of orbital periods of all
  planet pairs among Kepler's candidate multi-planetary systems
  detected from Quarters 1 to 6 (x-axis) versus the orbital period of
  the inner planet in each pair (in days, y-axis). Bins are $1/24$
  wide.  From left to right, vertical dashed lines show the location
  of the 5:4, 4:3, 3:2, 5:3 and 2:1 mean-motion resonances. Top:
  histograms of the orbital period ratio for all planet pairs (black),
  and for planet pairs where the inner planet's period is greater than
  10 days. Data were extracted from
  http://planetquest.jpl.nasa.gov/kepler/.}
\end{figure}

A striking feature of Kepler's multi-planetary systems is the large
number of planet pairs far from mean-motion resonances
\citep[hereafter MMR, ][]{Lissauer11statmultis}.  Planet pairs near
resonances show some tendency to have period ratios slightly greater
than resonant values (see Figure~\ref{fig:Kepler}; period ratios are
defined as the outer planet's orbital period divided by the inner
planet's).  For instance, there are almost twice as many planet pairs
with period ratios between 2.0 and 2.1 than between 1.9 to 2.0. This
intriguing feature could naturally arise from tidal orbital
circularization of close-in resonant planet pairs \citep{Papa11, LW12,
  Baty12}. However, as stressed by the histograms in
Figure~\ref{fig:Kepler}, the same trend holds for planet pairs where
the orbital period of the inner planet exceeds 10 days, most
particularly near first-order resonances.  For such systems, tidal
circularization is unlikely to have caused significant divergent
evolution of the planets orbits, as shown in Section~\ref{sec:tidal}
of the appendix.

An example of a planetary system with a period ratio slightly greater
than resonant is the Kepler-46 system (previously known as KOI-872).
The 0.9 Solar-mass star is surrounded by a $0.8 R_{\rm J}$ transiting
planet on a 33.6 day period (Kepler-46b). Large transit time
variations (TTV) of Kepler-46b are caused by the presence of an outer
planet, Kepler-46c, whose mass ($0.37 M_{\rm J}$) and orbital period
(57.0 days) have been determined by \cite{Nes12}. The same authors
have also reported the detection of a $1.7 R_{\oplus}$ planet
candidate on a 6.8 day period (Kepler-46d, not yet confirmed). The
main properties of the Kepler-46 planetary system are summarized in
Table~\ref{tab:tableKOI}. The low-eccentricity and quasi-coplanar
orbits in this system suggest that the planets reached their present
location most probably through disk-planet interactions. Still, the
ratio or orbital periods between Kepler-46c and Kepler-46b is $\approx
1.696$, which is slightly greater than $5/3$. Capture in the 5:3 MMR
is not a natural outcome of planet migration driven by disk-planet
interactions. It requires a rate of convergent migration that is large
enough to cross the 2:1 MMR, but not too large so that the planets do
not reach their 3:2 MMR. Capture into a second-order resonance also
requires small but not zero eccentricities, which disk-planet
interactions tend to damp quite efficiently. Should disk-planet
interactions have led Kepler-46b and Kepler-46c to reach their 5:3
MMR, it is unclear how their period ratio would have then increased to
their present value.

The initial motivation of this work is to investigate under what
circumstances disk-planets interactions may account for the orbital
period ratio between Kepler-46c and Kepler-46b.  In
Section~\ref{sec:hydro}, we present results of hydrodynamical
simulations modeling the early evolution of both planets as they were
embedded in their parent protoplanetary disk.  These Saturn-mass
planets are expected to open a partial gap around their orbit. A rapid
convergent migration causes the planets to merge their gap and to
evolve into a common gap. The planets' convergent migration is found
to stall with period ratios between 1.5 and 1.7, depending on the
disk's density profile inside the common gap. The observed period
ratio can be reproduced without the planets being in
resonance. Furthermore, we show that disk-driven migration of partial
gap-opening planets may lead to significant divergent evolution of
planet pairs over typical disk lifetimes. Such divergent evolution is
shown to arise from the interaction between planets and the wakes of
their companions.  This mechanism is illustrated for super-Earth-mass
planets in Section~\ref{sec:results_nbody} with both hydrodynamical
simulations and customized three-body integrations. Divergent
evolution of gap-opening planet pairs mediated by wake-planet
interactions could partly account for the diversity of period ratios
in Kepler's multi-planetary systems.  Concluding remarks are provided
in Section~\ref{sec:conclusion}.

\begin{table}
  \centering
  \caption{\label{tab:tableKOI}Kepler-46 planetary system \citep{Nes12}}
  \smallskip
  \begin{minipage}{\hsize}
    \centering
    \begin{tabular}{l l l l}
      ~ & Mass or radius & Period [days] & Eccentricity\\
      \hline
      \hline
      Kepler-46 & $0.9 M_{\odot}$ & $-$ & $-$\\
      Kepler-46d\footnote{Unconfirmed.} & $\approx 1.7 R_{\oplus}$ & 6.8 & 0 (assumed)\\
      Kepler-46b & $\approx 0.8 R_{\rm J}$ & 33.6 & $\leq 0.02$\\
      Kepler-46c\footnote{Inferred from TTV signal of Kepler-46b.} & $0.37 M_{\rm J}$ & 57.0 &  $\approx 0.015$\\
      \hline
    \end{tabular}\par
  \end{minipage}
\end{table}

\section{Hydrodynamical simulations of the Kepler-46 planetary system}
\label{sec:hydro}
We model in this section the early evolution of the Kepler-46
planetary system when the planets were embedded in their parent
protoplanetary disk. Our study does not address the formation of the
planets. We focus instead on the planets orbital evolution due to
disk-planet and planet-planet interactions. Two-dimensional
hydrodynamical simulations of disk-planets interactions were carried
out using the code FARGO \citep{fargo1}. The physical model and
numerical setup of the hydrodynamical simulations are described in
Section~\ref{sec:setup}. Results of simulations follow in
Section~\ref{sec:results_hydro}.

\subsection{Physical model and numerical setup}
\label{sec:setup}

\par\noindent\emph{Disk model---}
We adopt a two-dimensional disk model for the protoplanetary disk in
which the planets orbiting Kepler-46 formed. Disk-planets interactions
determine the rate of convergent migration of the planets and the
damping rate of their eccentricity. Both quantities are sensitive to
the disk's surface density, temperature, turbulent viscosity and
radiative properties. The parameter space is therefore particularly
large, and the need to assess the planets' orbital evolution over
typically a thousand planet orbits has led us to make simplifying
assumptions for the disk model and to fix some of the disk parameters.

Disk self-gravity is discarded given the low disk masses adopted in
this work (the Toomre-Q parameter associated with the unperturbed
density profile does not go below 7). A locally isothermal equation of
state is used where the vertically-integrated pressure $P$ and density
$\Sigma$ satisfy $P = \Sigma\,c^2_{\rm s}$, with the sound speed
$c_{\rm s}$ being specified as a fixed function of the cylindrical
radius $r$. The sound speed is related to the disk's pressure scale
height $H$ through $H = c_{\rm s} / \Omega_{\rm K}$ with $\Omega_{\rm
  K}$ the Keplerian angular velocity. The (fixed) temperature profile
is taken proportional to $r^{-1}$ so that the disk's aspect ratio
$h=H/r$ is uniform. We take the standard value $h=0.05$. A locally
isothermal equation of state is generally not appropriate to model the
migration of an embedded low-mass planet, as it underestimates the
magnitude of the corotation torque that the planet experiences from
the disk; see \citet{BM13} for a recent review on planet
migration. This is not an issue in this study, however, since we
consider partial gap-opening planets for which the corotation torque
has a rather weak effect on migration. This may be quantified through
the dimensionless parameter $q/h^3$ ($q$ denotes the planet-to-star
mass ratio) which is $\gtrsim 2$ in our simulations. This lower limit
implies that the direction and speed of planet migration are weakly
sensitive to the choice for the equation of state \citep{KleyCrida08}.

The effects of turbulence are modeled by a constant shear kinematic
viscosity $\nu$.  We take $\nu = 1.1 \times 10^{-5}$ in standard code
units (defined below) which translates into a viscous alpha parameter
$\alpha = 4.5\times 10^{-3}$ at the initial location of the inner
planet.  This value of $\alpha$ is typical of the magnetically active
regions of protoplanetary disks with magneto-hydrodynamic turbulence
driven by the magneto-rotational instability. The present orbital
periods of the planets in the Kepler-46 system imply that the planets
have probably interacted with such active regions in their parent
disk. Note, however, that the disk's magnetic field is not included in
our simulations. We also point out that the planets that we consider
are of large enough mass so that small-scale turbulent fluctuations
can be safely ignored when modeling the interaction with their parent
disk. In other words, the viscous diffusion approximation is expected
to be a reasonable approximation for the large planet masses we
consider \citep{qmwmhd2, bfnm11}.

The disk is set up in radial equilibrium, with the centrifugal
acceleration and the radial acceleration related to the pressure
gradient balancing the gravitational acceleration due to the central
star. The initial gas surface density is taken proportional to
$r^{-1/2}$.  Its value at the initial location of the inner planet is
taken as a free parameter, which we denote by $\Sigma_0$.
\\
\par\noindent\emph{Planet  parameters---}
We model the orbital evolution of the outer two planets in the
Kepler-46 system: Kepler-46b (hereafter refereed to as the inner
planet) and Kepler-46c (as the outer planet). The $1.7R_{\oplus}$
unconfirmed planet candidate with a period of 6.8 days is assumed to
have had a negligible impact on the orbital evolution of Kepler-46b
and Kepler-46c. The planets are assumed to have already formed at the
beginning of our simulations, they therefore take their present mass.
Kepler-46b has a physical radius $\approx 0.8 R_{\rm J}.$ However, its
mass could not be constrained from TTV; see \citet{Nes12}.  The
mass-radius diagram of exoplanets detected by transiting methods
indicates that the mass of Kepler-46b most probably lies between $0.2
M_{\rm J}$ and $0.6 M_{\rm J}$. Its planet-to-primary mass ratio
($q_{\rm inner}$) is therefore varied in the range
$[2.2\times10^{-4},6.6\times 10^{-4}]$. The TTV signal of Kepler-46b
constrains the mass of Kepler-46c to be $\approx 0.37 M_{\rm J}$
\citep{Nes12}. The planet-to-primary mass ratio of Kepler-46c is thus
fixed to $q_{\rm outer} = 4\times 10^{-4}$.  The initial orbital
radius of Kepler-46b, $r_{\rm inner}$, is taken to be the code's unit
of length. That of Kepler-46c is arbitrarily set to $r_{\rm outer} =
1.8 r_{\rm inner}$, so that the ratio of orbital periods $\approx 2.4$
initially. For the disk aspect ratio and viscosity taken in our study,
the two planets are expected to open a partial gap around their orbit
\citep{PPIII,crida06}.
\\
\par\noindent\emph{Numerical setup---}
The hydrodynamical equations are solved in a cylindrical coordinate
system $\{r, \varphi\}$ centered on to the star with $r \in
[0.2-2.6]r_{\rm inner}$ and $\varphi \in [0,2\pi]$. The computational
grid has $N_{\rm r} = 400$ zones evenly spaced along the radial
direction and $N_{\rm s} = 800$ azimuthal sectors. The frame rotates
with the Keplerian frequency at the inner planet's location, and the
indirect terms that account for the acceleration of the central star
by the disk and the planets are included in the equations of motion.
A standard outflow boundary condition is taken at the grid's inner
edge, while damping is used in a so-called wave killing-zone extending
from $r=2.2$ to $r=2.6$ (disk quantities are damped towards their
instantaneous axisymmetric profile). To avoid a violent relaxation of
the disk due to the sudden introduction of the planets, the mass of
the planets is gradually increased over 10 orbital periods. The
gravitational potential of the planets is smoothed over a softening
length, $\varepsilon$, equal to $0.6 H$ with $H$ evaluated at the
planets location. The calculation of the force exerted by the disk on
the planets excludes the content of the planets' circumplanetary disk,
the size of which being about $60\%$ of the planets' Hill radius
\citep{cbkm09}. The Hill radius of each planet is resolved by about 10
grid cells along each direction initially.
\\
\par\noindent\emph{Code units---}
Results of simulations are expressed in the standard units of
disk-planet calculations: the mass unit is the mass of the central
star (denoted by $M_{\star}$ and equal to $0.9 M_{\odot}$ for the
Kepler-46 system). The length unit is the initial orbital radius of
the inner planet ($r_{\rm inner}$), and the time unit is the initial
orbital period of the inner planet ($T_{\rm orb}$) divided by
$2\pi$. Whenever time is expressed in orbits, it refers to the orbital
period at the initial location of the inner planet.
\begin{figure*}
  \centering
  \resizebox{\hsize}{!}
  {
    \includegraphics{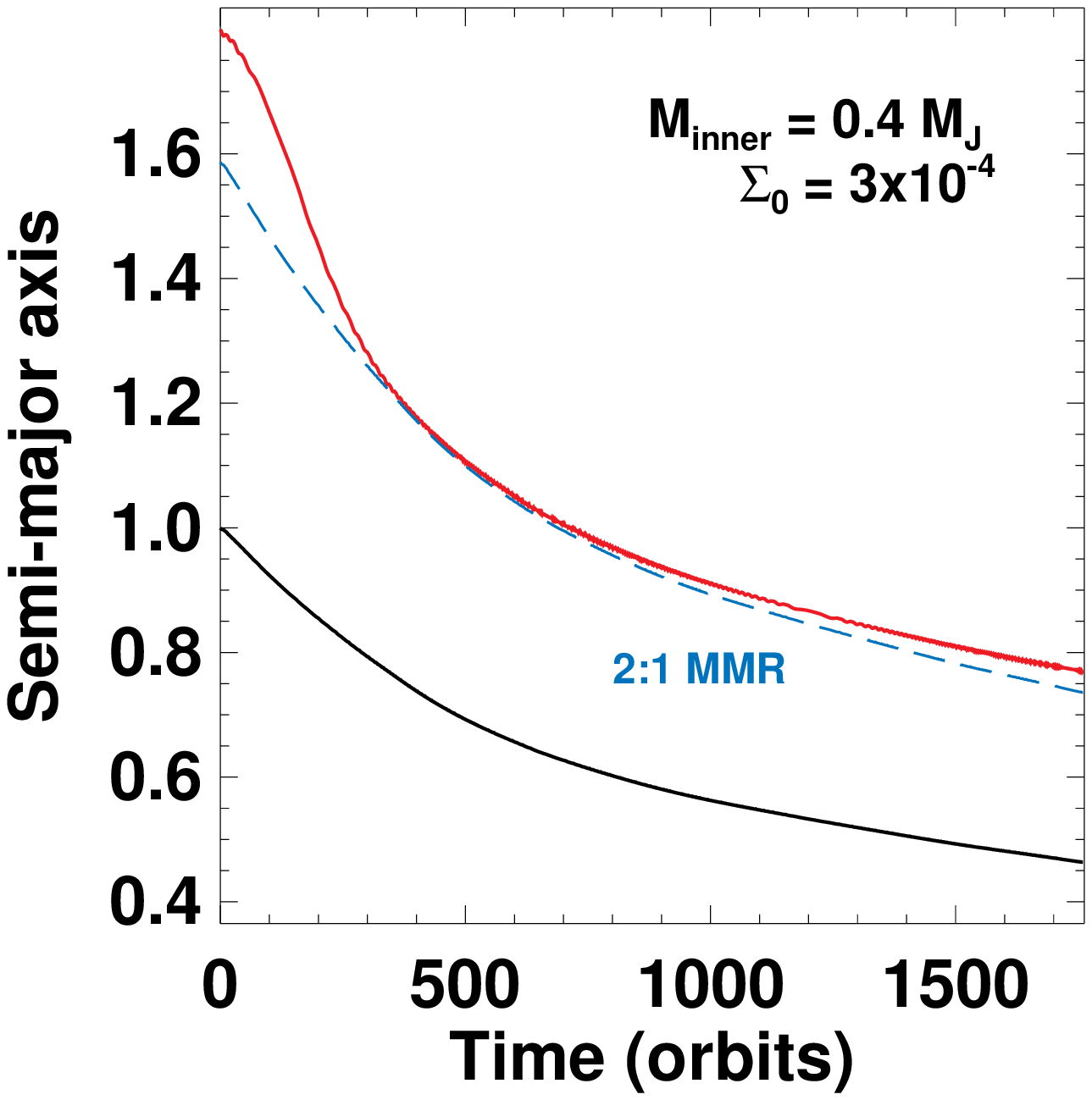}
    \includegraphics{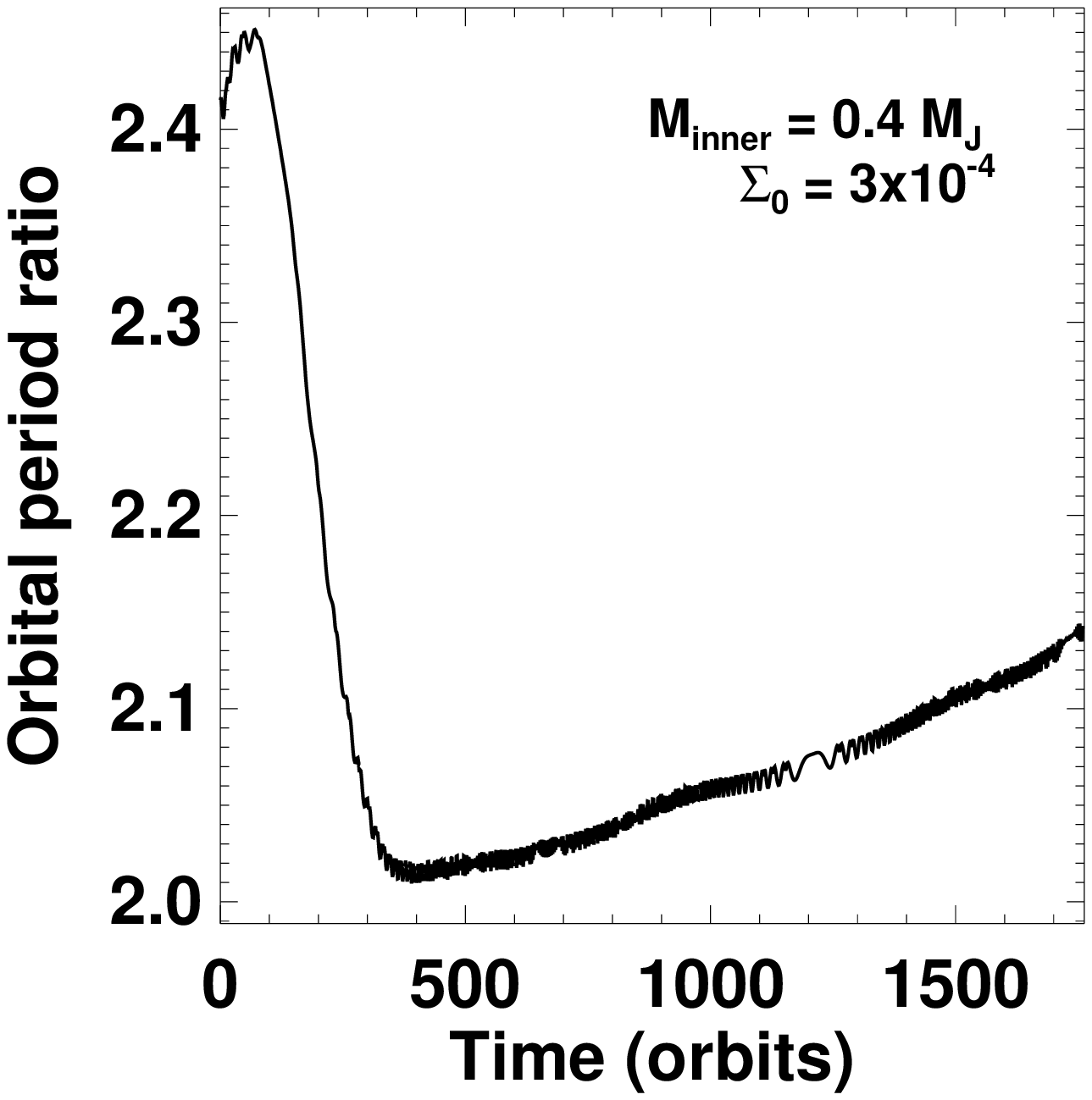}
    \includegraphics{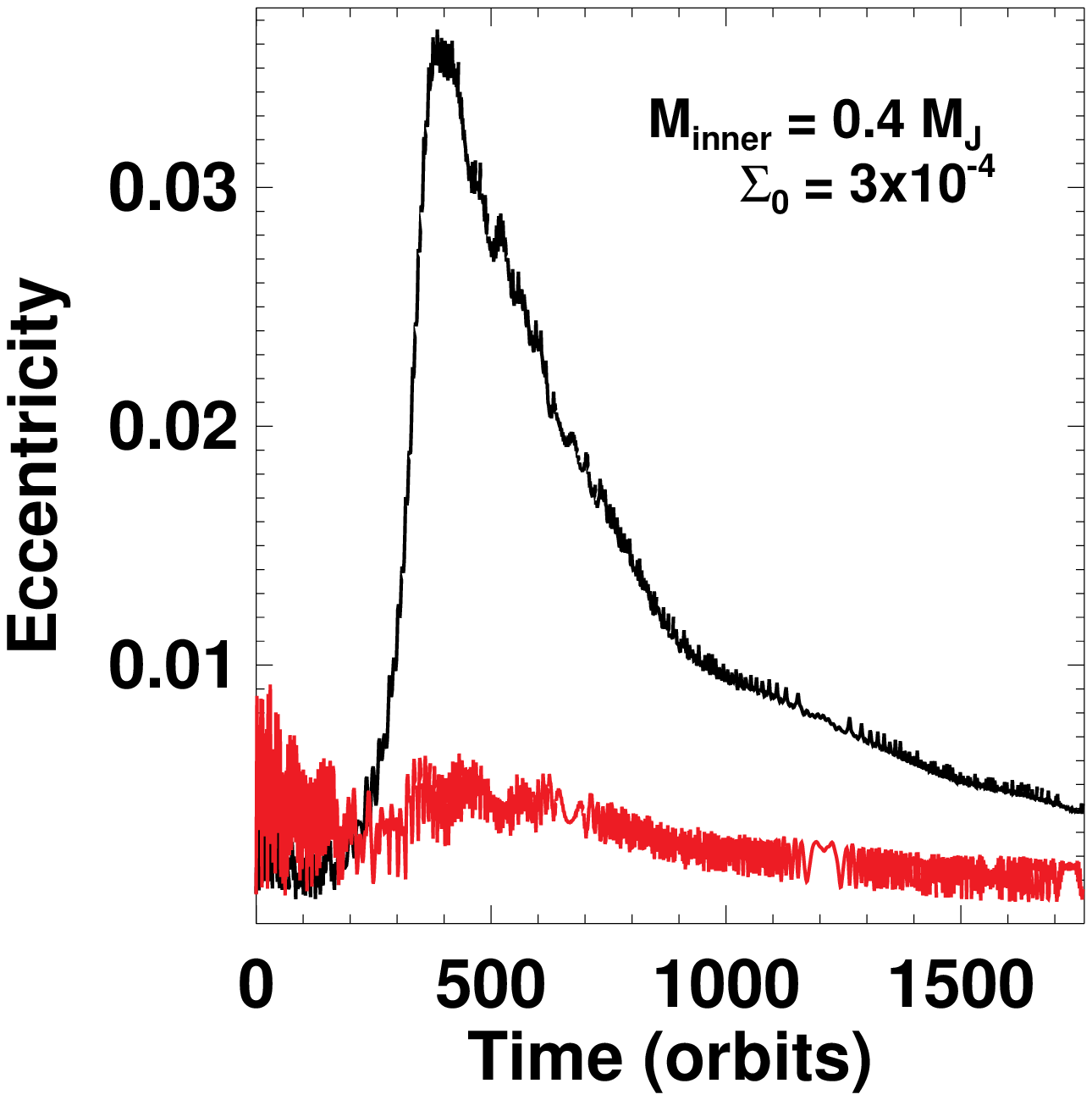}
    \includegraphics{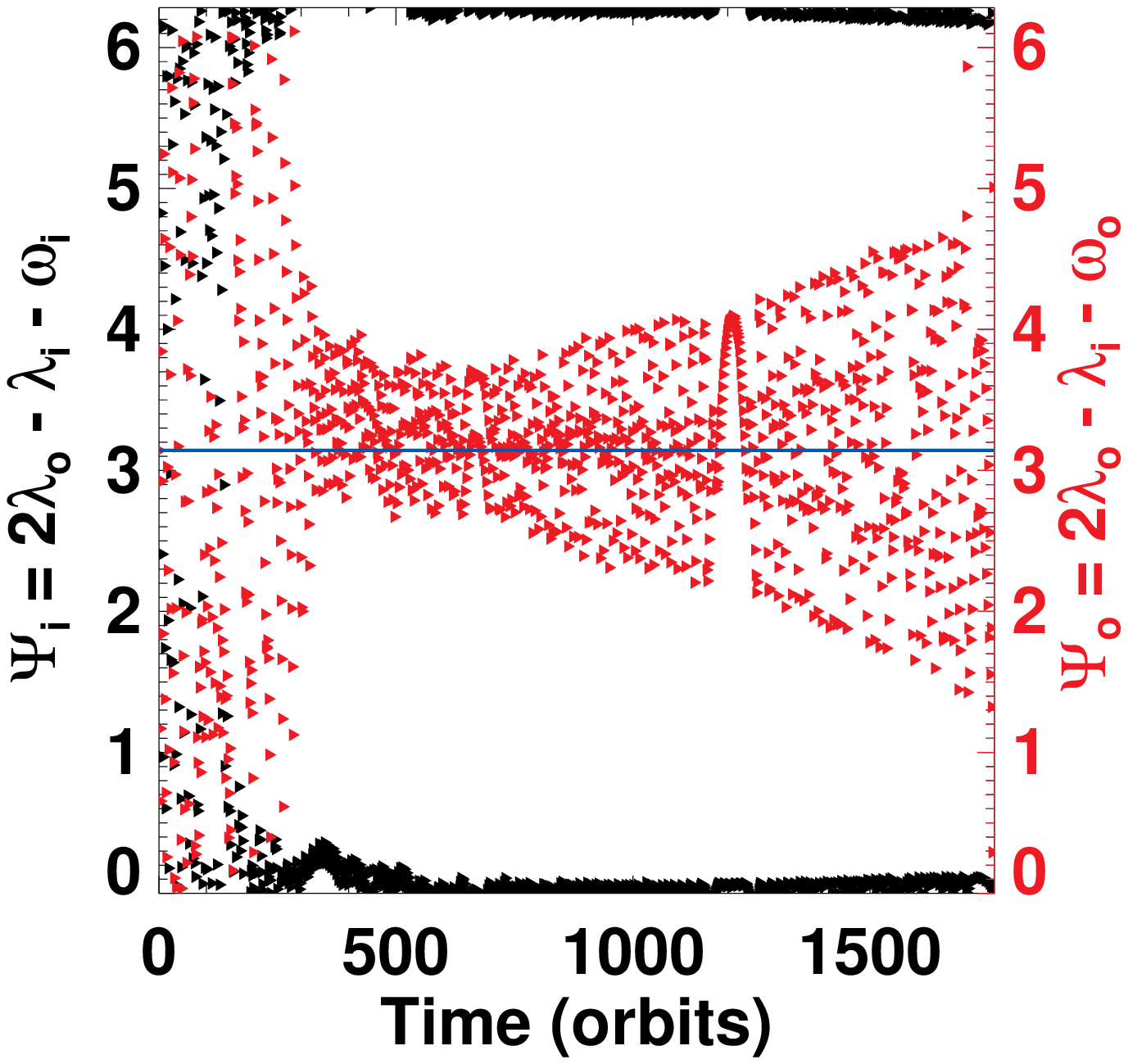}
  } 
  \resizebox{\hsize}{!}
  {
    \includegraphics{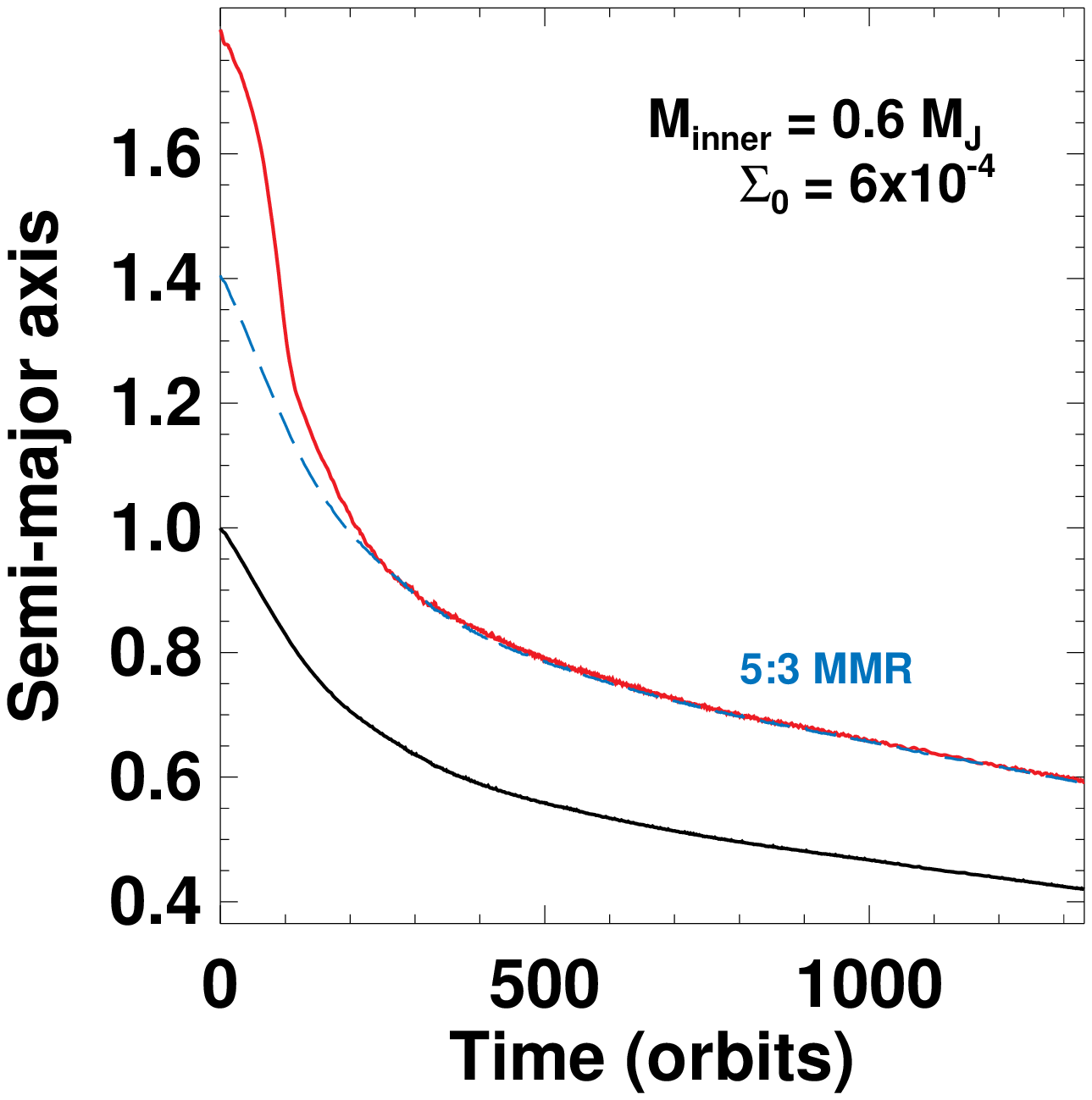}
    \includegraphics{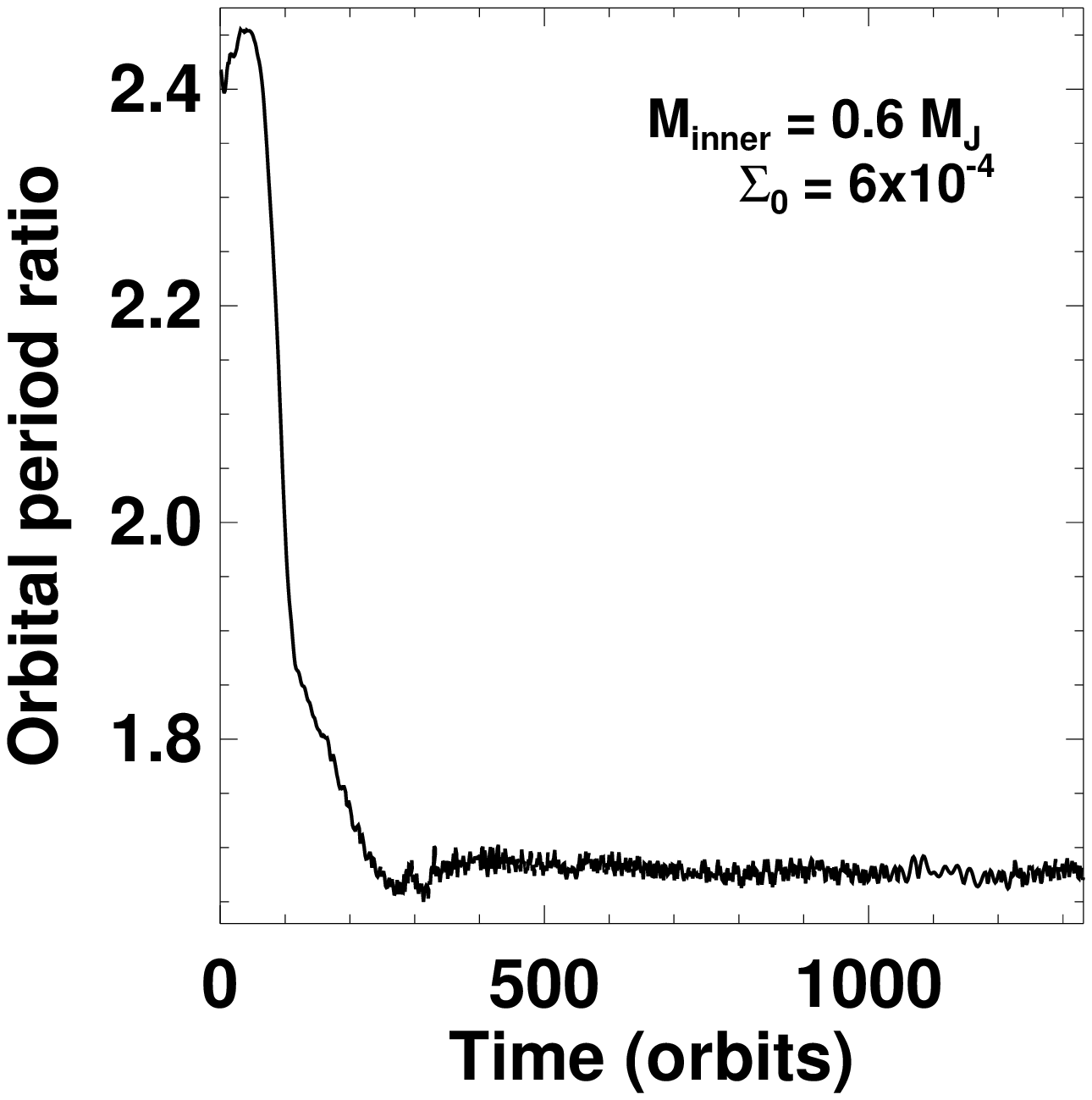}
    \includegraphics{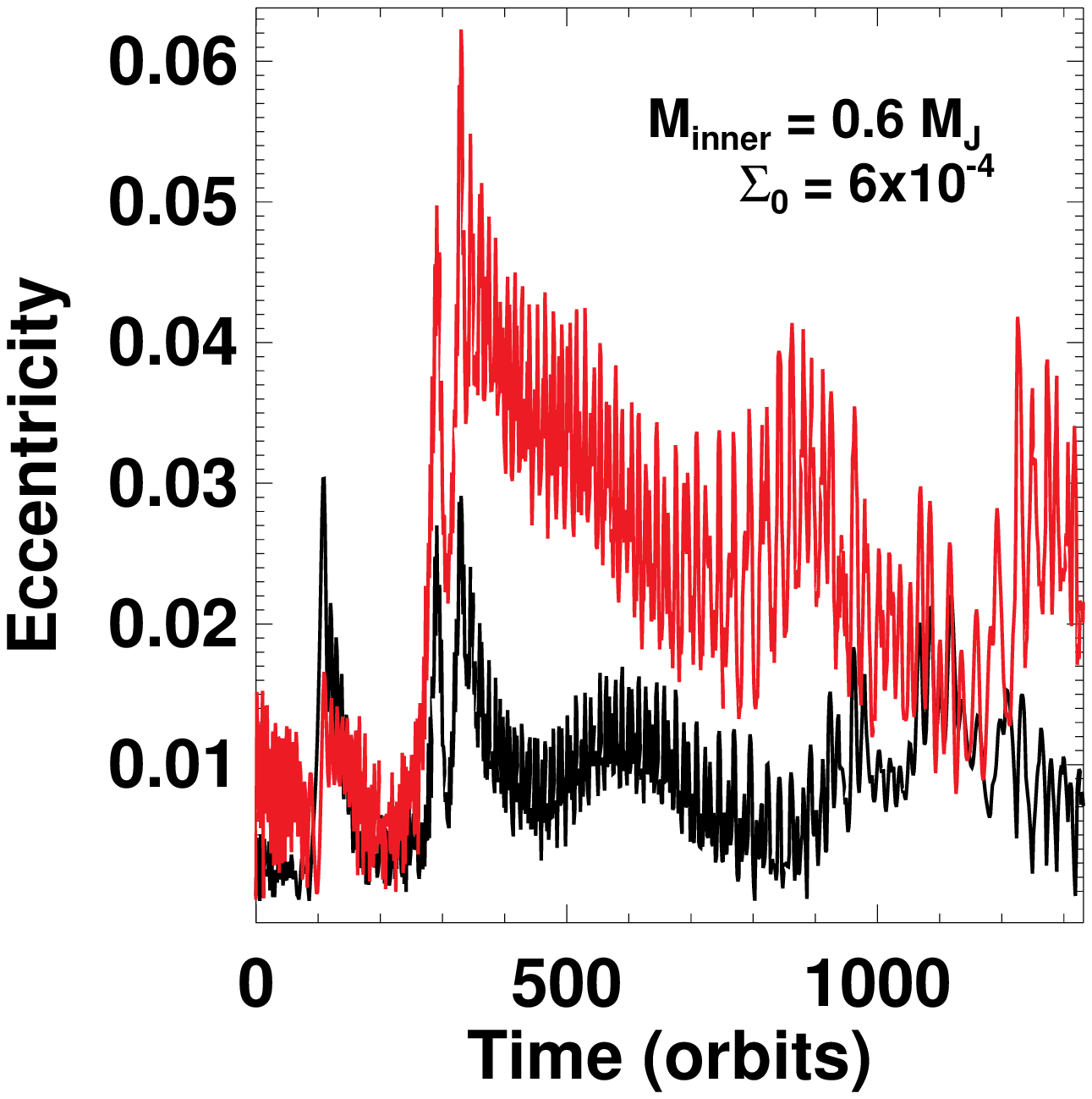}
    \includegraphics{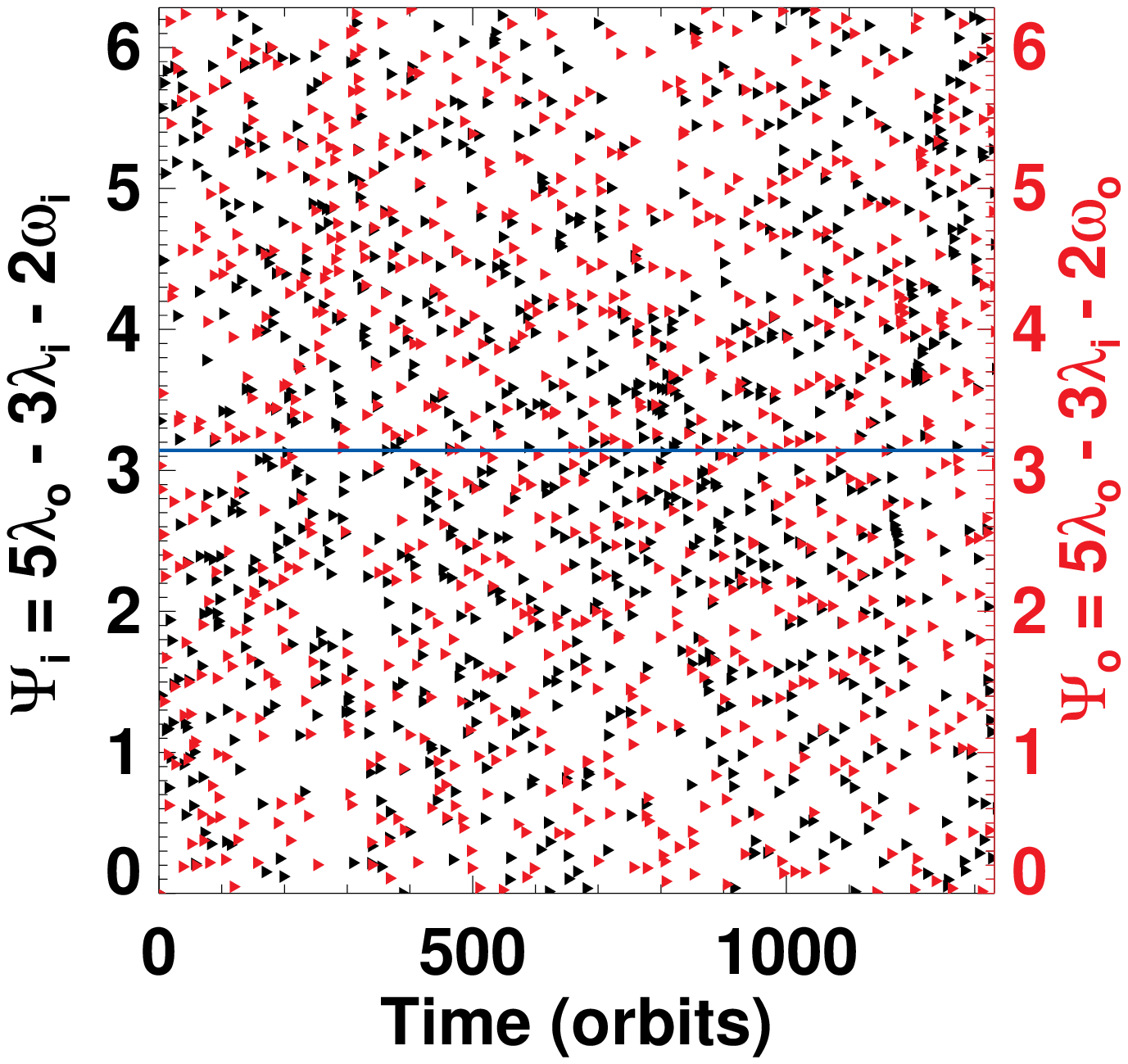}
  }
    \resizebox{\hsize}{!}
  {
    \includegraphics{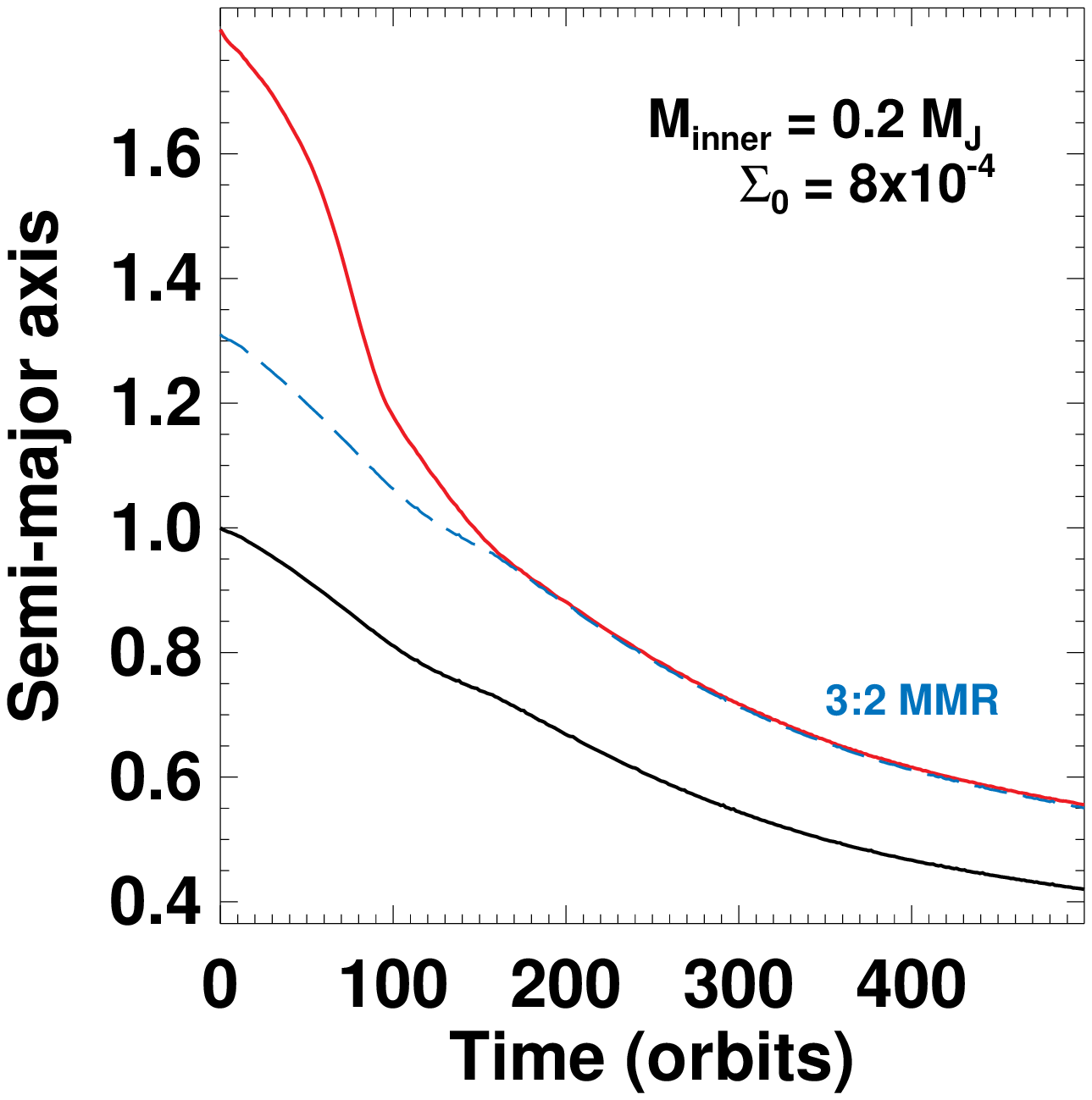}
    \includegraphics{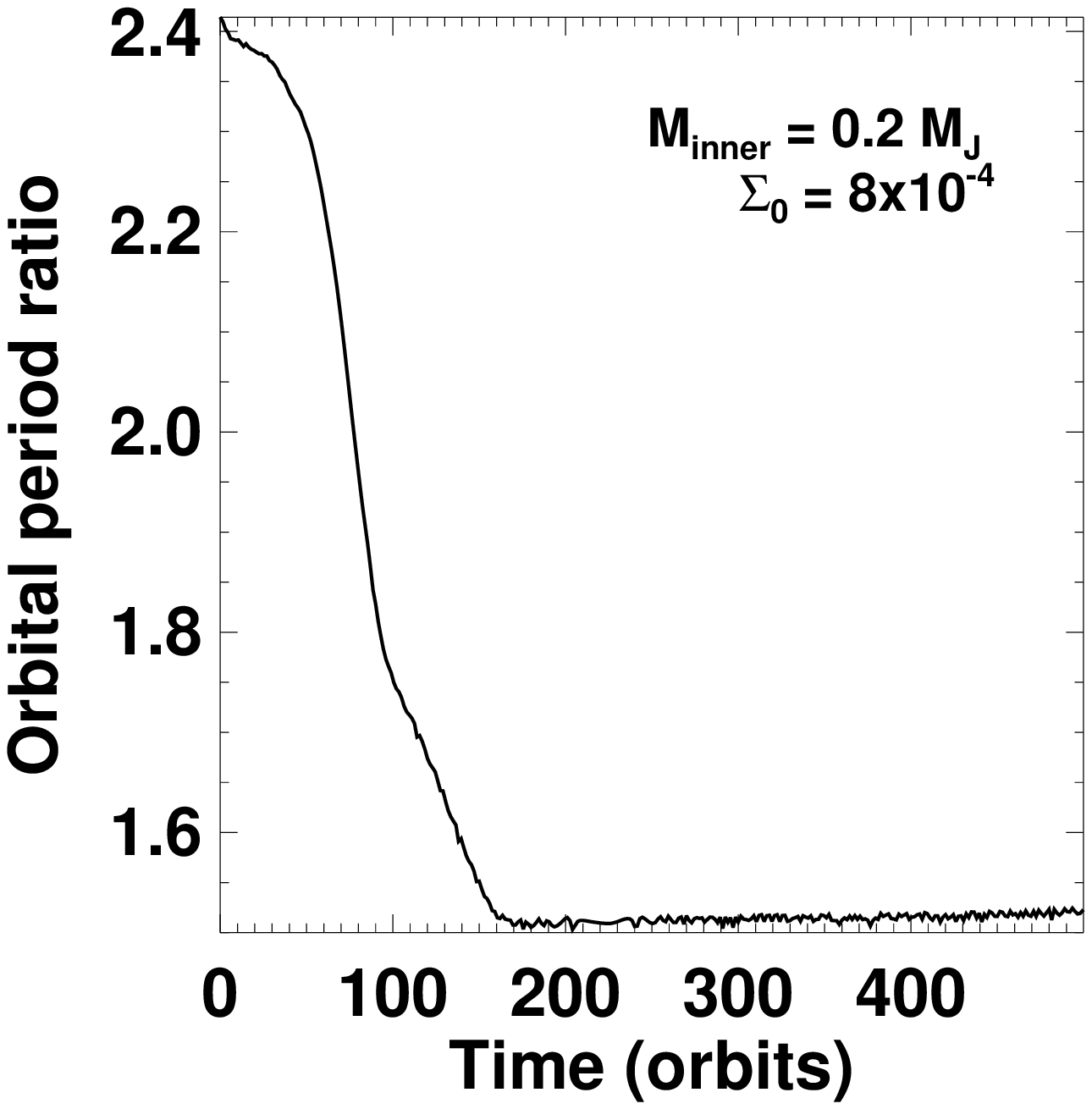}
    \includegraphics{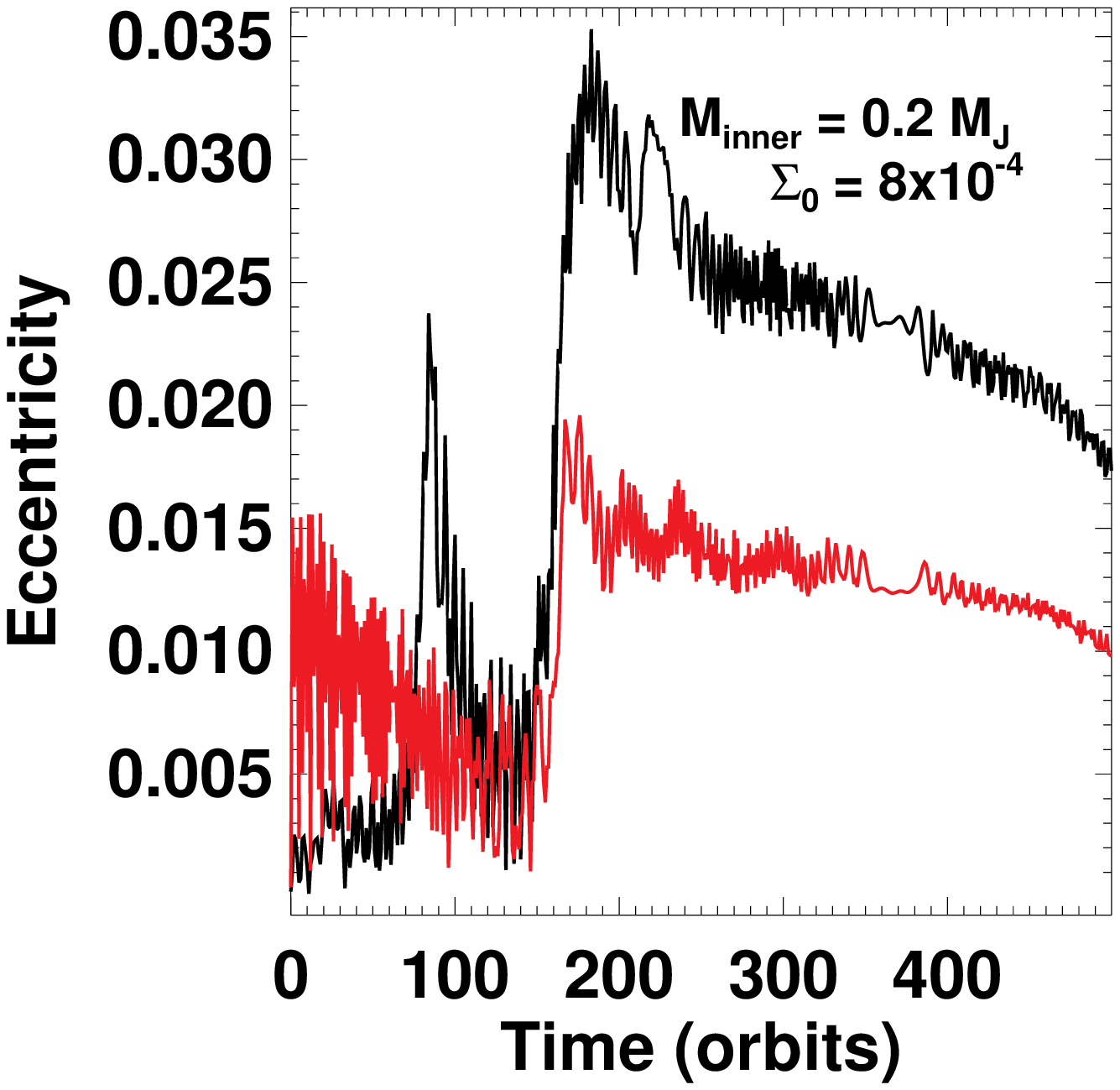}
    \includegraphics{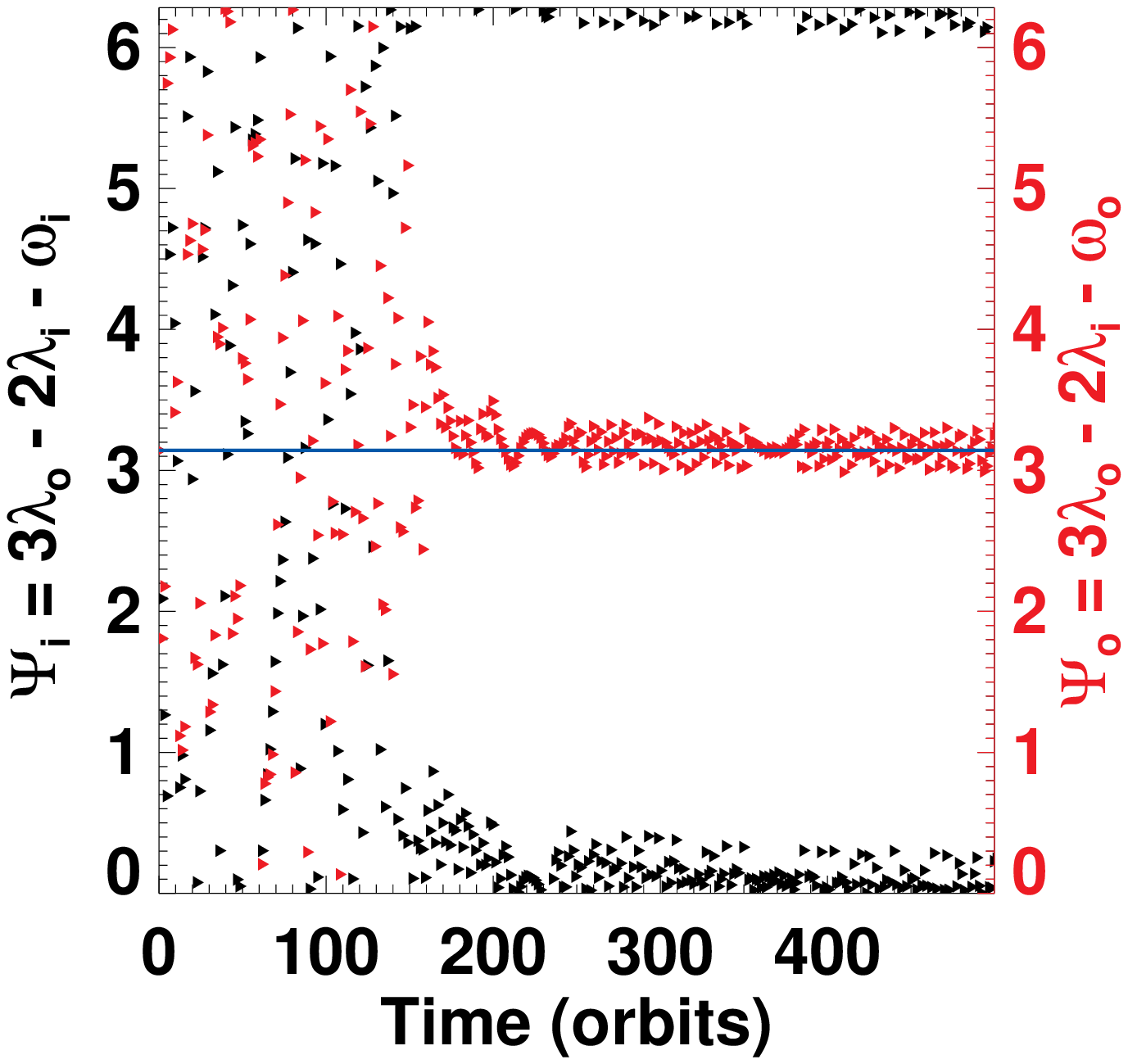}
  } 
  \caption{\label{fig:ex}Results of hydrodynamical simulations of the
    Kepler-46 system. From left to right: time evolution of the
    planets' semi-major axis, orbital period ratio, eccentricity and
    of the resonant angles relevant to the planets' period ratio (the
    horizontal line marks $\pi$). Each row corresponds to a different
    simulation, for which the inner planet's mass ($M_{\rm inner}$)
    and the disk's unperturbed surface density at $r=1$ ($\Sigma_0$)
    are indicated in the top-right corner of the panels. The outer
    planet's mass is 0.37 $M_{\rm J}$. The location of some
    mean-motion resonances (MMR) are displayed by dashed curves in the
    semi-major axes panels. The first row shows an example of
    convergent migration towards the 2:1 MMR followed by rapid
    divergent evolution. The second row shows a case where convergent
    migration stalls with an orbital period ratio near 1.7 (as
    determined by TTV). In the third row, capture into the 3:2 MMR is
    followed by slow divergent evolution.}
\end{figure*}

\subsection{Results of hydrodynamical simulations}
\label{sec:results_hydro}
The free parameters in our simulations are the mass of the inner
planet and the initial surface density of the disk. Both quantities
determine the rate of convergent migration and the evolution of the
orbital eccentricities. We consider three different masses for the
inner planet: $0.2 M_{\rm J}$, $0.4 M_{\rm J}$ and $0.6 M_{\rm J}$
(recall that the mass of the outer planet is fixed at $0.37 M_{\rm
  J}$). For each planet mass, a set of 9 simulations was carried out
varying $\Sigma_0$, the unperturbed surface density at the initial
orbital radius of the inner planet. We took $\Sigma_0$ in the range
$[2\times 10^{-4} , 10^{-3}]$, which corresponds to an initial disk
mass between $3.5\times 10^{-3} M_{\star}$ and $1.8 \times 10^{-2}
M_{\star}$. Typical outcomes of the simulations are described in
Section~\ref{sec:illust}. We find that disk-planets interactions may
lead to significant divergent evolution of the planet orbits, which we
interpret in Section~\ref{sec:divergent}. To obtain evolution of the
planets' orbital elements to a steady state, a simple model for disk
dispersal is included and its results presented in
Section~\ref{sec:disp}.  The robustness of our results to varying some
physical and numerical parameters is discussed in
Section~\ref{sec:robust} of the appendix.

\subsubsection{A few illustrative cases}
\label{sec:illust}
Figure~\ref{fig:ex} displays typical outcomes of our hydrodynamical
simulations of the Kepler-46 system. From left to right in the figure
we show the time evolution of the planets' semi-major axes, orbital
period ratio, eccentricities and resonant angles. Here $\lambda_{\rm
  i,o}$ and $\omega_{\rm i,o}$ denote the mean longitude and longitude
of periastron, with the subscripts i and o referring to the inner and
outer planets, respectively.

The top row shows results with $M_{\rm inner} = 0.4 M_{\rm J}$ and
$\Sigma_0 = 3\times 10^{-4}$. As the planets progressively clear a
partial gap around their orbit, convergent migration does not occur at
a constant rate. In particular, the inward migration of the outer
planet slightly accelerates at early times due to a stage of type III
migration \citep{mp03} before slowing down upon approaching the 2:1
MMR with the inner planet. The planets are locked in the 2:1 MMR from
about 300 orbits onwards, as can be seen from the libration of the
resonant angles associated with this resonance. The resonance capture
increases the inner planet's eccentricity to about 0.035. The increase
in the outer planet's eccentricity is much more modest, reaching about
$5 \times 10^{-3}$, even though the mass ratio of the planets is near
unity. This modest increase in the outer planet's eccentricity arises
from the contribution of the indirect term nearly cancelling the
contribution from the direct term to the term that is first order in
the eccentricities in the development of the disturbing function,
which governs the response of the outer planet at a 2:1 MMR
\citep[see, e.g.,][]{MCC05, PT10}.  Damping of the eccentricities
follows from the gravitational interaction with the background gas
disk. Interestingly, the ratio of orbital periods rises from 2.02 to
2.15 in only 1500 orbits, while the eccentricities are damped. The
physical origin of this divergent evolution is discussed in
Section~\ref{sec:divergent}. Note that resonant coupling is maintained
throughout this divergent evolution, the width of the 2:1 MMR
increasing as the eccentricities decrease.
\begin{figure}
  \centering
  \resizebox{\hsize}{!}
  {
    \includegraphics{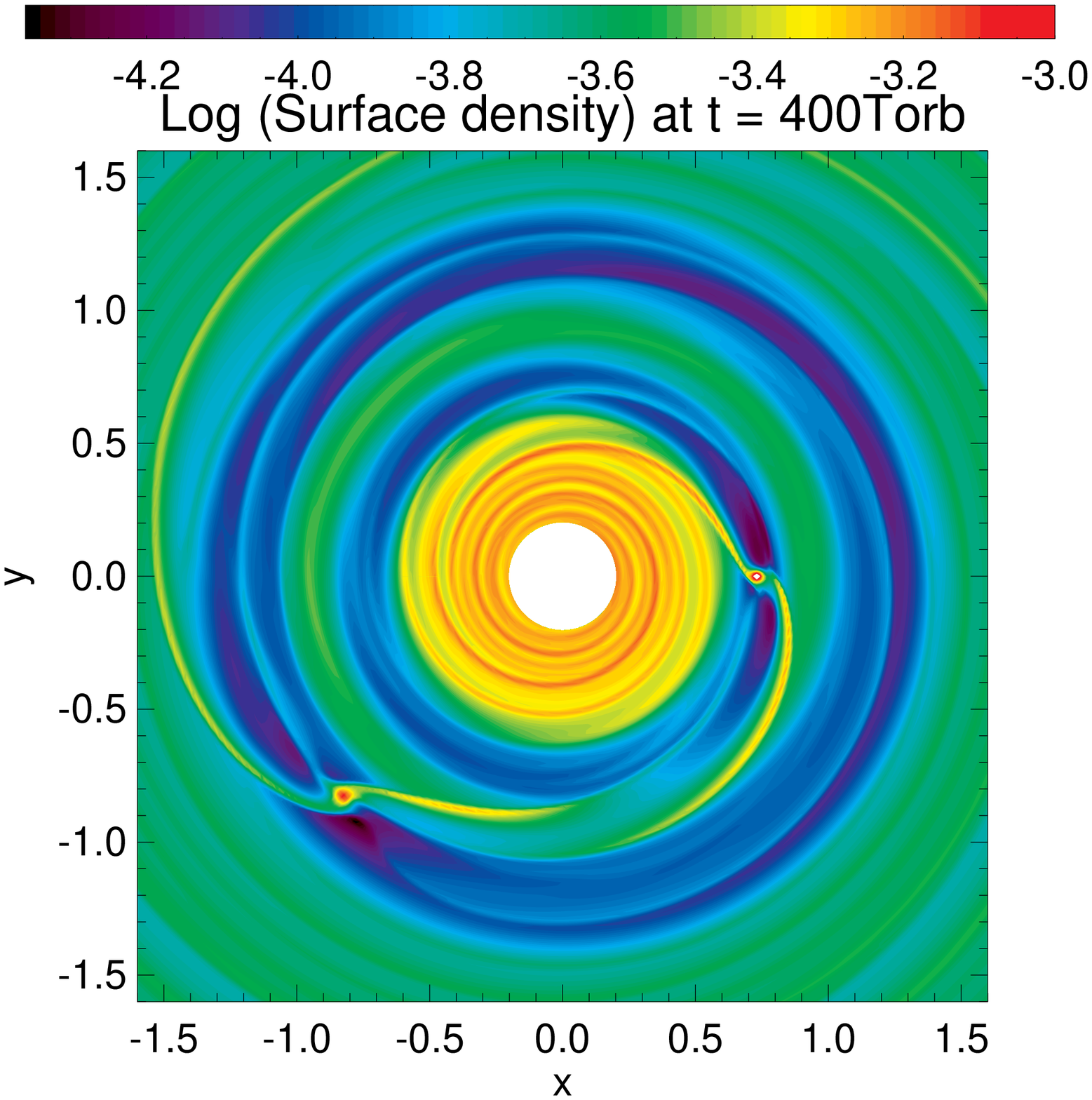}
  }
  \resizebox{\hsize}{!}
  {
    \includegraphics{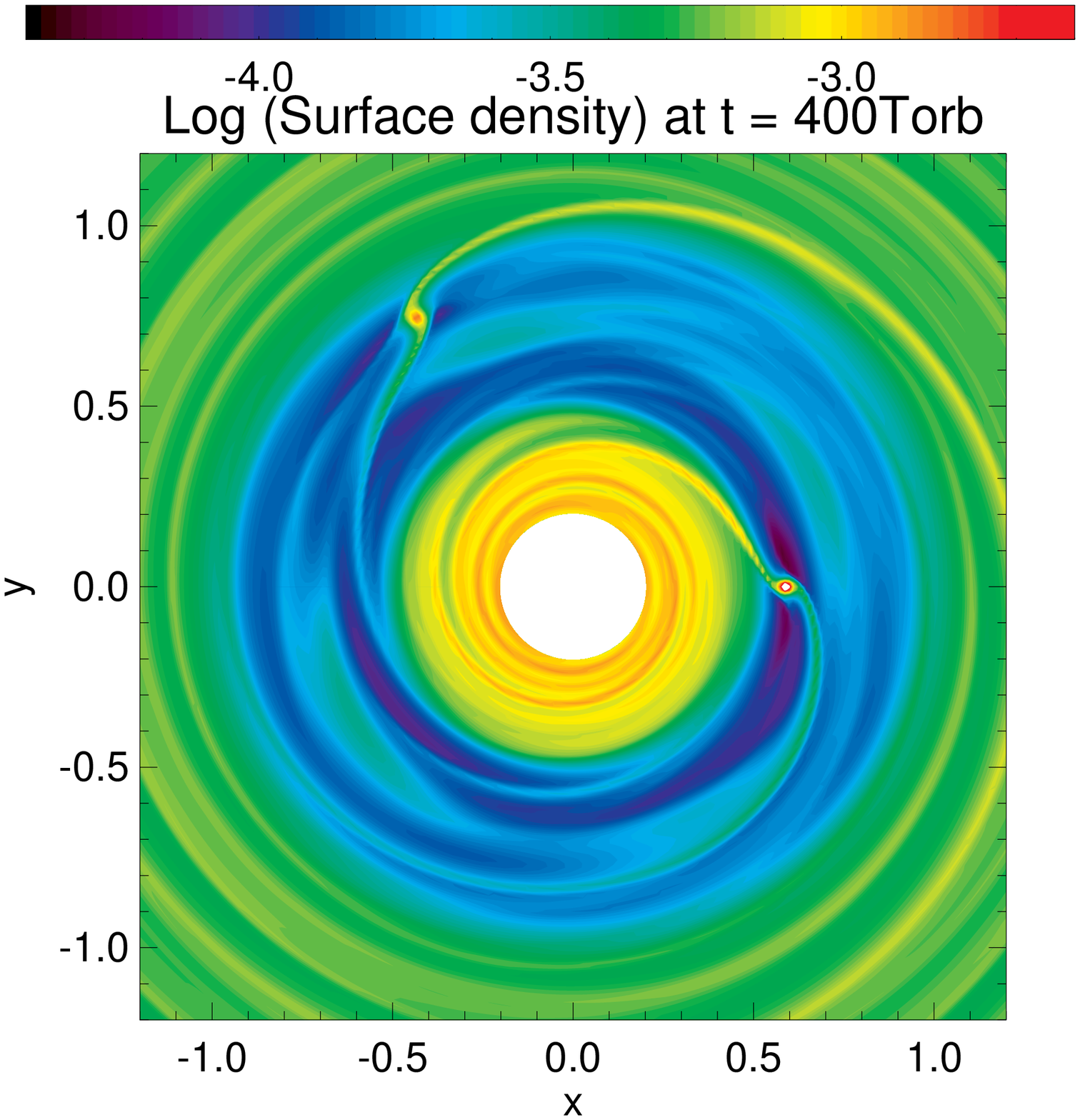}  
  }
  \caption{\label{fig:dens}Disk's surface density at 400 orbits for
    $M_{\rm inner} = 0.4M_{\rm J}$ and $\Sigma_0 = 3\times 10^{-4}$
    (top), and for $M_{\rm inner} = 0.6M_{\rm J}$ and $\Sigma_0 =
    6\times 10^{-4}$ (bottom). The outer planet's mass is $M_{\rm
      outer} = 0.37 M_{\rm J}$.}
\end{figure}

The second row of panels in Figure~\ref{fig:ex} displays results for
$M_{\rm inner} = 0.6 M_{\rm J}$ and $\Sigma_0 = 6\times 10^{-4}$. This
case is particularly interesting as it shows that convergent migration
may stall with an orbital period ratio very close to the observed
value ($\approx 1.7$). The accelerating inward migration of the outer
planet is particularly visible in the first 100 orbits. This runaway
migration is fueled by the dense gas region located between the two
planet gaps. This dense material is continuously funneled beyond the
outer planet's orbit upon embarking on horseshoe U-turns with respect
to the outer planet. The density between the two planet gaps therefore
decreases, and so does the migration rate of the outer
planet. Convergent migration then proceeds at a slower pace and the
planets end up merging their gap, thereby forming a common gap in the
disk. This is illustrated in Figure~\ref{fig:dens}, which compares the
disk's surface density at 400 orbits for the two aforementioned
simulations which have $M_{\rm inner} = 0.4 M_{\rm J}$ and $\Sigma_0 =
3\times 10^{-4}$ (top panel), and $M_{\rm inner} = 0.6 M_{\rm J}$ and
$\Sigma_0 = 6\times 10^{-4}$ (bottom panel). While in the former case
a narrow ring of gas is left between the two planet gaps, the latter
case shows that both planets end up in a common gap.  Back to the
second row of panels in Figure~\ref{fig:ex}, we see that the orbital
period ratio remains approximately stationary from about 400 orbits,
and so do the eccentricities. The time-averaged eccentricity of the
inner planet is close to 0.01; that of the outer planet is $\approx
0.025$. These values are consistent, albeit slight larger than those
inferred from observations (see Table~\ref{tab:tableKOI}). Note from
the critical angles that the planets are not locked in the 5:3 MMR.

Lastly, the third row of panels in Figure~\ref{fig:ex} is for $M_{\rm
  inner} = 0.2 M_{\rm J}$ and $\Sigma_0 = 8\times 10^{-4}$. Here,
convergent migration is rapid enough that the planets approximately
reach the nominal location of their 3:2 MMR (the minimum value of the
period ratio is $1.508$ at 170 orbits). The planet eccentricities
increase to about 0.03 (inner planet) and 0.015 (outer
planet). Subsequent damping of the eccentricities occurs along with a
slight divergent evolution of the planets. The rate of divergent
evolution is much reduced compared to the 2:1 MMR case shown in the
top row of panels in Figure~\ref{fig:ex}. We will analyse this result
further in Section~\ref{sec:divergent}. We also point out that no
steady-state has been reached at the end of the simulation. For this
simulation, as for all simulations of the Kepler-46 system, results
are shown until the inner planet reaches $r \sim 0.4$, below which the
proximity of the grid's inner edge (at $r=0.2$) starts affecting the
planets orbital evolution.

\begin{figure}
  \centering
  \resizebox{\hsize}{!}
  {
    \includegraphics{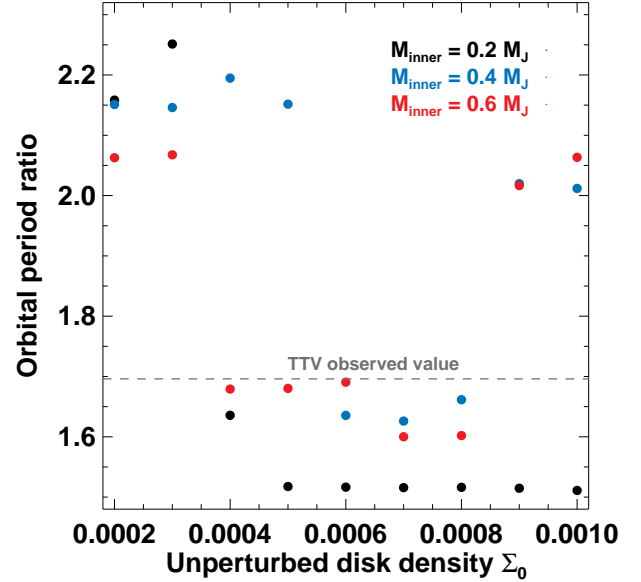}
  }
  \caption{\label{fig:finalopr}Summary of the results of
    hydrodynamical simulations for Kepler-46: ratio of orbital periods
    versus the unperturbed disk's surface density at the initial
    location of the inner planet ($\Sigma_0$). Results are shown for
    the three masses assumed for the inner planet (Kepler-46b).}
\end{figure}
The final period ratio obtained in all our simulations is displayed in
Figure~\ref{fig:finalopr} as a function of $\Sigma_0$ for the three
masses assumed for Kepler-46b. By final period ratio, we mean either
its value obtained at the end of the simulation (after between 1500
and 2000 planet orbits), or that obtained when the inner planet's
orbital radius falls below $r=0.4$. Figure~\ref{fig:finalopr}
highlights the three major outcomes of our simulations: (i) capture in
2:1 MMR generally followed by rapid divergent evolution, (ii) capture
in 3:2 MMR followed by slow divergent evolution, and (iii) convergent
migration stalling with a period ratio between 1.6 and 1.7, depending
on the disk's density distribution inside the planets' common
gap. Several simulations can reproduce the observed period ratio of
Kepler-46b and Kepler-46c. The best agreement with observations is
obtained for $M_{\rm inner} = 0.6 M_{\rm J}$, our upper mass
value. This mass would imply a mean density for Kepler-46b about 20\%
larger than Jupiter's (that is, about 1.6 g cm$^{-3}$).

\subsubsection{Disk-driven repulsion of a planet pair}
\label{sec:divergent}
\begin{figure*}
  \centering
  \resizebox{\hsize}{!}
  {
    \includegraphics{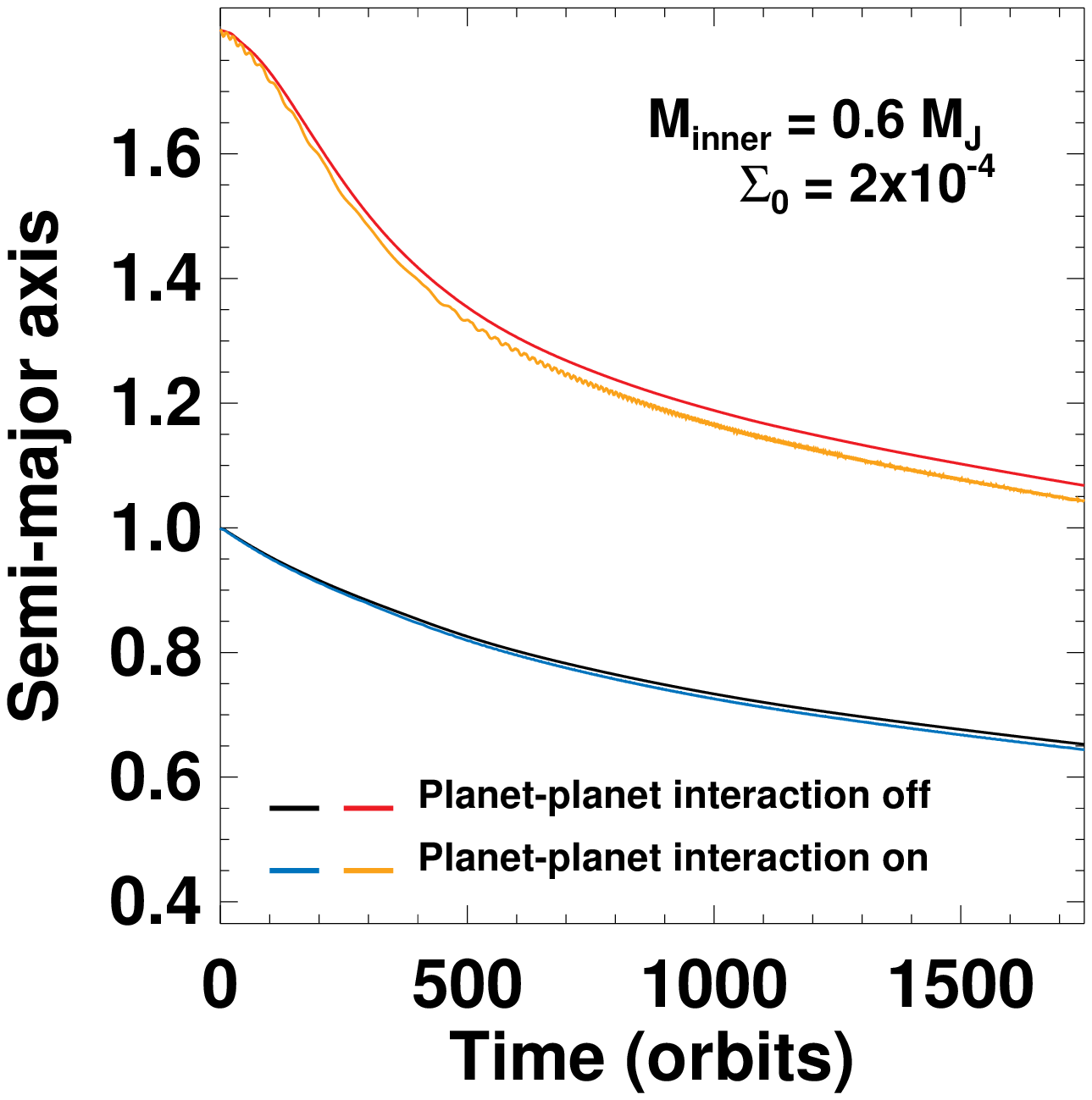}
    \includegraphics{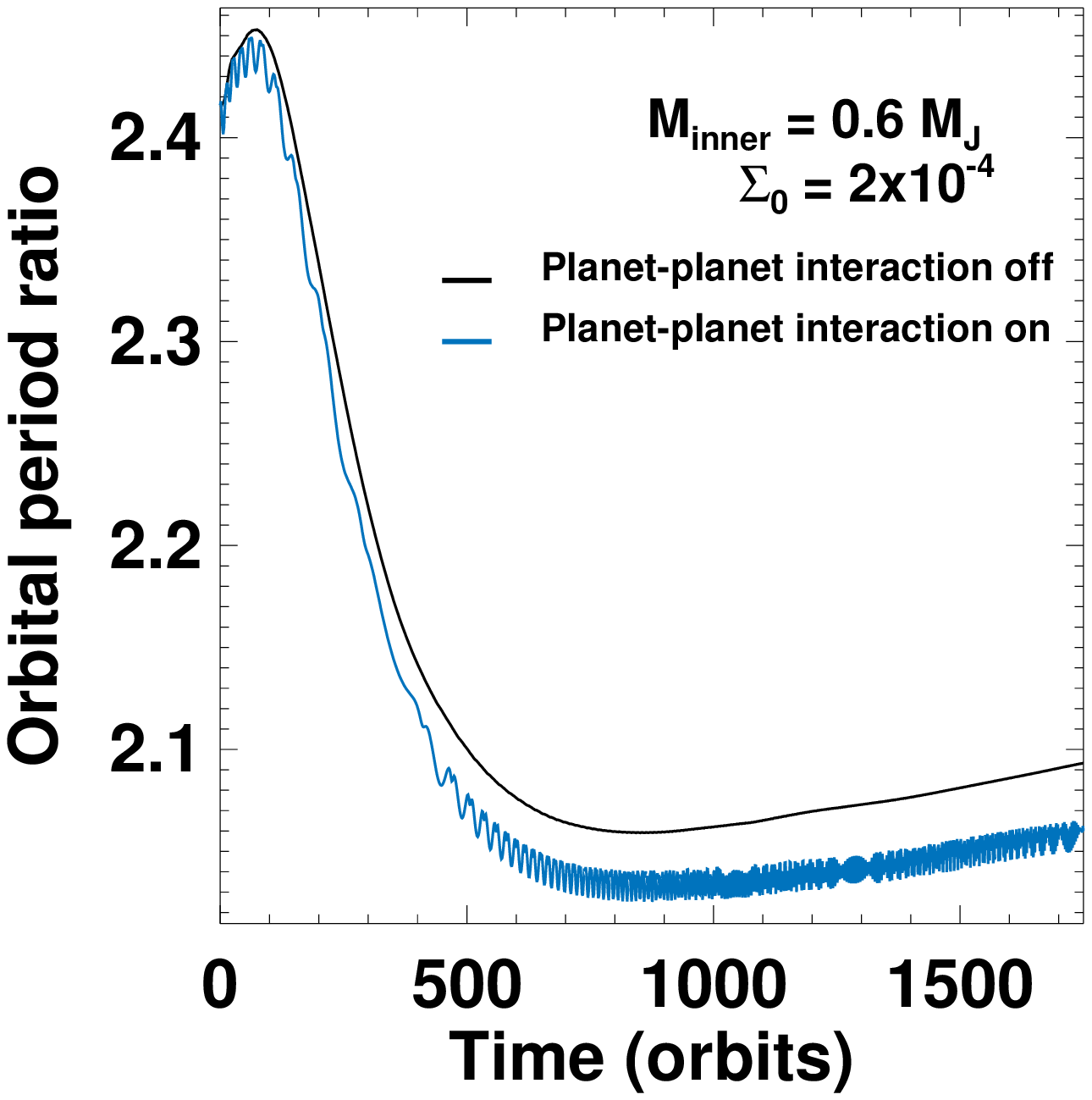}
    \includegraphics{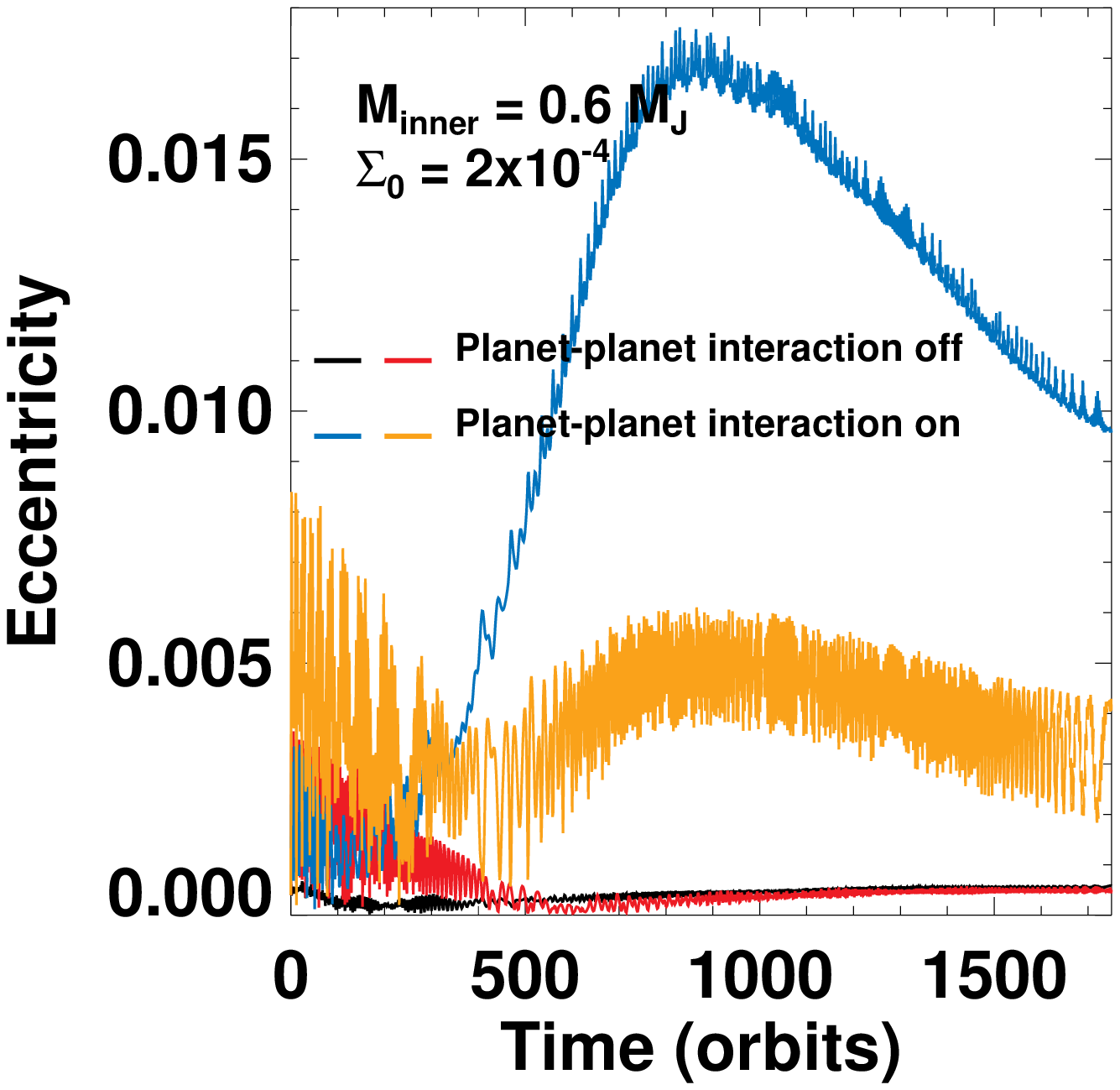}
    \includegraphics{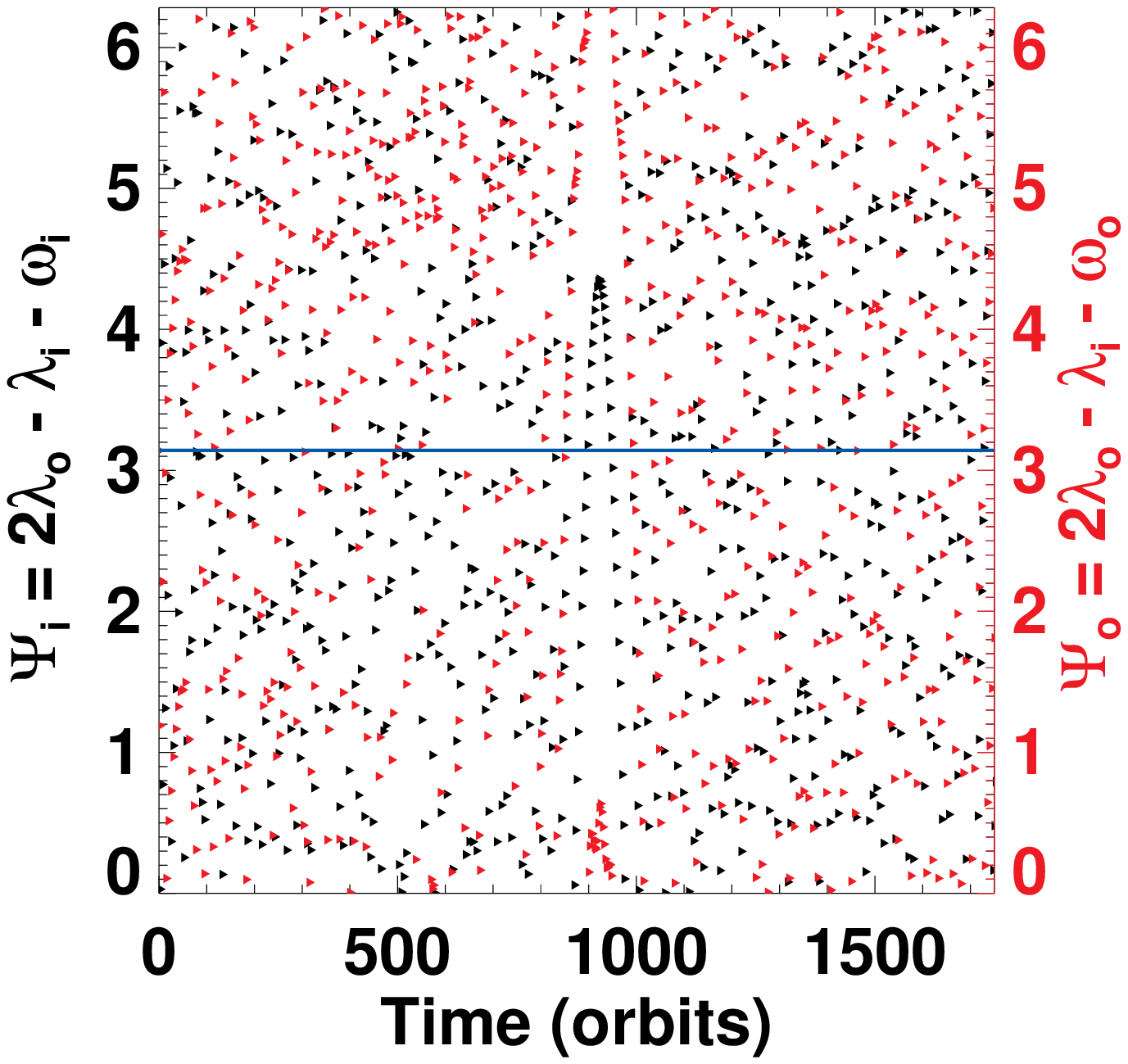}
  }
  \caption{\label{fig:fono}Time evolution of the semi-major axes, the
    orbital period ratio, the eccentricities and the $2:1$ resonant
    angles for $M_{\rm inner}=0.6 M_{\rm J}$ and $\Sigma_0 = 2\times
    10^{-4}$. Two situations are compared: (i) the planets do not feel
    each other's gravity (planet-planet interaction off) and (ii) the
    planets feel each other's gravity (planet-planet interaction on;
    our fiducial case). The resonant angles are those obtained with
    planet-planet interaction off.}
\end{figure*}
The results of hydrodynamical simulations in Section~\ref{sec:illust}
show that the orbital period ratio of two partial gap-opening planets
may increase from a near resonant value. This divergent evolution is
reminiscent of the late evolution of short-period planet pairs through
star-planet tidal interactions. In that case, tidal circularization
causes a slow divergent evolution of the orbits \citep{PT10, Papa11,
  LW12, Baty12}, a mechanism known as tidally-driven resonant
repulsion \citep{LW12}. It is based on energy dissipation of the
planet pair while its total angular momentum remains constant
\citep{Papa11}.

Do we have an analogous resonant repulsion mechanism due to
disk-planets interactions?  To address this question, it is
instructive to first look at the evolution of two embedded planets
that do not interact gravitationally. In Figure~\ref{fig:fono} we
compare the results of two hydrodynamical simulations: one in which
the gravitational interaction between the planets is switched off
(both the direct and the indirect interaction terms are discarded) and
another where the planet-planet interaction is fully accounted for.
The inner planet's mass is $M_{\rm inner}=0.6 M_{\rm J}$ and the
initial disk surface density at $r=1$ is $\Sigma_0 = 2\times
10^{-4}$. The other disk and planet parameters are those described in
Section~\ref{sec:setup}. When planet-planet interactions are on,
capture into the 2:1 MMR is followed by rapid divergent evolution and
damping of the eccentricities. This case is similar to that presented
in the top row of panels in Figure~\ref{fig:ex}.  Quite surprisingly,
we see that a very similar divergent evolution is also obtained when
planet-planet interactions are switched off, while the planets are not
resonantly coupled (the resonant angles do not librate) and their
eccentricities take vanishingly small stationary values.  This
comparison highlights that, when disk-planet interactions are
considered, eccentricity damping is not necessarily responsible for
the planets divergent evolution. The divergent evolution of a planet
pair may instead be driven by changes to the disk-planet interaction
brought about through the proximity of the planets to each other, such
as the development of an interaction between each planet and the wake
of its companion.  We argue below that the damping of the
eccentricities can actually be a consequence of divergent evolution 
driven by wake-planet interaction.

Before describing the process of wake-planet interaction in more
detail, we briefly report that, in some other simulations, not
illustrated here, with planet-planet interactions switched off, a tiny
increase in the orbital eccentricities is observed when the orbital
period ratio takes near resonant values. Eccentricities may reach a
few $\times 10^{-3}$ before being damped by the disk. Although this is
a small effect, it suggests that planets could be weakly resonantly
coupled through interaction with the wakes and/or the circumplanetary
disk of their companions. In our simulations, the mass of the
circumplanetary disks does not exceed one percent of the planets mass.
\\
\par\noindent\emph{Wake-planet interactions---} 
The results of simulations shown in Figure~\ref{fig:fono} make clear
that disk-driven divergent evolution of a planet pair is not
necessarily driven by eccentricity damping and it may continue to
operate even when the eccentricities are very small.  We here indicate
how the interaction between a planet and the wake of its companion
could act as a driving mechanism for divergent evolution.  Consider
for instance the inner planet.  Its outer wake carries positive fluxes
of angular momentum and energy.  A fraction of this angular momentum
flux can be deposited in the coorbital region of the outer planet
through the dissipation of shocks, and then transferred to the outer
planet via an effective positive corotation torque
\citep[see][]{PPS12}. Associated with the angular momentum flux
$F_{\rm J}$ deposited in the coorbital region of the outer planet is
an energy flux $F_{\rm E} = n_{\rm i} F_{\rm J}$, where $n_{\rm i}$
denotes the inner planet's mean motion (or angular velocity). From
this energy flux, $n_{\rm o} F_{\rm J}$ is used to keep the outer
planet and its coorbital region nearly circular (see
Figure~\ref{fig:fono}; here, $n_{\rm o}$ is the outer planet's mean
motion). The remaining energy flux, $(n_{\rm i}-n_{\rm o}) F_{\rm J}$,
is dissipated in the outer planet's coorbital region.  Reciprocally, a
fraction of the negative inward flux of angular momentum carried by
the inner wake of the outer planet can be deposited in the inner
planet's coorbital region and subsequently transmitted to the inner
planet. Again, this exchange is associated with energy dissipation in
the disk at a rate equal to the (absolute value of) the product of the
flux of angular momentum transmitted to the inner planet by the outer
one and the difference in the planets angular velocities. This
dissipation occurs, as is usual in disk-planet interactions, through
the tidally excited density waves steepening into shocks
\citep[e.g.,][]{Goodman:2001ul}.  Numerically, shocks are dealt with
through a combination of numerical diffusion and the application of an
artificial viscosity. As a locally isothermal equation of state is
adopted in our simulations, this dissipated energy is effectively lost
to the system through cooling.  While the total orbital energy of the
two planets is dissipated due to the wake-planet interactions
described above, their total angular momentum remains unaltered. As
studied in the context of star-planet tidal interactions, energy
dissipation at constant angular momentum leads to the divergent
evolution of a planet pair \citep{Papa11, LW12, Baty12}.

We now derive an equation satisfied by the period ratio of the two
planets as a direct consequence of considering the evolution of their
total orbital energy and angular momentum. We denote by $a_{\rm i}$,
$e_{\rm i}$ and $M_{\rm i}$ the semi-major axis, eccentricity and mass
of the inner planet (with the subscript changed to o denoting the same
quantities for the outer planet). The total angular momentum of the
two planets is
$$
J = J_{\rm i} + J_{\rm o} = M_{\rm i} \sqrt{GM_{\star}a_{\rm i} (1-e_{\rm i}^2)} + M_{\rm o} \sqrt{GM_{\star}a_{\rm o} (1-e_{\rm o}^2)},
$$
which gives, correct to second order in the eccentricities, 
$$
\dot{J} \equiv \frac{dJ}{dt} = 
J_{\rm i} \left( \frac{\dot{a}_{\rm i}}{2a_{\rm i}} - e_{\rm i} \dot{e}_{\rm i}\right)
+
J_{\rm o} \left( \frac{\dot{a}_{\rm o}}{2a_{\rm o}} - e_{\rm o} \dot{e}_{\rm o}\right).
$$
We are looking for changes in the planets' semi-major axes and
eccentricities that occur on similar timescales, we may therefore
discard the eccentricity terms in the expressions for $\dot{J},$
$J_{\rm i}$ and $J_{\rm o} $ . Denoting by $\Gamma_{\rm i}$ and
$\Gamma_{\rm o}$ the torques exerted by the disk on the inner and
outer planets, respectively, conservation of angular momentum reads
\begin{equation}
\dot{J} = J_{\rm i} \frac{\dot{a}_{\rm i}}{2a_{\rm i}} + J_{\rm o} \frac{\dot{a}_{\rm o}}{2a_{\rm o}} = \Gamma_{\rm i} + \Gamma_{\rm o}.
\label{eq_J}
\end{equation}
The total orbital energy is given by
$$
E = -\frac{GM_{\star}M_{\rm i}}{2a_{\rm i}} -\frac{GM_{\star}M_{\rm o}}{2a_{\rm o}},
$$
and conservation of energy reads
\begin{eqnarray}
  \dot{E} & = & \frac{GM_{\star}M_{\rm i}}{2a_{\rm i}^2} \dot{a}_{\rm i} + \frac{GM_{\star}M_{\rm o}}{2a_{\rm o}^2} \dot{a}_{\rm o}\nonumber \\
  & = & n_{\rm i} \Gamma_{\rm i} + n_{\rm o} \Gamma_{\rm o} 
  - \frac{GM_{\star}M_{\rm i} e_{\rm i}^2}{a_{\rm i} \tau_{\rm c,i}} 
  - \frac{GM_{\star}M_{\rm o} e_{\rm o}^2}{a_{\rm o} \tau_{\rm c,o}}
  - |\dot{E}|_{\rm wake}, 
\label{eq_E}
\end{eqnarray}
where $n_{\rm i,o} = \sqrt{GM_{\star} / a^3_{\rm i,o}}$ the planets
mean motion. Here the right hand side of Eq.~(\ref{eq_E}) contains
three contributions.  The first arises from the energy dissipation in
the disk resulting from the action of the torques $\Gamma_{\rm i,o}.$
As these are negative this energy is removed from the orbital motion.
The second originates from orbital circularization due to disk-planet
interactions.  The quantities $\tau_{\rm c,i}$ and $\tau_{\rm c,o}$
are the circularization times for the planets that arise from the disk
torque and energy dissipation acting on each planet through the
effects of its own wake \citep{nelson02,papszusz2005}.  The third and
final contribution $-|\dot{E}|_{\rm wake}$ stems from the energy
dissipation due to wake-planet interactions. We remark that wake
dissipation in the coorbital region of the outer planet contributes
$(n_{\rm i}-n_{\rm o}) |F_{\rm J}|$ to $|{\dot E}|_{\rm wake}$, in
addition to a corresponding contribution arising from the coorbital
region of the inner planet, as indicated above. Note that terms of
order $e_{\rm i,o}^2$ are retained above even though the
eccentricities are small as it is assumed that $\tau_{\rm i,o}$ are
small enough so that dissipation through circularization could be as
important as, or even dominate the other effects under some
circumstances.

Combining Eqs.~(\ref{eq_J}) and~(\ref{eq_E}), we find
\begin{eqnarray}
\lefteqn{ 
\frac{1}{3}\frac{n_{\rm o}}{n_{\rm i}}\frac{d}{dt}\left( \frac{n_{\rm i}}{n_{\rm o}} \right) = 
\left( \frac{\Gamma_{\rm o}}{J_{\rm o}} - \frac{\Gamma_{\rm i}}{J_{\rm i}} \right)
}\nonumber\\
&&
+
\frac{J_{\rm i} + J_{\rm o}}{J_{\rm i} J_{\rm o} (n_{\rm i} - n_{\rm o})} 
\left( 
n_{\rm i} J_{\rm i} \frac{e_{\rm i}^2}{\tau_{\rm c,i}} 
+ n_{\rm o} J_{\rm o} \frac{e_{\rm o}^2}{\tau_{\rm c,o}}
+ |\dot{E}|_{\rm wake}
\right).
\label{eq_pr}
\end{eqnarray}
When wake-planet interactions are negligible ($|\dot{E}|_{\rm
  wake}=0$), Eq.~(\ref{eq_pr}) shows that, far from resonance, where
the eccentricities are very small, the planet's period ratio $n_{\rm
  i} / n_{\rm o}$ decreases as a consequence of convergent migration,
provided that $\Gamma_{\rm o}/J_{\rm o} < \Gamma_{\rm i}/J_{\rm i}.$
As expected, the latter condition is that the inward migration rate of
the outer planet should exceed the inward migration rate of the inner
planet. As the eccentricities increase as resonance is approached, a
steady state may be reached in which the planets period ratio and
eccentricities take stationary values. If the eccentricity damping
timescales are sufficiently short, disk-driven circularization of the
orbits can reverse convergent migration, that is the planets period
ratio can increase.

Eq.~(\ref{eq_pr}) highlights that wake-planet interactions may also
lead to divergent evolution of a planet pair. Divergent evolution
driven by wake-planet interactions does not require the planets to be
resonantly coupled, as already inferred from the simulations shown in
Figure~\ref{fig:fono}. Similar divergent evolution of a non-resonant
planet pair embedded in a disk, comprising a Jupiter and a
super-Earth, was reported by \citet{PPS12}. When divergent evolution
is primarily due to wake-planet interactions, as the period ratio of
the planet pair increases away from a resonant value, the distance to
resonance increases, and the forced eccentricities therefore decrease
\citep[e.g.,][]{Papa11}. It implies that orbital circularization is a
natural consequence of divergent evolution driven by wake-planet
interactions. This is actually a major difference between the
divergent evolutions driven by wake-planet interactions and by orbital
circularization (through the disk or stellar tides): eccentricity
damping is the consequence of the former mechanism, while it is the
cause of the latter. Note that, with both mechanisms, the divergent
evolution of a resonant planet pair maintains the libration of the
resonant angles by damping the eccentricities.

We call {\it disk-driven resonant repulsion} the divergent evolution
of a resonant planet pair embedded in a disk, following the
terminology introduced by \cite{LW12} for the divergent evolution of a
resonant planet pair due to star-planet tidal interactions. We stress
that disk-driven resonant repulsion can have two origins: (i) disk
circularization of the orbits and (ii) wake-planet interactions. In
other words, disk-driven repulsion comprises (i) {\it
  circularization-driven repulsion} and (ii) {\it wake-driven
  repulsion}. The results of hydrodynamical simulations shown above
indicate that, for the disk and planet parameters that we have
considered, wake-planet repulsion dominates circularization-driven
repulsion. This point will be further illustrated in
Sect.~\ref{sec:nbody_method}.
\\
\par\noindent\emph{When is wake-driven repulsion important?---} 
We have seen above that the period ratio of a planet pair embedded in
a disk may increase due to angular momentum transfer between the
planets through their wakes. We further discuss in this paragraph
under what conditions wake-planet interactions may lead to efficient
divergent evolution of a planet pair. We separate the planet pair into
a donor and a recipient. The bigger the donor's mass, the stronger its
wake, and the easier for the wake to deposit angular momentum in the
coorbital region of the recipient planet through the dissipation of
shocks.  This requires the dimensionless parameter $q/h^3$ of the
donor planet to typically exceed unity ($q$ is the planet-to-star mass
ratio and $h$ the disk's aspect ratio). The donor planet can thus be a
partial or a deep gap-opening planet. The angular momentum that is
deposited by the donor's wake is (at least partly) transferred to the
recipient planet through an effective corotation torque
\citep{PPS12}. If the recipient's mass is too big, the density in its
coorbital region will be too low for this effective corotation torque
to matter. In other words, the recipient could be a partial
gap-opening planet or even a type-I migrating planet. These
considerations show why divergent evolution is neither expected for
two type-I migrating planets (the donor's wake being too weak), nor
for two type-II migrating planets (the recipient's gap being too
deep). We have checked this with hydrodynamical simulations, not
illustrated here.

The above considerations also help understand the different rates of
divergent evolution obtained in Figure~\ref{fig:ex}. The fact that
both planets open a partial gap is a favorable situation for
wake-planet interactions to play a significant role. In the top row of
panels, where the planets become locked in the 2:1 MMR, the
high-density region between the two gaps (see upper panel in
Figure~\ref{fig:dens}) causes significant deposition of angular
momentum in the planets coorbital region and therefore very efficient
divergent evolution. In contrast, in the bottom row of panels, where
the planets merge their gap and become locked in the 3:2 MMR, the
moderate disk density left between the planets can only trigger slow
divergent evolution. In the middle row of panels, where the period
ratio stalls near 1.7, a steady state is reached with background
convergent migration balanced by wake-planet divergent evolution.
\begin{figure*}
  \centering
  \resizebox{0.85\hsize}{!}
  {
    \includegraphics{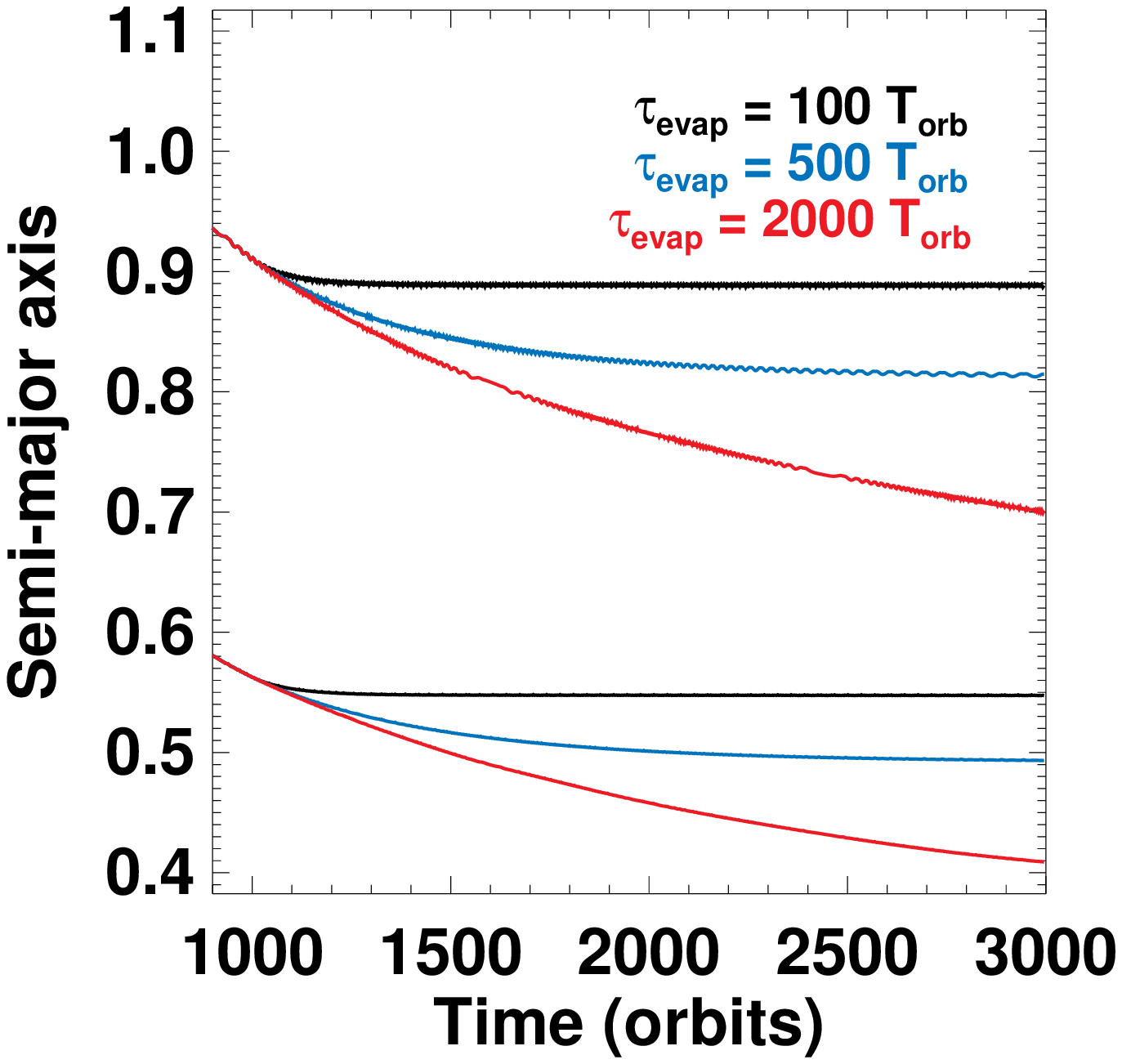}
    \includegraphics{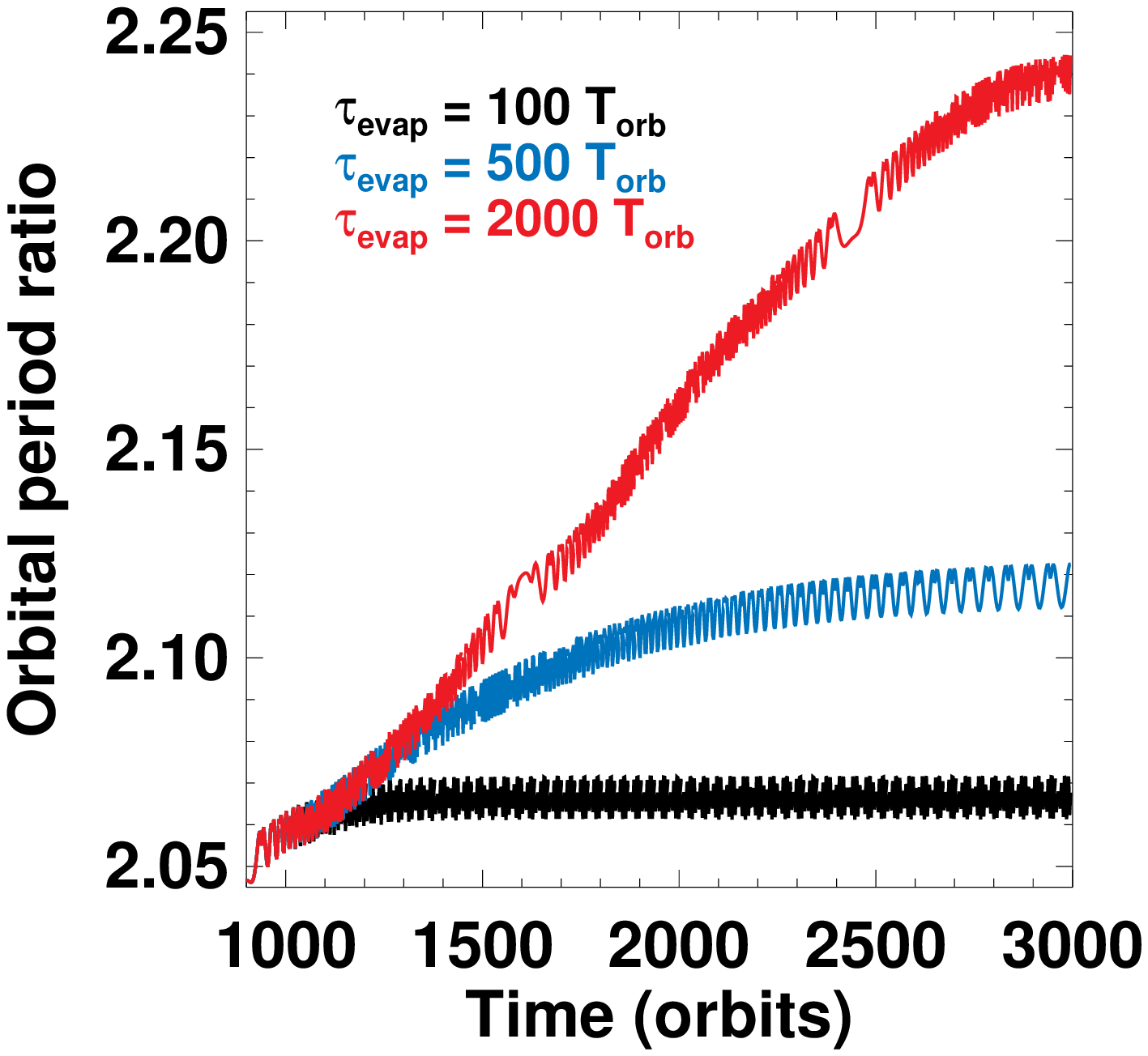}
    \includegraphics{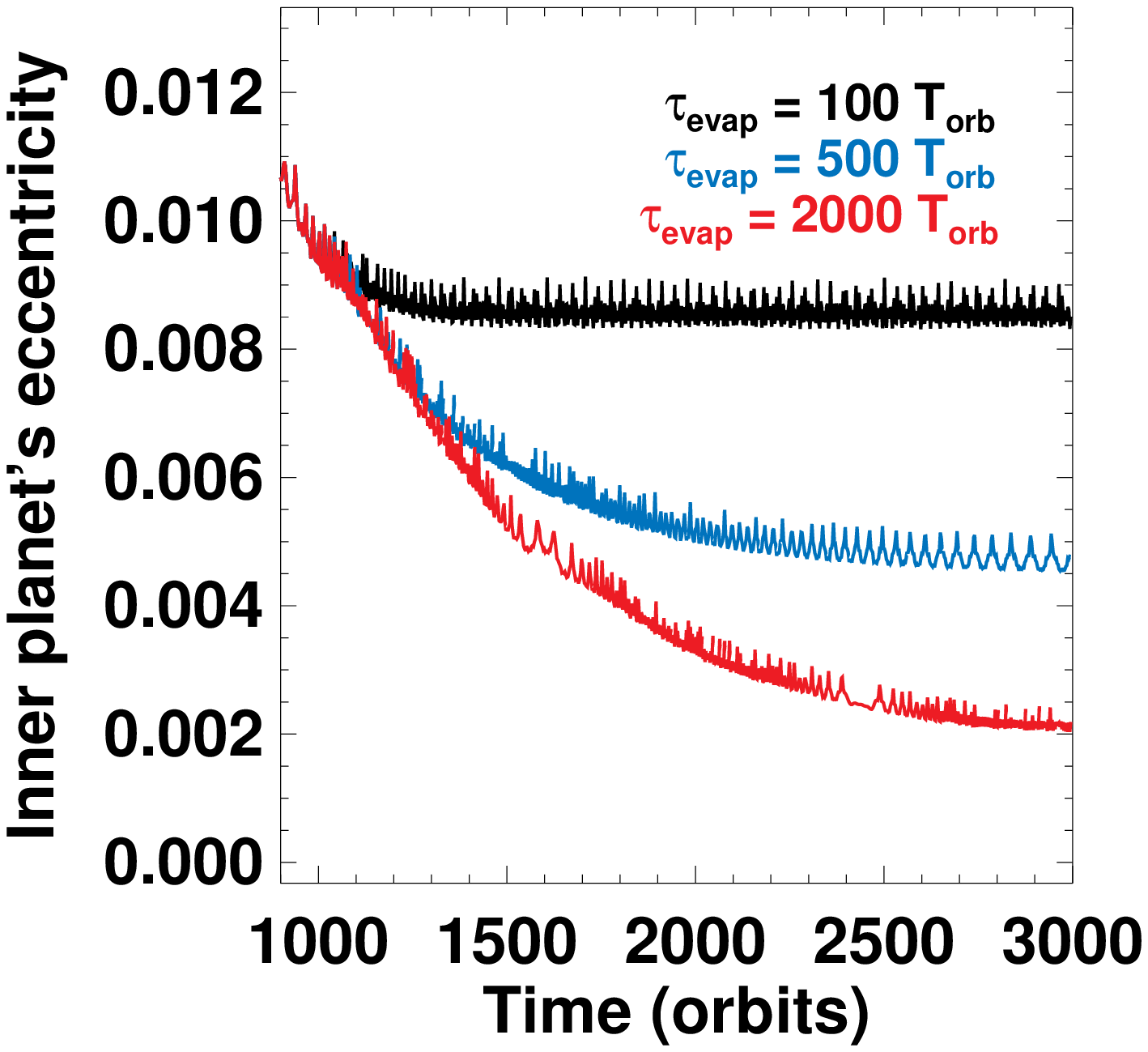}
  }
  \resizebox{0.85\hsize}{!}
  {
    \includegraphics{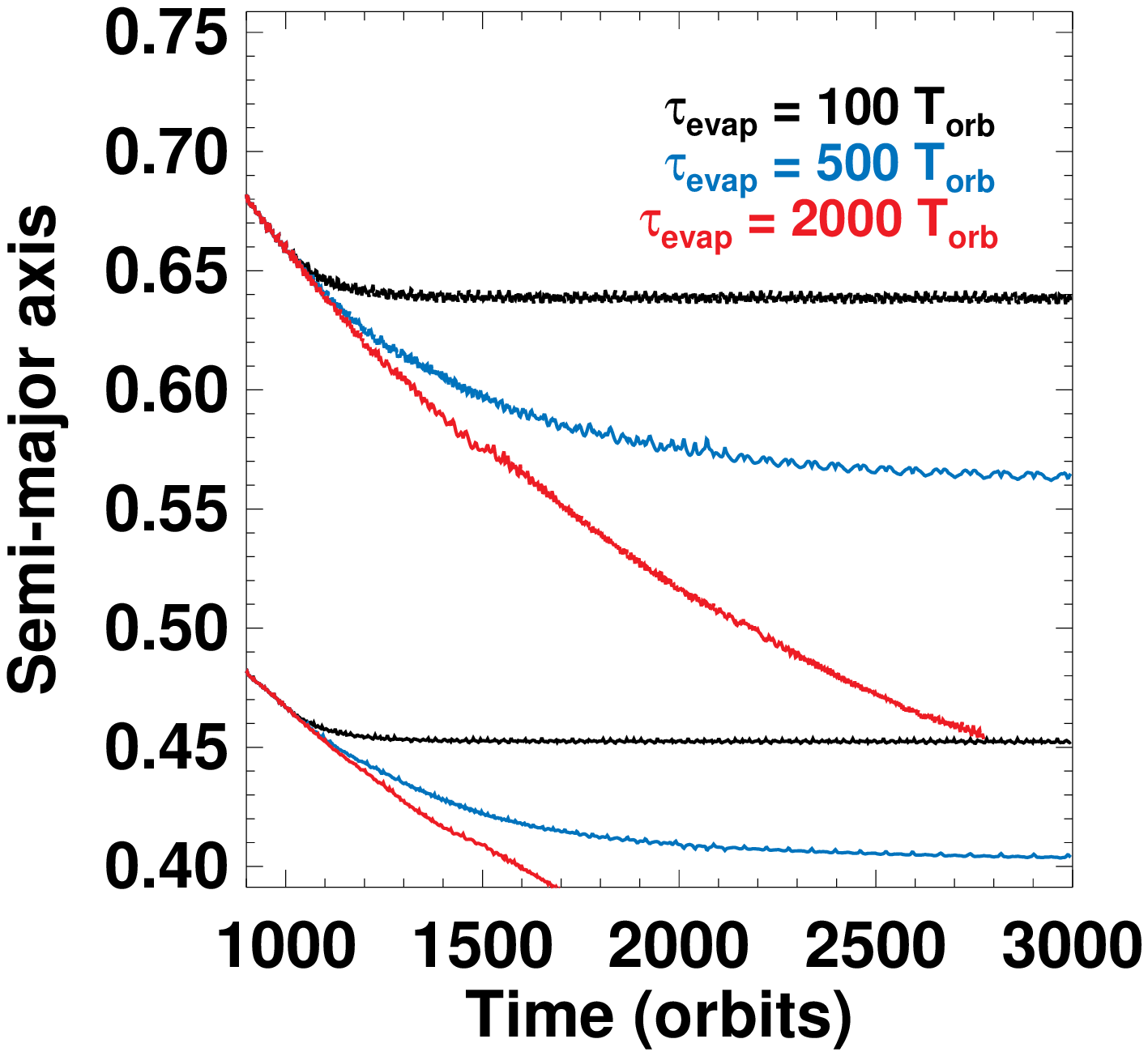}
    \includegraphics{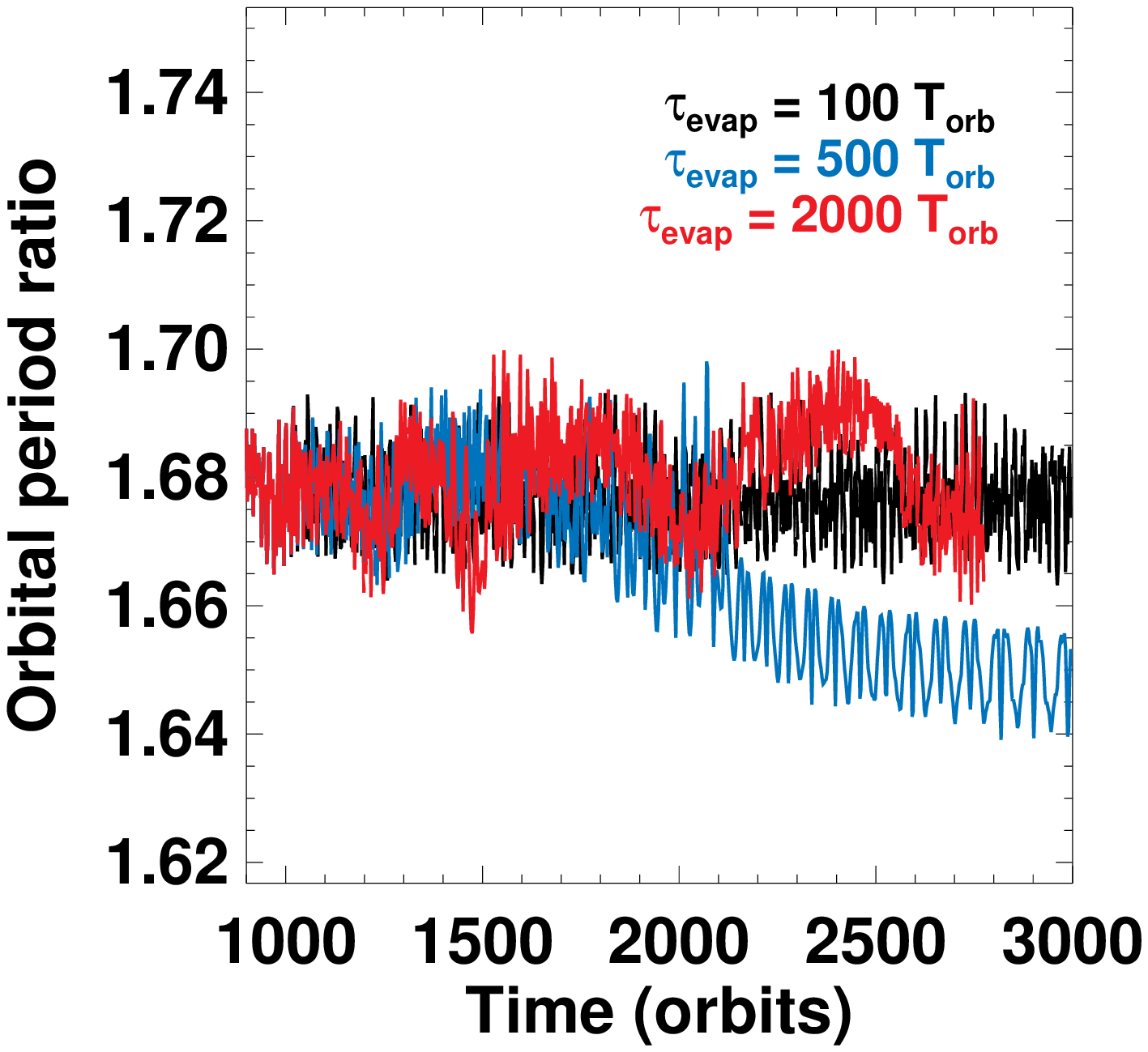}
    \includegraphics{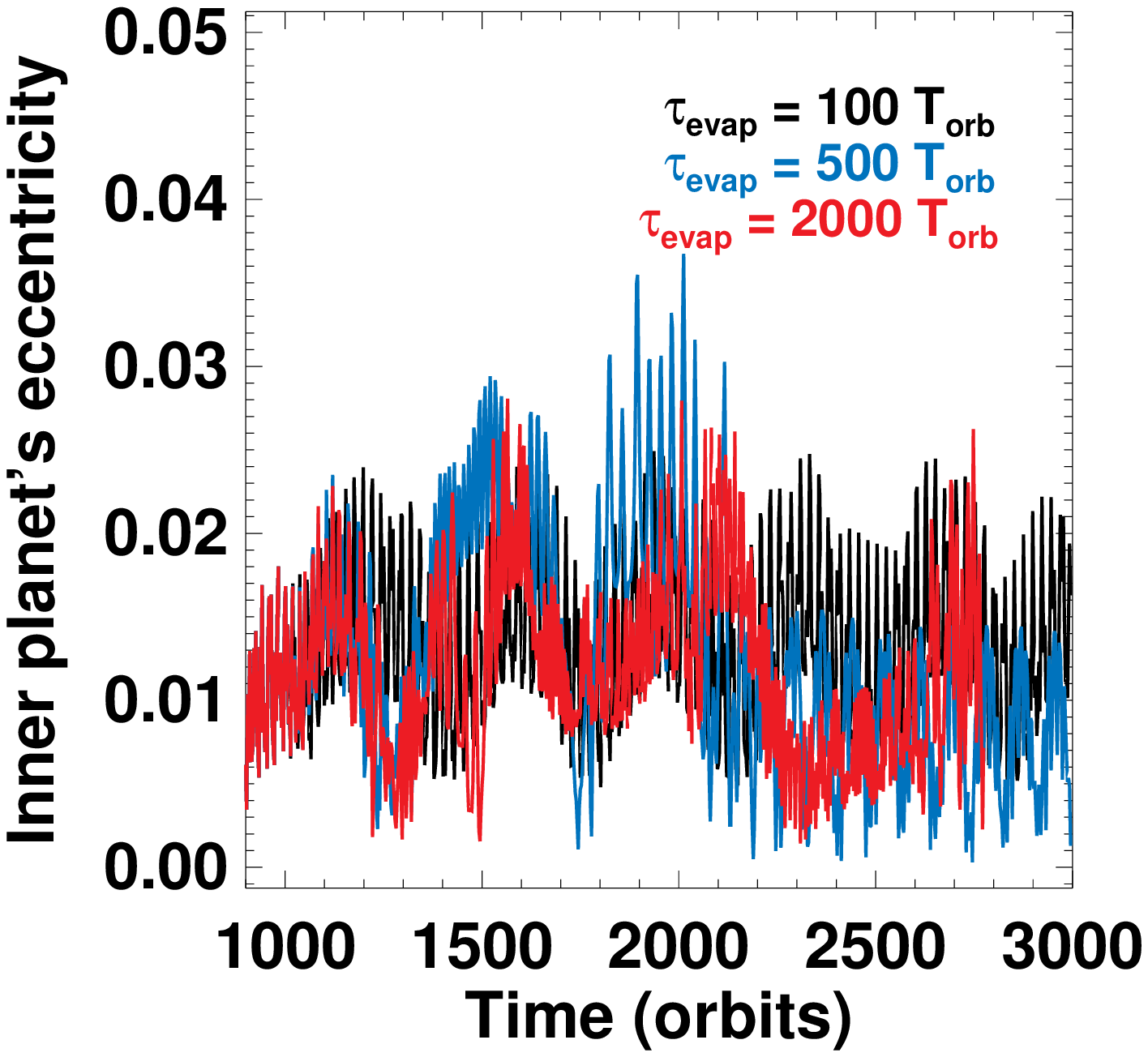}
  }
  \caption{\label{fig:disp}Time evolution of the planets' semi-major
    axis, period ratio and eccentricity (for the inner planet only)
    with a simple model for disk dispersal (see text).  Disk dispersal
    is switched on at 1000 orbits and the evaporation timescale
    ($\tau_{\rm evap}$) is varied from 100 to 2000 orbits. Results in
    the upper panels are for $M_{\rm inner}=0.4 M_{\rm J}$ and
    $\Sigma_0 = 3\times 10^{-4}$. Results in the lower panels are for
    $M_{\rm inner}=0.6 M_{\rm J}$ and $\Sigma_0 = 6\times 10^{-4}$.}
\end{figure*}
\\
\par\noindent\emph{Evolution towards a steady state---} 
When the planets are widely separated, they are expected to interact
with the disk independently of each other and they can then enter a
convergent migration phase.  As they approach each other both
planet-planet interactions, and wake-planet interactions where the
planets interact with each others wakes can occur. Both these
processes involve exchanges of orbital energy and angular momentum
between the planets. Planet-planet interactions can result in resonant
trapping associated with the growth of eccentricities, while
wake-planet interactions act to produce divergent migration without
the growth of eccentricities.  From Eq.~(\ref{eq_pr}) we see that both
circularization- and wake-driven resonant repulsions act together to
increase the period ratio. If wake-driven repulsion dominates after
the eccentricities have damped to low values, with $|{\dot E}_{{\rm
    wake}}|$ remaining constant, the period ratio should increase
linearly with time, which is roughly what is obtained in the
simulation presented in the top row of panels in Figure~\ref{fig:ex}.
However, in the longer term $|{\dot E}_{{\rm wake}}|$ should decrease
with time as the planets separate, which indicates that a steady state
should be ultimately reached with convergent migration driven by
background disk torques balancing wake-driven repulsion. We stress
that wake-planet interactions cause a non-linear evolution of the
disk's density profile near the planets, which may affect the time
evolution of the background torques responsible for convergent
migration. At least in the simulation without planet-planet
interaction of Figure~\ref{fig:fono}, a slow decrease in the
background torques could explain why the period ratio decreases then
increases, instead of reaching a stationary non-resonant
value. Another possibility is that convergent migration and
disk-driven repulsion behave differently as the planet pair moves
closer to the star, if the torques responsible for both mechanisms
have different radial dependences. Investigation of these aspects is
deferred to a future study.
 
\subsubsection{Disk dispersal model}
\label{sec:disp}
The above simulations of the Kepler-46 system show that the planets'
period ratio can take a range of values, depending on the convergent
migration rate during the early evolution of the system.  The limited
duration of the simulations (up to a few thousand orbits) and the fact
that planets still migrate at the end of the simulations however raise
the question of the long-term evolution of the system. Even when the
inward migration of the planets gets stalled\footnote{This could be
  the case for instance when the mass of the disk inside the inner
  planet becomes smaller than the mass of the two gap-opening planets,
  a migration regime known as planet-dominated type II migration.},
further evolution of the period ratio can be maintained by disk-driven
resonant repulsion. A steady state may {\it a priori} be achieved
after depletion of the protoplanetary disk.  Disk dispersal typically
occurs between $10^6$ and $10^7$ years after formation of the central
star. Photoevaporation driven by extreme ultraviolet radiation from
the star opens a gap in the disk at separations of typically a few AU
for Sun-like stars \citep[e.g.,][]{Owen10,AP12}. For planets below
that separation the background disk will be depleted on up to a
viscous timescale at the orbital separation where photoevaporation
sets in. Assuming a local viscous alpha parameter of a few $\times
10^{-3}$, viscous draining of the inner disk could operate in a few
$10^4$ years, that is typically a few $10^5$ orbital periods at the
present location of Kepler-46b and Kepler-46c.

To assess the impact of disk evaporation on our results, we have
restarted two simulations adopting a simple exponential decay of the
surface density profile. This is done by solving $\partial_t \Sigma =
-(\overline{\Sigma} - \Sigma_{\rm target})/\tau_{\rm evap}$ in
addition to the hydrodynamical equations, where $\overline{\Sigma}$ is
the azimuthally-averaged density profile at restart time (from which
evaporation switches on), $\Sigma_{\rm target} =
10^{-3}\overline{\Sigma}$ is an arbitrarily small density profile in a
steady state (taken to be not zero for numerical convenience), and
$\tau_{\rm evap}$ is the evaporation timescale. For illustrative
purposes, we have considered three short evaporation timescales: 100,
500 and 2000 orbits, simulations being restarted at 1000 orbits.
Results are shown in Figure~\ref{fig:disp}.

The upper panels are obtained with $M_{\rm inner}=0.4 M_{\rm J}$ and
$\Sigma_0 = 3\times 10^{-4}$, for which a rapid divergent evolution
occurs after capture in the 2:1 MMR.  We see that the evolution is
frozen out at a few evaporation timescales after restart, the planets
reaching stationary eccentricities and semi-major axes.  Similarly,
the lower panels in Figure~\ref{fig:disp} show the results for $M_{\rm
  inner}=0.6 M_{\rm J}$ and $\Sigma_0 = 6\times 10^{-4}$, for which
convergent migration stalls at about the observed period ratio in the
Kepler-46 system. In that case, including disk evaporation stalls the
migration of both planets, as expected, but it does not significantly
affect the planets eccentricities and period ratio. The final period
ratio remains very close to the observed value.

\section{Application to Kepler's multi-planetary systems}
\label{sec:results_nbody}

\subsection{Introduction and strategy}
In Section~\ref{sec:hydro} we have presented results of hydrodynamical
simulations modeling the early evolution of Kepler-46b and Kepler-46c
as they were embedded in their parent protoplanetary disk.  The
variety of period ratios obtained in our simulations raises the
question of whether disk-planets interactions could partly explain the
diversity of period ratios in Kepler's multi-planetary systems.  We
review below some of the processes that might account for such
diversity.

\subsubsection{Tidally-driven resonant repulsion?}
As highlighted in Figure~\ref{fig:Kepler}, Kepler's multi-planetary
systems show a clear tendency for planet pairs near resonances to
feature period ratios slightly greater than strict commensurability.
Divergent evolution of a short-period resonant planet pair due to
star-planet tidal interactions could partly explain this feature
\citep{LW12, Baty12}. Although the efficiency of tidal dissipation
remains largely uncertain, it is thought that efficient tidal
circularization requires the inner planet to orbit its star in less
than a few days. As can be seen in Figure~\ref{fig:Kepler} many of
Kepler's candidate systems have an inner planet beyond 10 days, for
which tidal circularization is unlikely to have caused significant
resonant repulsion (see Section~\ref{sec:tidal} of the appendix). For
these systems we propose that resonant repulsion driven by
disk-planets interactions could have played a prominent role in
determining the observed period ratios.

\subsubsection{Disk-driven type I migration?}
The vast majority of the planets in Kepler's multi-planetary systems
have physical radii between 0.1 and 0.3 Jupiter radii, which places
them in the super-Earth to Neptune-mass range. From the mass-radius
diagram of Kepler candidates followed up by radial velocity, the
median mass of a 0.2 Jupiter-radius planet is about 10 to 15
Earth-masses (see, e.g., http://exoplanets.org). Based on generally
considered temperatures and viscosities in protoplanetary disks
(corresponding to $h = 0.05$ and $\alpha \sim$ a few $\times 10^{-3}$)
such planets are expected to experience type I migration. Recent
N-body experiments by \cite{ReinKepler} have shown that convergent
migration of type-I migrating planets yields very little departure
from strict commensurability in contrast with Kepler's data. He showed
that the inclusion of stochastic forces in addition to type I
migration could reproduce the observed distribution of period ratios
quite nicely. Such agreement has been obtained assuming that the
initial period ratio distribution is the observed one, and taking the
same convergent migration timescale and the same amplitude of
stochastic forces for all Kepler's multiple systems, which remains
quite uncertain given the expected diversity of physical properties in
protoplanetary disks. Still, his results indicate that type I
migration alone may not be able to account for Kepler's diversity of
period ratios.

\begin{figure*}
  \centering
  \resizebox{\hsize}{!}
  {
    \includegraphics{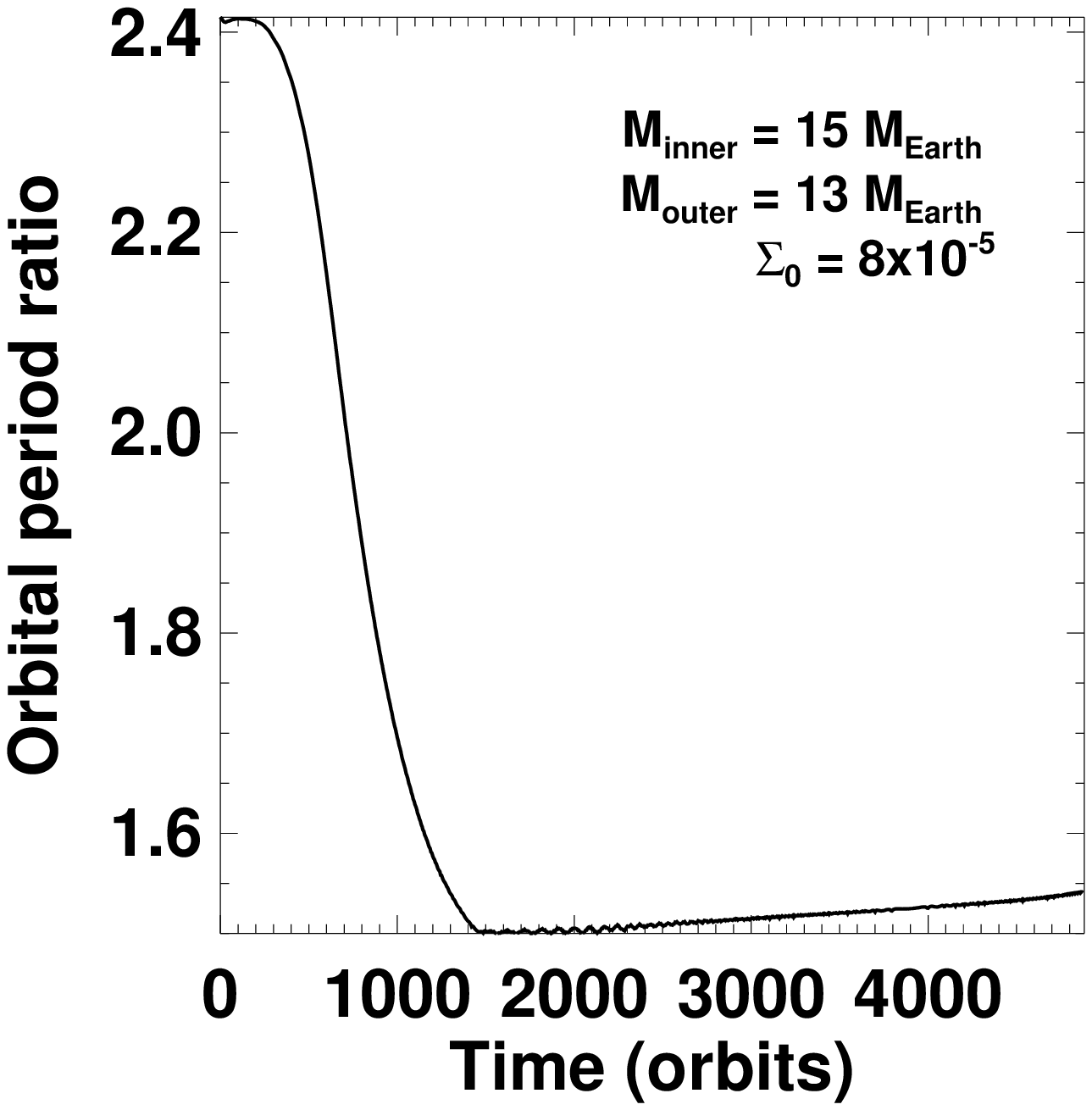}
    \includegraphics{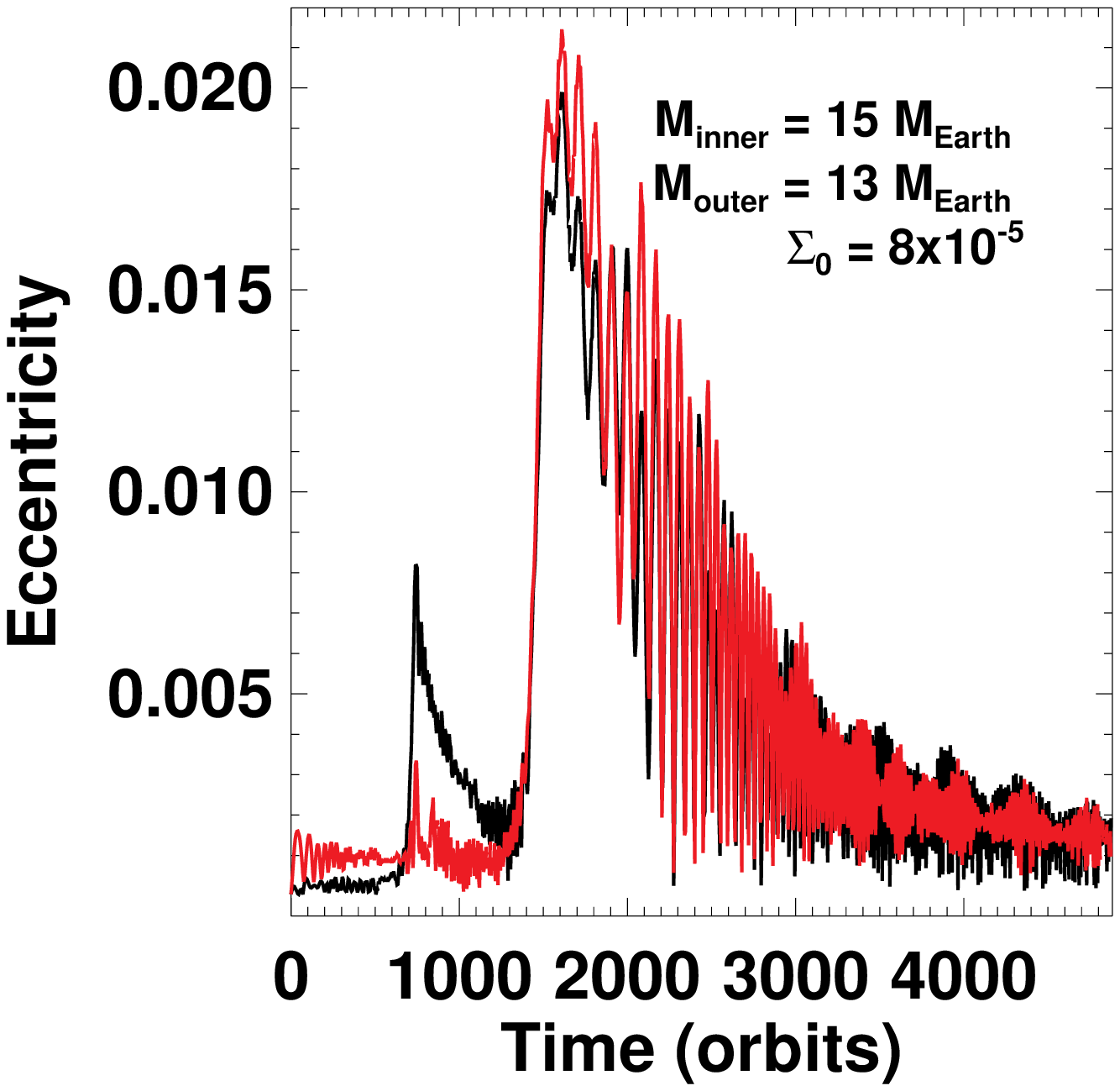}
    \includegraphics{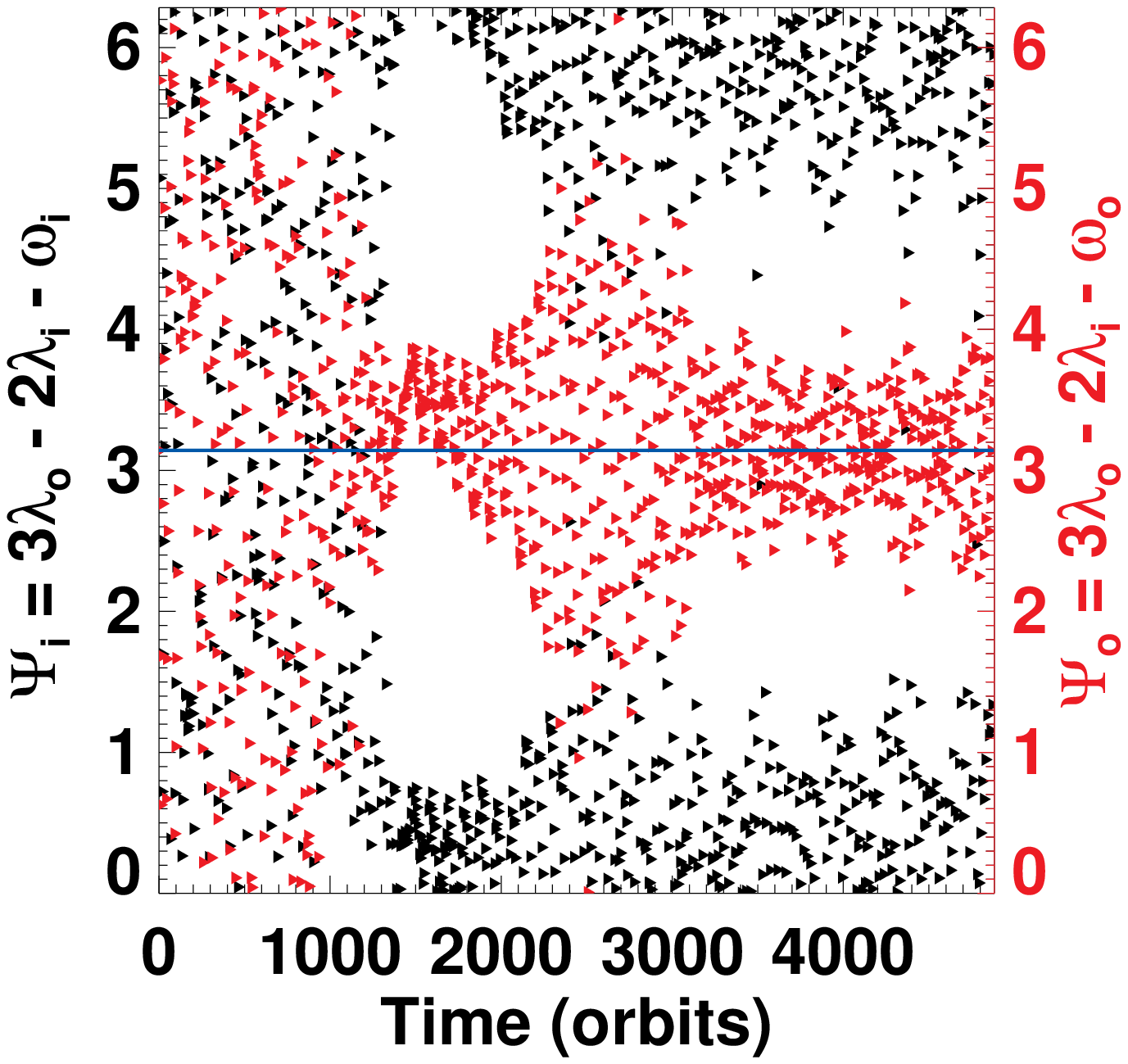}
    \includegraphics{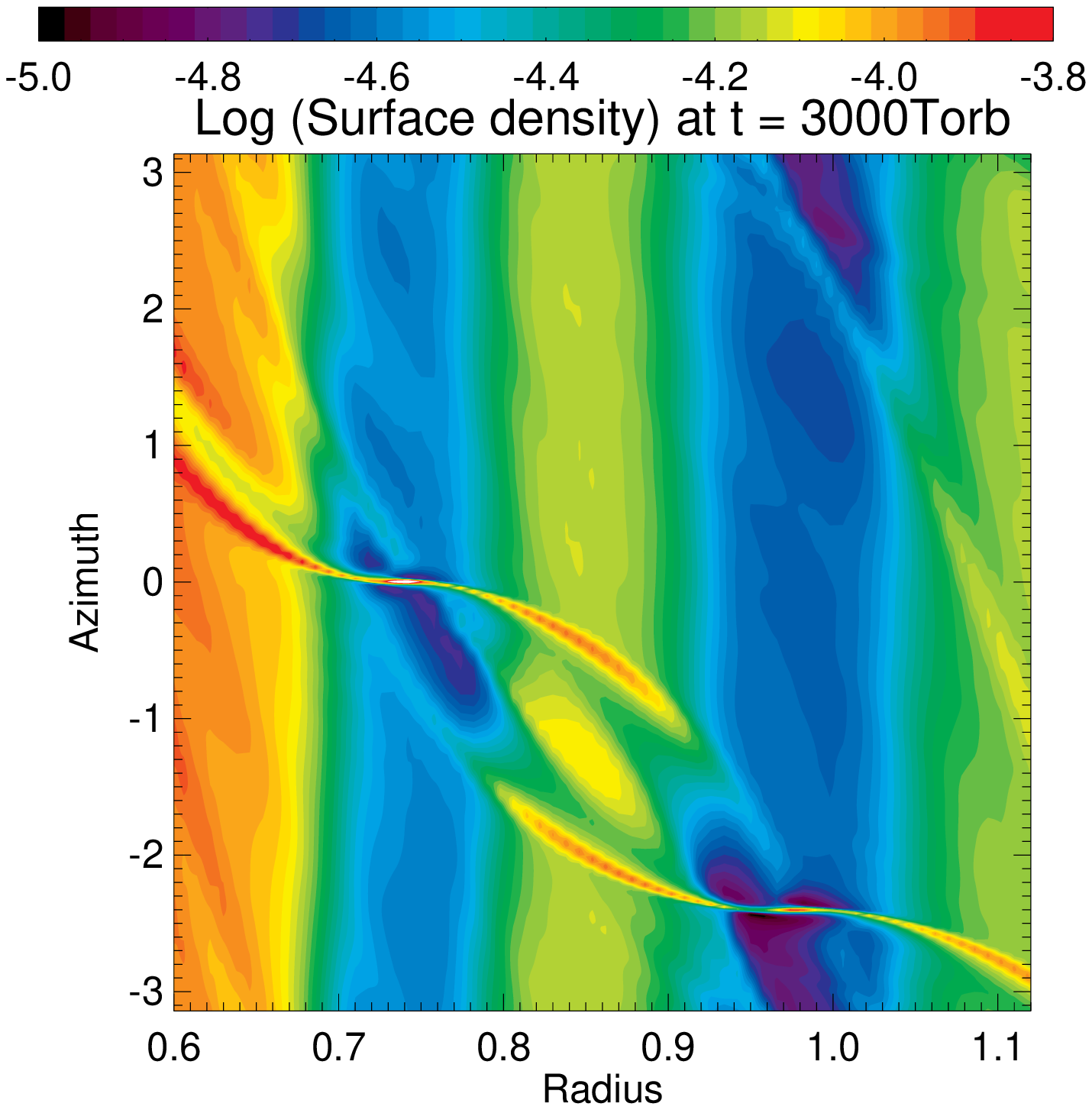}
  }
  \resizebox{\hsize}{!}
  {
    \includegraphics{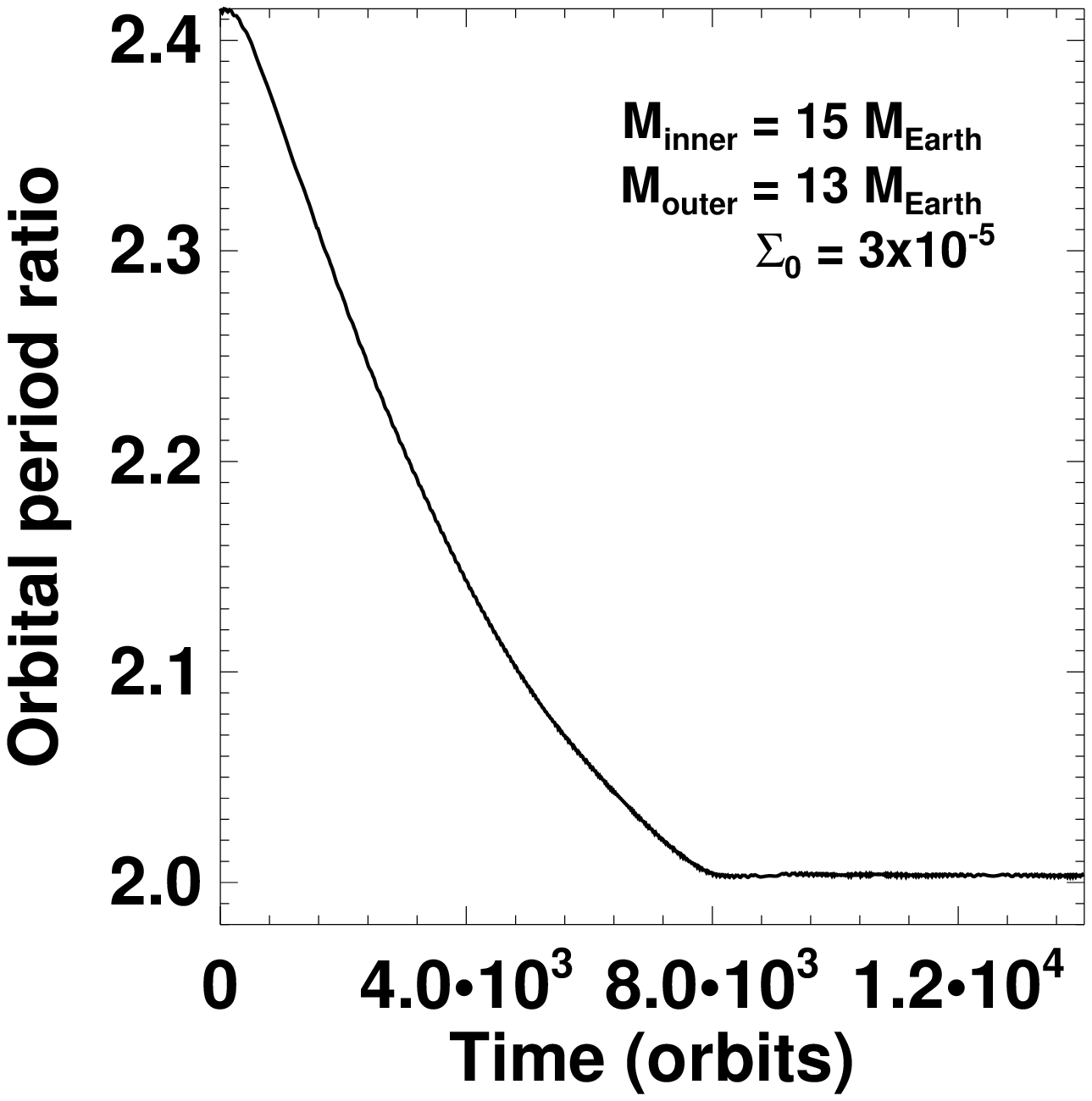}
    \includegraphics{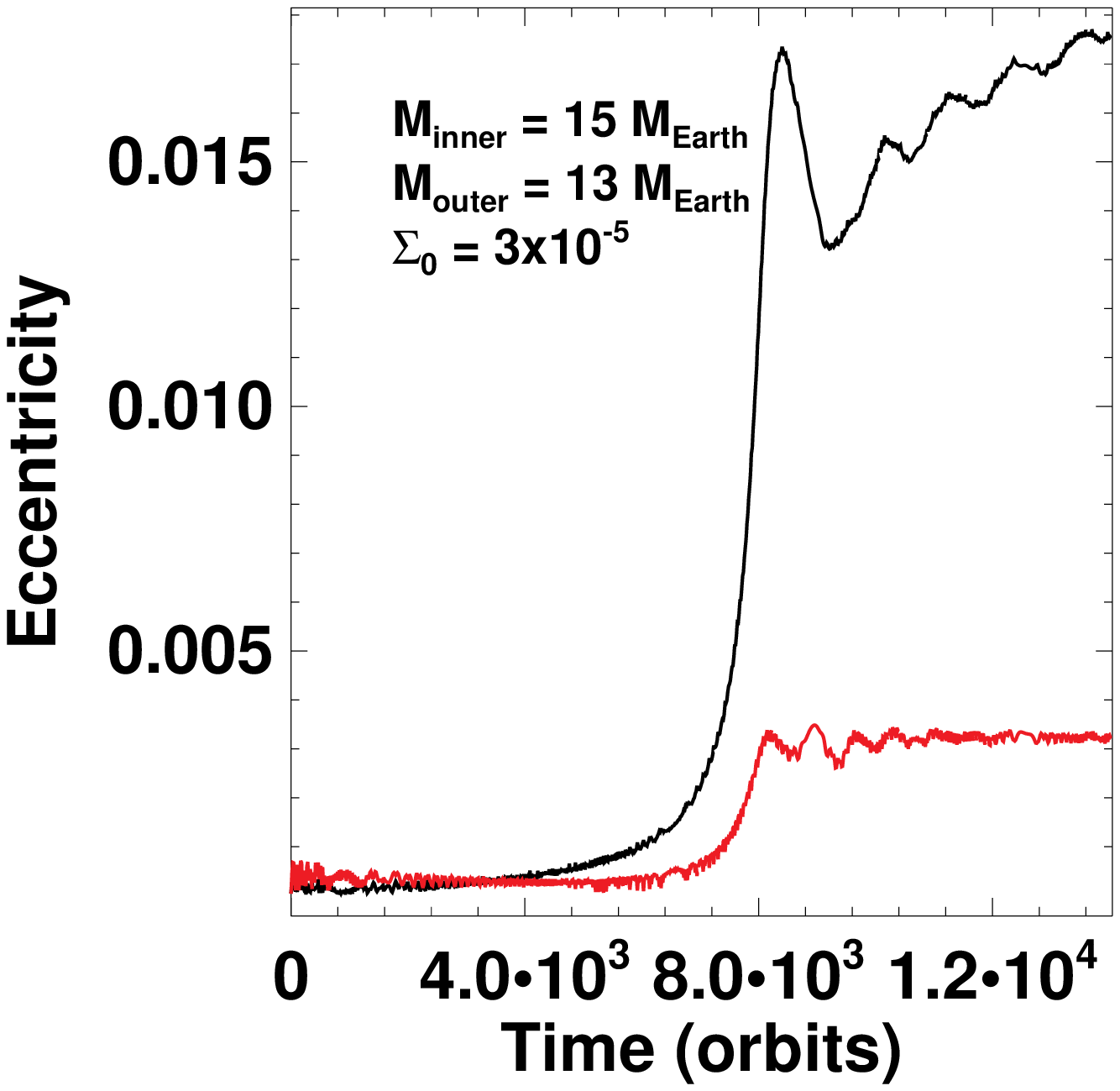}
   \includegraphics{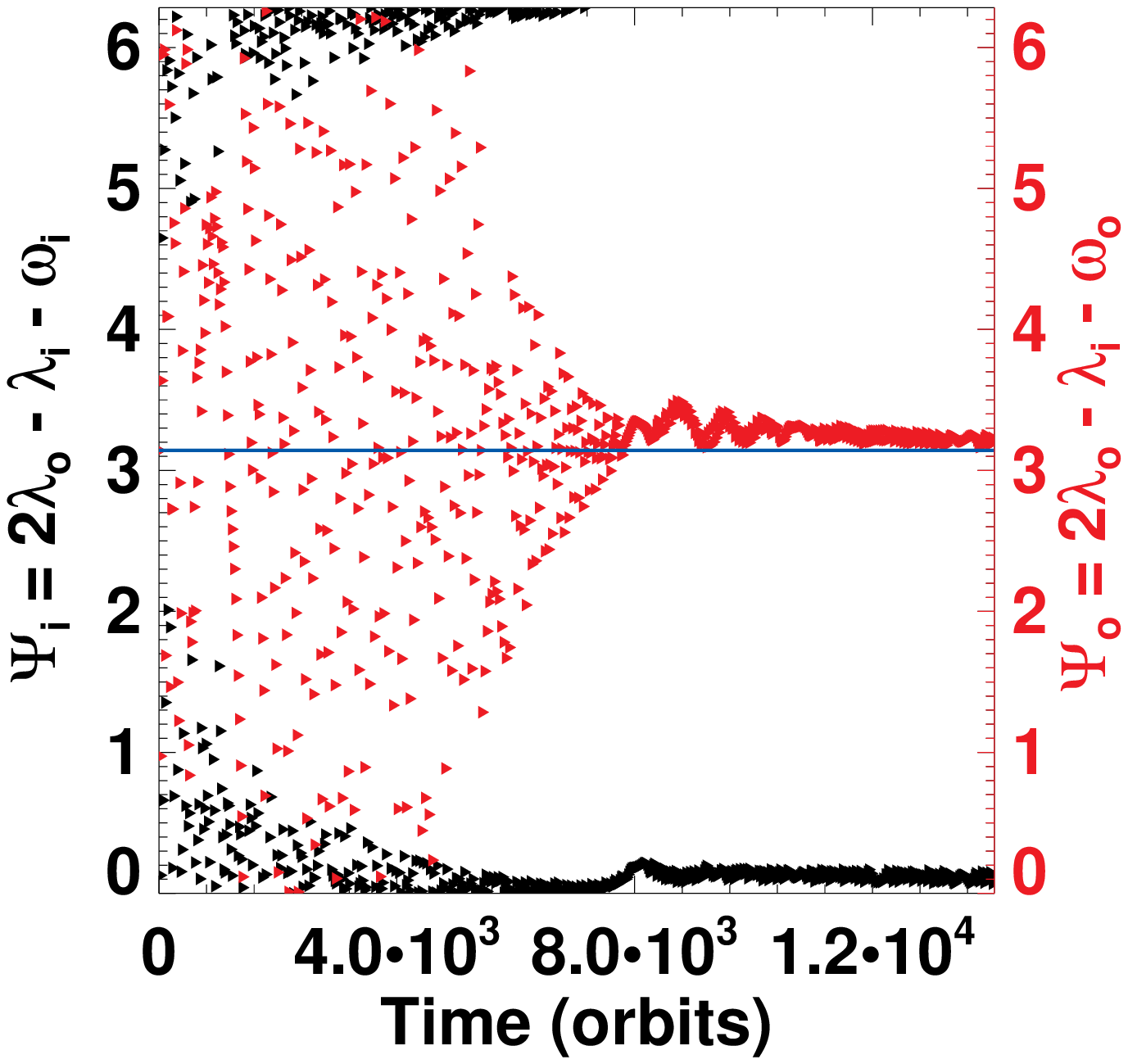}
   \includegraphics{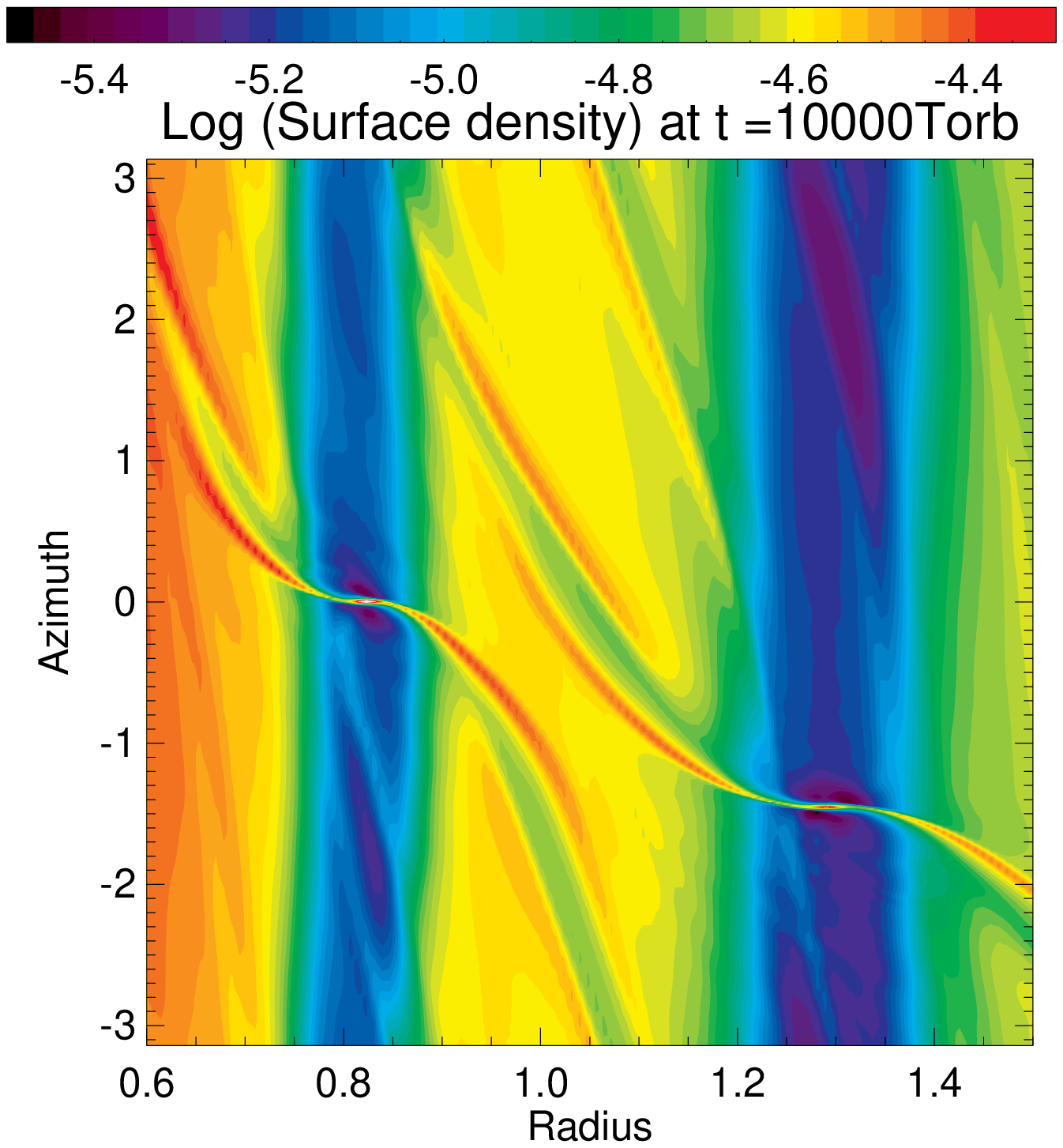}
  }
  \caption{\label{fig:M10}Results of hydrodynamical simulations with
    two super-Earths opening partial gaps in their disk. Top row: case
    where convergent migration leads to capture into the 3:2 MMR
    followed by rapid divergent evolution.  Bottom row: slow
    convergent migration leading to capture into the 2:1 MMR with no
    subsequent divergent evolution. From left to right, the panels
    display the time evolution of the planets orbital period ratio,
    eccentricities, relevant resonant angles, and a snapshot of the
    disk's surface density.}
\end{figure*}
\subsubsection{Disk-driven repulsion of partial gap-opening planets?}
\label{sec:M10}
Based on our results of hydrodynamical simulations for the Kepler-46
system, we propose that some of Kepler's planet candidates in
multi-planetary systems could have opened partial gaps in their parent
disk, thereby escaping the type I migration regime.  This could be the
case if, for instance, planets formed and/or migrated in regions of
low turbulent activity \citep[dead zones; see,
e.g.,][]{flemingstone03} or if the disk's aspect ratio takes smaller
values than commonly adopted. Aspect ratios $\lesssim 3\%$ are likely
typical of the short orbital separations at which Kepler's candidate
systems are detected. To address this possibility, we carried out
hydrodynamical simulations with two planets in the super-Earth mass
range. The inner planet's mass $M_{\rm inner} = 15 M_{\oplus}$, the
outer planet's mass $M_{\rm outer} = 13 M_{\oplus}$ and the central
star is one Solar mass.  The planet-to-primary mass ratios are
therefore $q_{\rm inner} = 4.4\times 10^{-5}$ and $q_{\rm outer} =
4\times 10^{-5}$, that is an order-of-magnitude smaller than those of
Kepler-46b (median value) and Kepler-46c. In order to get similar gap
depths in the super-Earths and Kepler-46 simulations, we adopted
values of $h$ and $\alpha$ such that the dimensionless parameters in
the gap-opening criterion of \cite{crida06} take the same values. For
the super-Earths simulations, this yields $h = 0.023$ and $\alpha =
2.3 \times 10^{-3}$. All other simulation parameters are otherwise
identical, except the grid resolution that we increased to
$600\times1200$.

Figure~\ref{fig:M10} displays the results of two simulations. In the
top panels, the unperturbed disk's surface density is $\Sigma_0 =
8\times 10^{-5}$.  We see that the planets lock themselves in the 3:2
MMR from about 1300 orbits. Divergent evolution proceeds with a
decrease in the planets eccentricities. After reaching exactly 1.5 at
about 1500 orbits, the period ratio increases to 1.54 at 5000
orbits. As illustrated by the screenshot of the disk surface density,
the proximity of the planets allows their wakes to penetrate into each
others coorbital region, where they can deposit part of their angular
momentum flux.  As we have shown in Section~\ref{sec:divergent},
transfer of angular momentum between planets through wake-planet
interactions will result in a loss rate of the total orbital energy of
the two planets.  This has the consequence of increasing the period
ratio and decreasing the eccentricities.  In the bottom panels of
Figure~\ref{fig:M10}, the unperturbed disk surface density parameter
is decreased to $\Sigma_0 = 3\times 10^{-5}$. Here, convergent
migration due to the background disk torques is slow enough to lock
the planets in the 2:1 MMR, but this time the resonant capture does
not lead to divergent evolution. Instead, a steady state is reached in
which both the eccentricities and the orbital period ratio take
stationary values. This steady state is a consequence of negligible
wake-planet interactions in this case. As illustrated in the
bottom-right panel of Figure~\ref{fig:M10}, the planets are
sufficiently far from each other that their wakes primarily transfer
energy and angular momentum to regions of the disk that are not
coorbital with the planets.
\begin{figure*}
  \centering
  \resizebox{0.85\hsize}{!}
  {
    \includegraphics{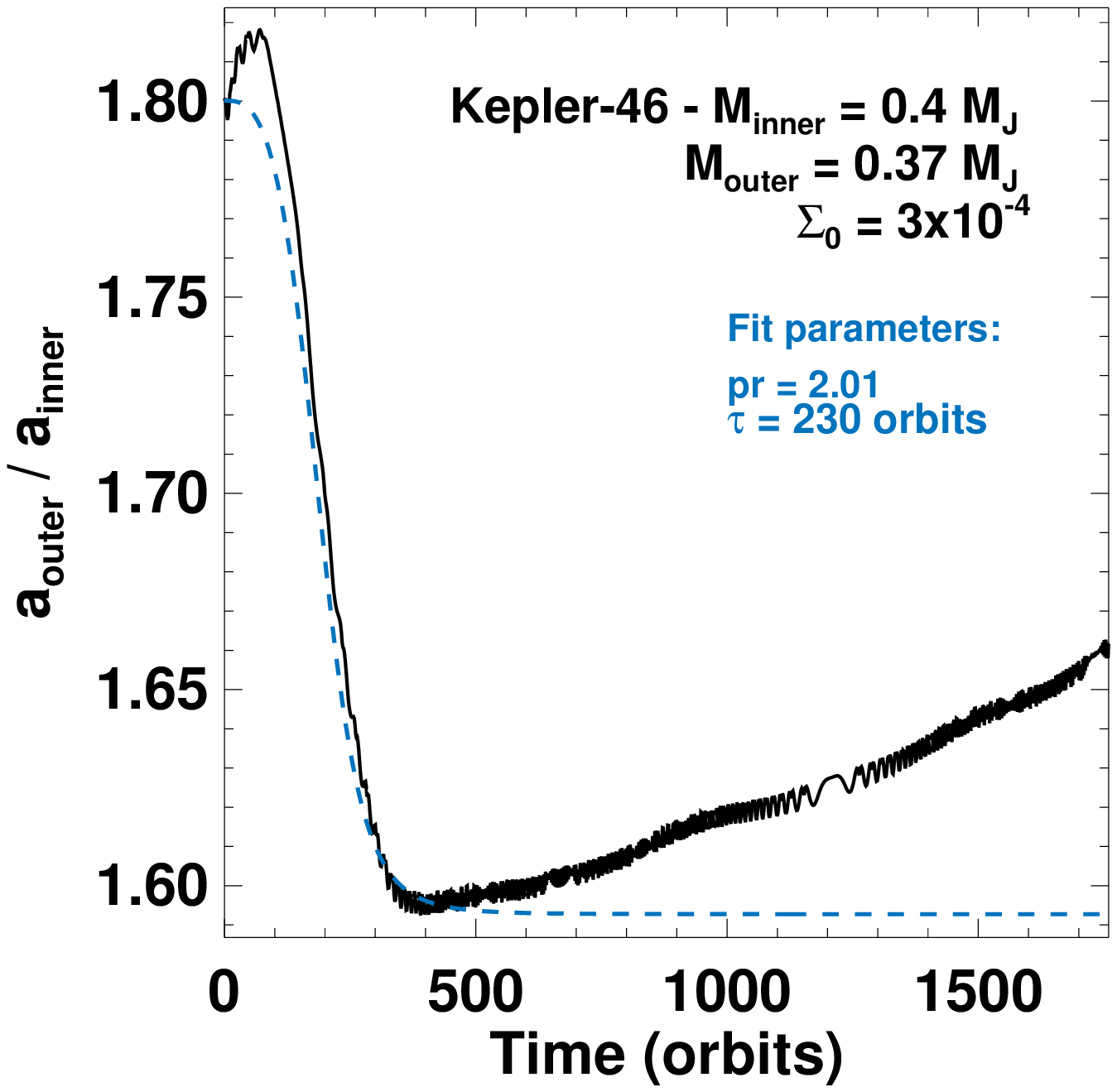}
    \includegraphics{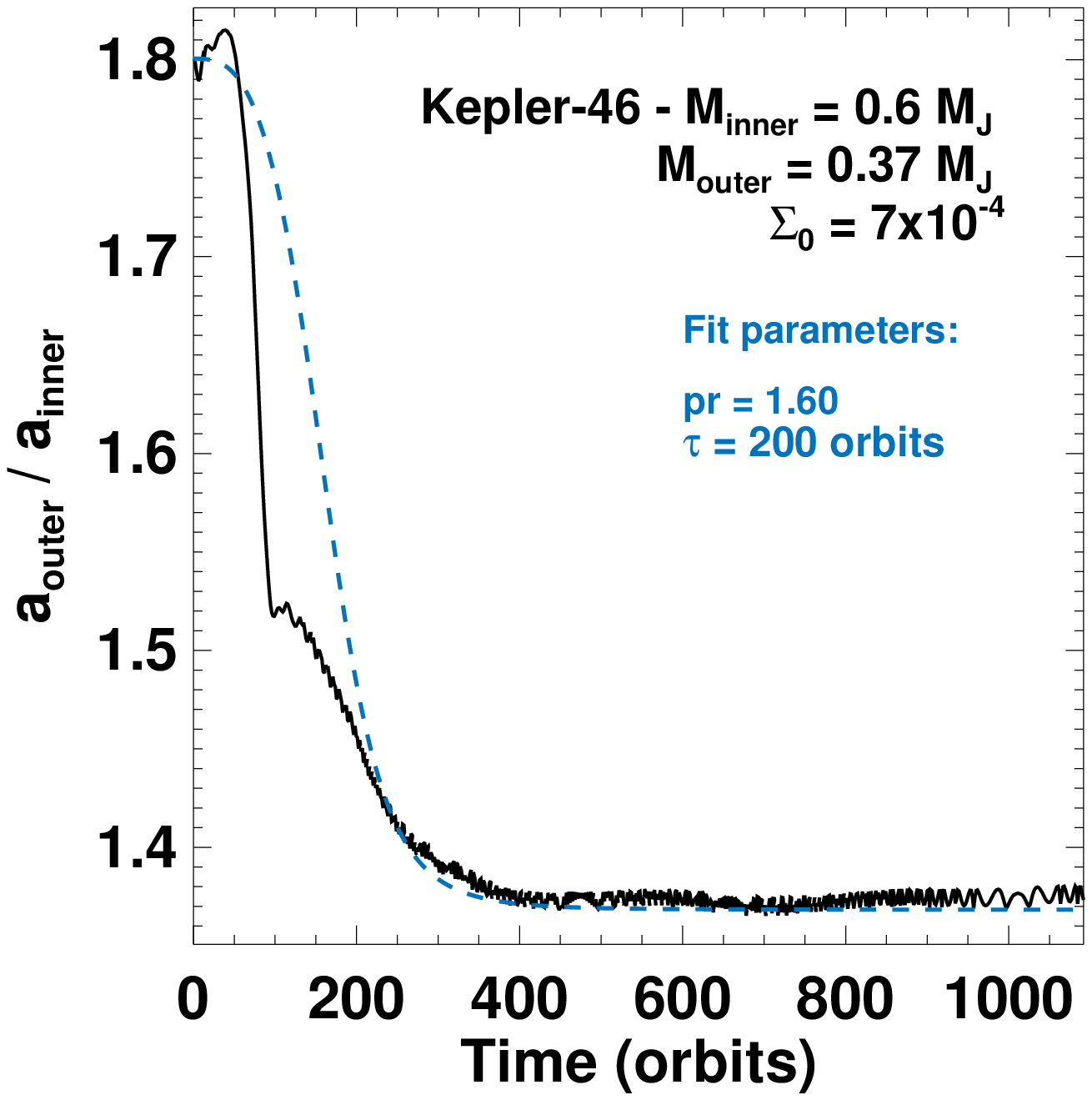}
    \includegraphics{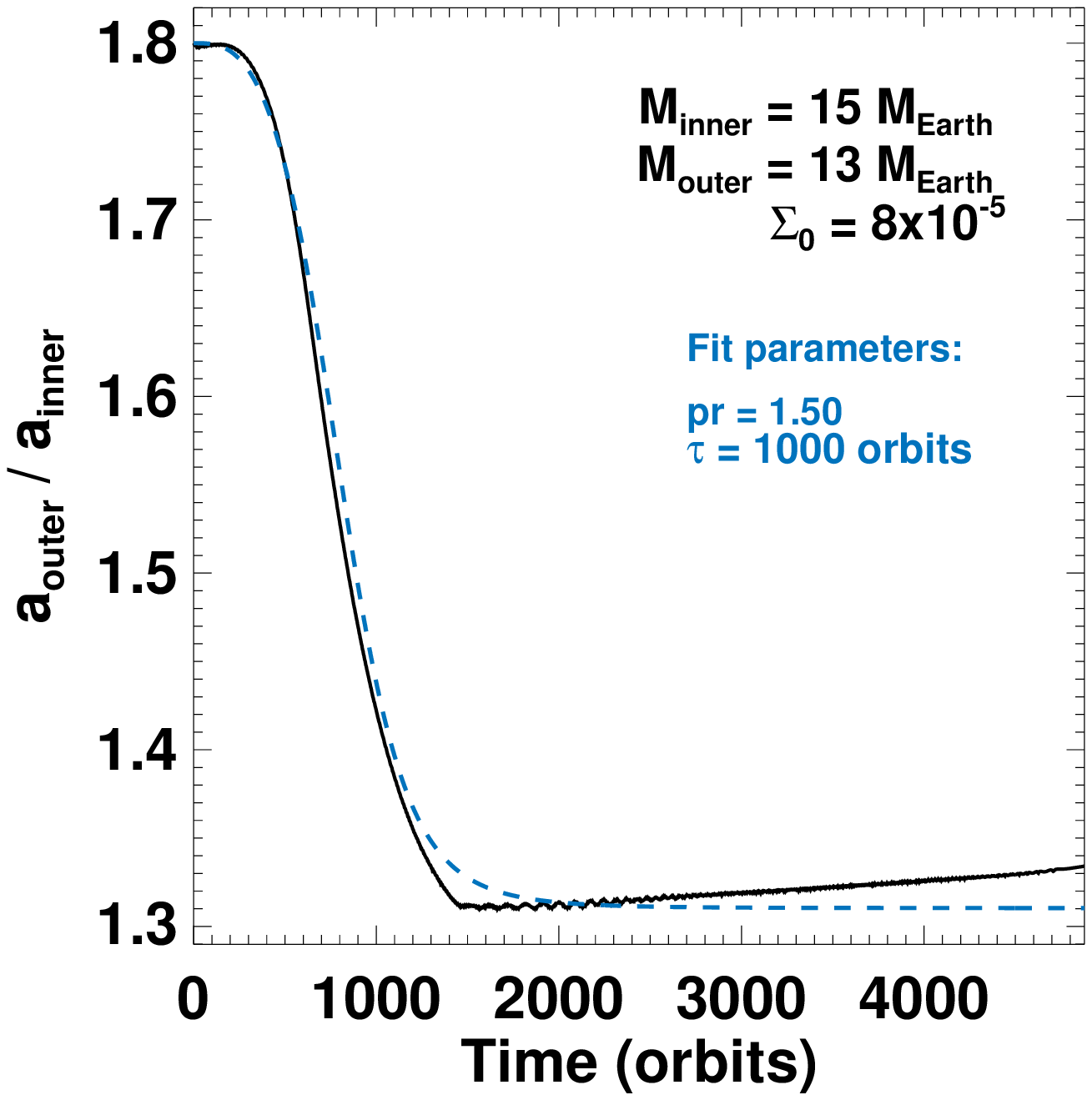}
   } 
   \caption{\label{fig:fit}Time evolution of the ratio of planets
     semi-major axis obtained in three hydrodynamical simulations
     (solid curves). The planet masses and the unperturbed surface
     density parameter $\Sigma_0$ are indicated in the top-right
     corner of each panel. The dashed curves display our fitting
     formula for the planets convergent migration prior to disk-driven
     resonant repulsion, given by Eq.~(\ref{eq:prescmig}).}
\end{figure*}

\subsection{Customized three-body simulations}
\label{sec:nbody}
The results of Section~\ref{sec:M10} support the idea that disk-driven
resonant repulsion of partial gap-opening super-Earths may lead to
period ratios that substantially differ from nominal resonant values.
As shown in Section~\ref{sec:divergent}, this mechanism has two
origins: circularization of the orbits and wake-planet interactions.
Its efficiency depends on the planets masses and the evolution of the
disk's surface density. The survey of such a large parameter space is
out of reach of hydrodynamical simulations, but three-body simulations
with customized prescriptions for migration, eccentricity damping and
disk dispersal can help illustrate the diversity of period ratios that
can be achieved. This is the strategy that we have adopted. For
illustration purposes we have considered fixed planet masses: $M_{\rm
  inner} = 15 M_{\oplus}$, $M_{\rm outer} = 13 M_{\oplus}$ and
$M_{\star} = M_{\odot}$ (as in Section~\ref{sec:M10}).

\subsubsection{Method}
\label{sec:nbody_method}
Our three-body simulations solve the equations of motion for the two
planets and the central star using a standard technique.  Planets are
assumed to be coplanar. Disk-planet interactions are incorporated by
applying appropriate dissipative forces \citep[for details,
see][]{Papa11}. Inspection at Eq.~(\ref{eq_pr}) shows that, in order
to model the planets evolution, we need to specify (i) the rate of
convergent migration before disk-driven repulsion sets in, (ii) the
damping rate of the eccentricities, (iii) the efficiency of
wake-driven divergent evolution, and (iv) how disk dispersal is
modeled. Point (iii) remains uncertain at this stage, as the term
$|\dot{E}_{\rm wake}|$ in the right-hand side of Eq.~(\ref{eq_pr}),
which models energy dissipation due to wake-planet interactions,
should depend sensitively on the planet masses, their mutual
separation and distance from the central star, and several disk
quantities (including its temperature and viscosity). A systematic
study of this dependence goes beyond the scope of this paper, and is
therefore left for future work. In order to do simple modeling, and to
minimize the number of free parameters, we do not attempt to
parameterize the $|\dot{E}_{\rm wake}|$ term. Instead, we model
disk-driven repulsion as circularization-driven repulsion only. In
other words, in Eq.~(\ref{eq_pr}) we compensate the absence of the
$|\dot{E}_{\rm wake}|$ term in our three-body integrations by
considering circularization timescales $\tau_{\rm c,i}$ and $\tau_{\rm
  c,o}$ that are shorter than in the hydrodynamical simulations.
Damping of the planets' eccentricity is modeled as an exponential
decay with a constant damping timescale to orbital period ratio, which
we take as a free parameter (specified below).

\begin{figure}
  \centering
  \resizebox{\hsize}{!}
  {
    \includegraphics{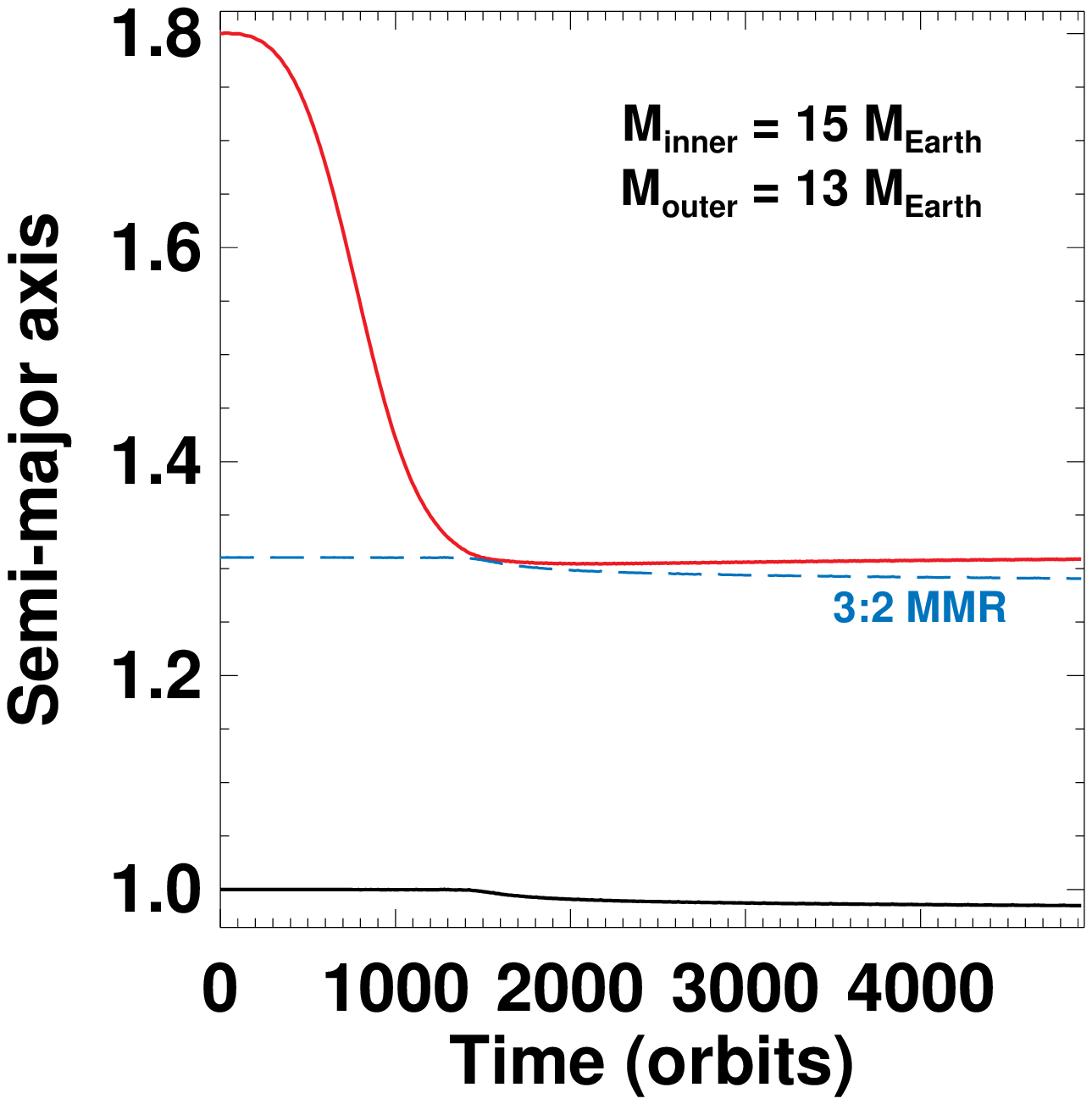}
    \includegraphics{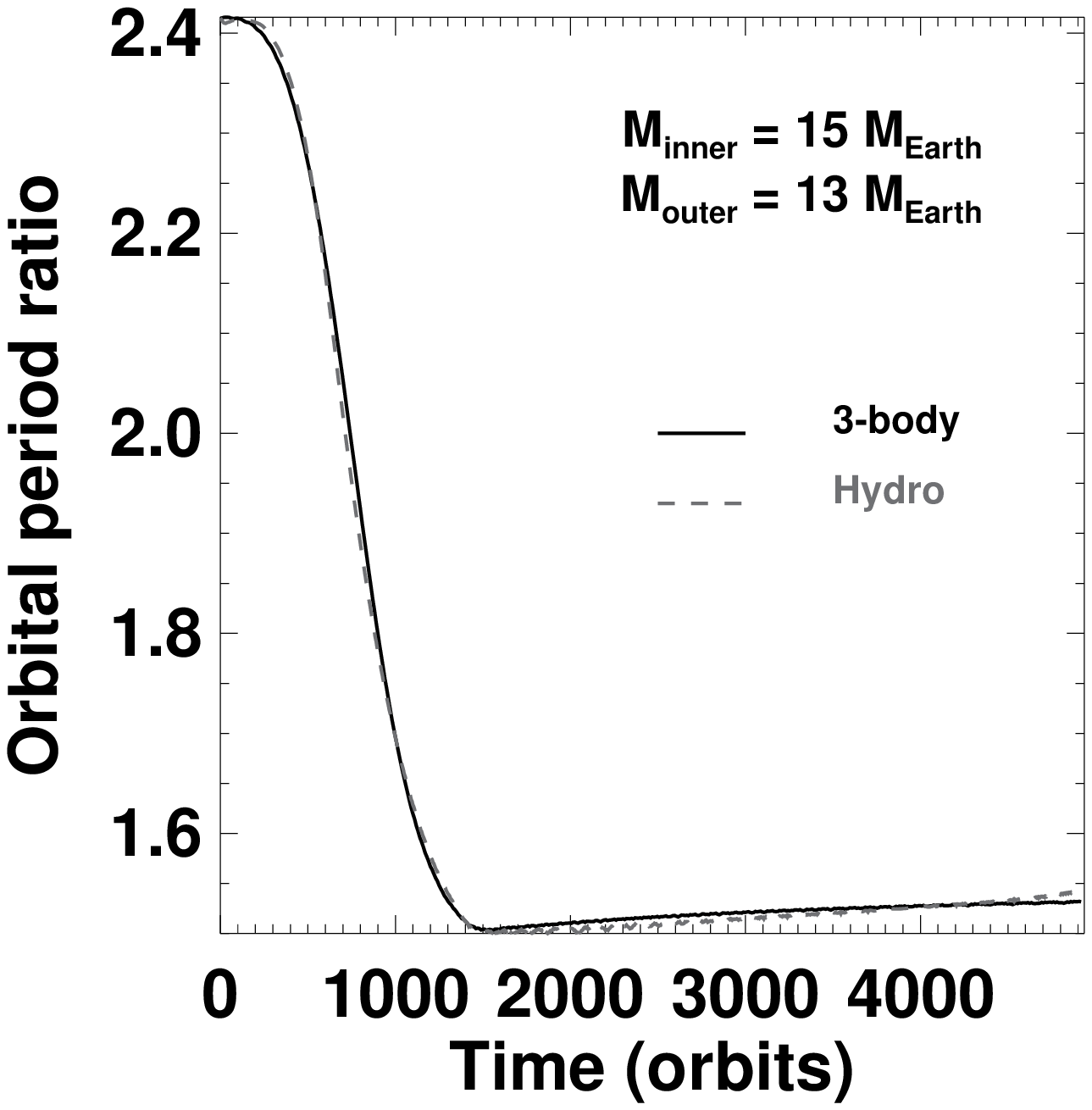}
  }
  \resizebox{\hsize}{!}
  {
    \includegraphics{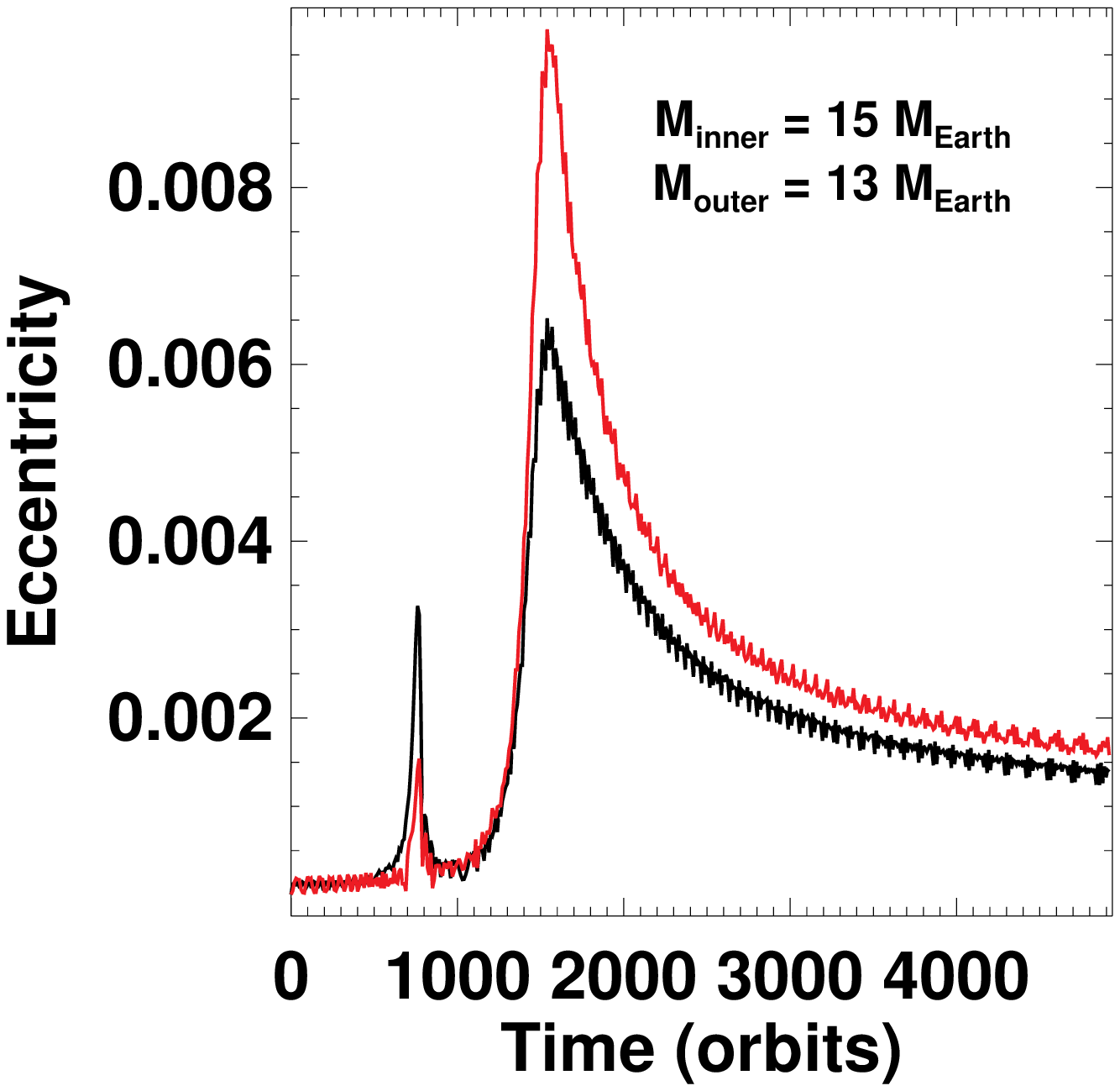}
    \includegraphics{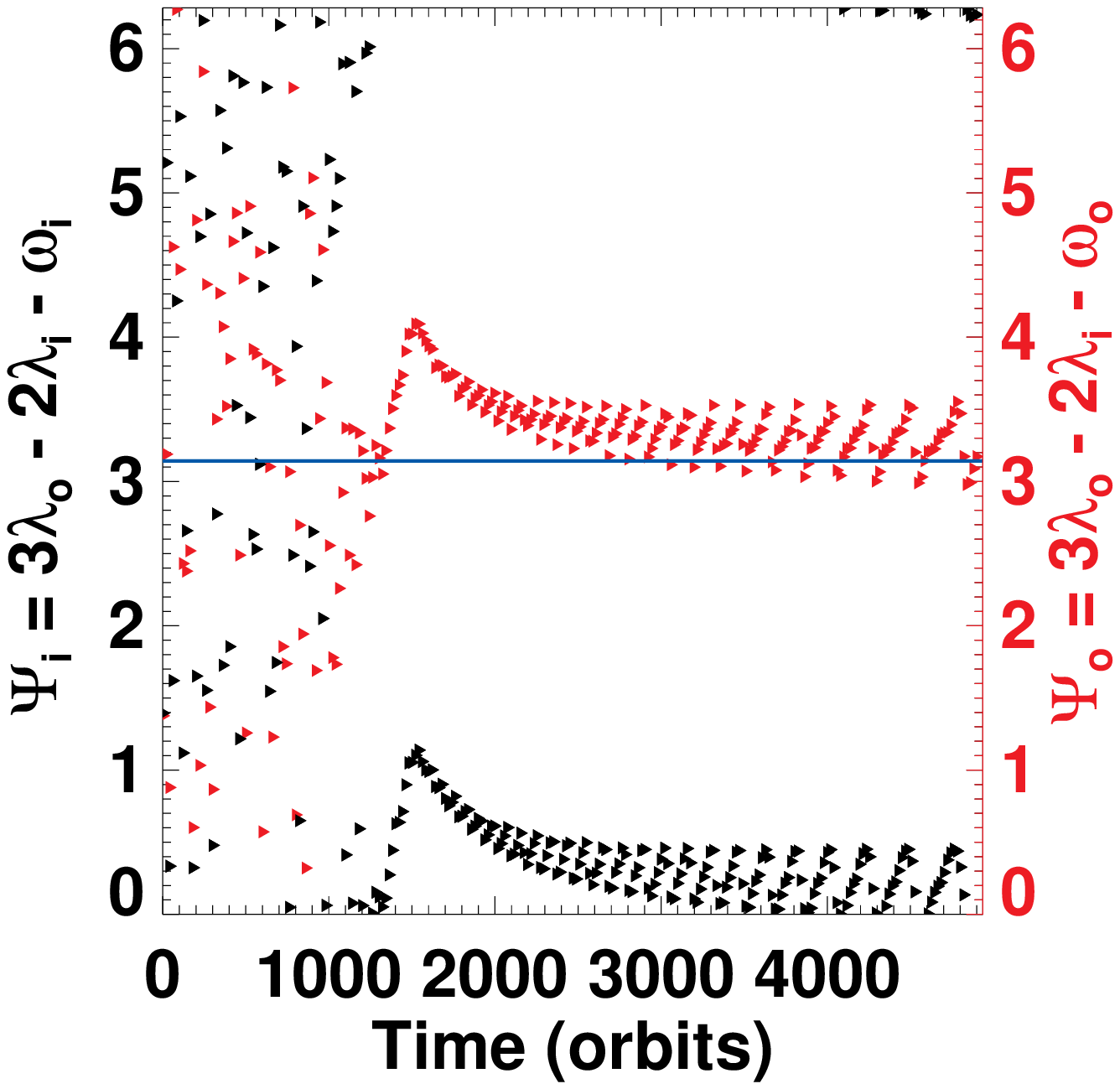}
  }
  \caption{\label{fig:john1}Results of a three-body simulation with
    prescribed convergent migration and eccentricity damping. The
    expression given by Eq.~(\ref{eq:prescmig}) is used for convergent
    migration with $\tau = 1000$ orbits and pr = 1.5. The eccentricity
    damping timescale is 20 orbital periods. The dashed grey curve in
    the top-right panel shows the period ratio obtained in the
    hydrodynamical simulation of \S~\ref{sec:M10} with $\Sigma_0 =
    8\times 10^{-5}$.}
\end{figure}
To model the convergent migration prior to resonant repulsion, we
adopt a simple prescription that reproduces the results of
hydrodynamical simulations. Denoting by $R$ the ratio of semi-major
axes ($R = a_{\rm outer} / a_{\rm inner}$), we found good agreement
using
\begin{equation}
R(t) = R(t=0) \times 10^{f} 
\label{eq:prescmig}
\end{equation}
with
\begin{equation}
  f = \log_{10}\left[ \frac{{\rm (pr)}^{2/3}}{R(t=0)} \right] \times \frac {(t / \tau)^3 }{\sqrt{1+(t/\tau)^6}},
\label{eq:f}
\end{equation}
where pr and $\tau$ are fitting parameters. The expression in
Eq.~(\ref{eq:prescmig}) tends to a constant value for $t \gg \tau$,
which mimics the fact that convergent migration is found to nearly
stall before resonant repulsion sets in.  In Eq.~(\ref{eq:f}), pr
denotes the period ratio at which convergent migration approximately
stalls, and $\tau$ the time at which this occurs. The good agreement
between this migration prescription and our hydrodynamical simulations
is illustrated in Figure~\ref{fig:fit}. Note that in the three-body
simulations, disk-driven migration is applied to the outer planet
only.

We display in Figure~\ref{fig:john1} the results of a three-body
simulation with ${\rm pr}=1.50$ and $\tau=1000$ orbits, which mimics
the convergent migration obtained in the hydrodynamical simulation of
Section~\ref{sec:M10} with $\Sigma_0 = 8\times 10^{-5}$ (see right
panel of Figure~\ref{fig:fit}). In the top-left panel of
Figure~\ref{fig:john1}, only the semi-major axis of the outer planet
varies in the first 1500 orbits as our prescription for disk-driven
migration is applied to the outer planet. Once resonant repulsion sets
in near 1500 orbits, the semi-major axis of the inner planet decreases
slightly, that of the outer planet increases slightly as well.  The
eccentricity damping timescale here is 20 planet orbits, and this
rather short timescale is found to give very similar divergent
evolutions in the three-body and hydrodynamical simulations (see
top-right panel in Figure~\ref{fig:john1}). At 5000 orbits, the period
ratio is 1.53 in the three-body run, and 1.54 in the hydrodynamical
run. This small difference can be qualitatively explained by the fact
that the repulsion occurs at a slightly larger pace in the
hydrodynamical simulation than has been modelled in the three-body
run. Furthermore, we point out that the eccentricity peaks near the
2:1 and 3:2 MMRs are about a factor of 2 to 3 smaller in the
three-body simulation (compare with the eccentricities of the
hydrodynamical run shown in the second upper panel of
Figure~\ref{fig:M10}). The reason for this difference is most likely a
reflection of the fact that the actual circularization time in the
hydrodynamical run is longer than that adopted in the three-body
integration, consistent with the operation of wake-planet repulsion in
addition to circularization-driven repulsion as discussed above.
  
We also incorporated a simple model for disk dispersal, which is taken
to occur between $10^4$ and $10^6$ orbits. Since planets are assumed
to be already formed at the beginning of the simulations, this range
of timescales is meant to account for different planet formation
timescales. The longest time ($10^6$ orbits) corresponds to a case
where both planet formation and capture into resonance following
convergent migration occur rapidly. The shortest one ($10^4$ orbits)
stands for the opposite. Once disk dispersal switches on, the
eccentricity damping timescale is progressively increased as an
exponential growth with characteristic timescale $\tau_{\rm evap} =
2000$ orbits. This timescale is shorter than the typical viscous
draining timescale of a disk inside the photoevaporation radius. It
implies that the efficiency of disk-driven resonant repulsion is
nearly entirely determined by the time from the beginning of the
simulations at which dispersal switches on.

\subsubsection{Results}
\label{sec:nbody_results}
\begin{figure}
  \centering
  \resizebox{\hsize}{!}
  {
    \includegraphics{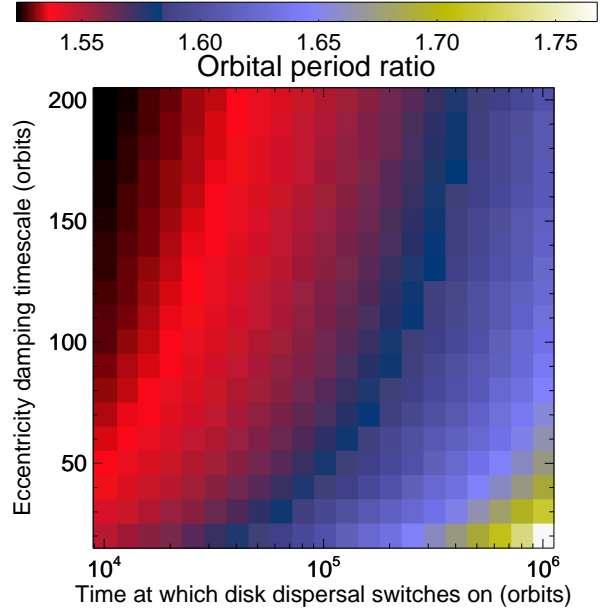}
  } 
  \caption{\label{fig:john2}Results of three-body simulations with
    prescribed convergent migration and eccentricity damping. The
    central star has a Solar mass. The mass of the inner and outer
    planets is $M_{\rm inner} = 15 M_{\rm Earth}$ and $M_{\rm inner} =
    13 M_{\rm Earth}$, respectively.  Convergent migration is modeled
    using the expression in Eq.~(\ref{eq:prescmig}) with pr = 1.5 and
    $\tau = 10^3$ orbits. The eccentricity damping timescale is varied
    from 20 to 200 orbits (y-axis), and it remains fixed until disk
    dispersal is switched on. The timescale at which disk dispersal
    switches on ranges from $10^4$ to $10^6$ orbits after the
    beginning of the simulations (x-axis), and the evaporation
    timescale is set to $2000$ orbits. The orbital period ratio
    obtained after $1.2\times 10^6$ orbits, and time-averaged over the
    last 1000 orbits, is colour-coded.}
\end{figure}

We display in Figure~\ref{fig:john2} the results of a series of
three-body simulations using the model described in
Section~\ref{sec:nbody_method}. Recall that $M_{\rm inner} = 15
M_{\oplus}$, $M_{\rm outer} = 13 M_{\oplus}$ and $M_{\star} =
M_{\odot}$, which can be thought of being representative of Kepler's
sample of multi-planetary systems (the planets' mass ratio is
arbitrary, it takes the same value as in the hydrodynamical
simulations in Section~\ref{sec:M10}).  Planets are initiated with a
period ratio of 2.4. The prescription for convergent migration given
at Eq.~(\ref{eq:prescmig}) is used with ${\rm pr} = 1.50$ and $\tau =
1000$ orbits. This value of pr is meant to explore the diversity of
period ratios that can be achieved by disk-driven resonant repulsion
away from the 3:2 MMR.  The eccentricity damping timescale is varied
from 20 to 200 orbits, and the time at which disk dispersal switches
on from $10^4$ to $10^6$ orbits (see Section~\ref{sec:nbody_method}).
Given the duration of disk-driven resonant repulsion, our results are
found to have very little dependence on the "convergent migration
timescale" $\tau$. Simulations were carried out over $1.2\times 10^6$
orbits, and the period ratio time-averaged over the last 1000 orbits
of the simulations is indicated by the color bar. For a given
eccentricity damping timescale, the later disk dispersal occurs, the
longer planets eccentricities are damped, and therefore the larger the
final period ratio. At a given time prior to disk dispersal, the
shorter the eccentricity damping timescale, the larger the final
period ratio, again as expected. For the shortest damping timescales
that we have considered, period ratios can reach $\sim$ 1.75. With
such short timescales, planets can evolve continuously from the 3:2
MMR to the 2:1 MMR (that is, the 2:1 resonant angles start to librate
when the 3:2 resonant angles no longer librate). Had we taken even
shorter damping timescales or delayed disk dispersal even more,
resonant repulsion would have led to period ratios exceeding 2.

The main conclusion that can be drawn from Figure~\ref{fig:john2} is
that partial gap-opening super-Earths may experience significant
disk-driven resonant repulsion away from the nominal 3:2 MMR. Under
very favorable circumstances (particularly efficient repulsion
maintained over a particularly long timescale) period ratios can
easily exceed nominal resonant values. Although we have considered
fixed planet masses for illustration purposes, we believe that
disk-driven resonant repulsion is a generic mechanism that may have
likely occurred among Kepler's multi-planetary systems.

\section{Concluding remarks}
\label{sec:conclusion}
The multi-planetary systems detected by the Kepler mission have very
diverse architectures, which reflect the wide range of conditions
under which planets may form and evolve. The quasi-coplanar orbits in
Kepler's multiple systems suggest that interactions between planets
and their parent protoplanetary disk should play a prominent role in
shaping such diverse architectures. In particular, close planet pairs
feature a variety of orbital period ratios and many of them are not in
mean-motion resonance.  Convergent migration of multiple planets due
to disk-planet interactions may lead to resonant systems in disks
where turbulence is low enough not to disrupt resonances
\citep{Ketchum11,PBH11,ReinKepler}.  Resonant planet pairs formed by
disk-planet interactions may experience divergent evolution of their
orbits due to tidal orbital circularization \citep{PT10, Papa11}. This
so-called {\it tidal resonant repulsion} may increase the orbital
period ratio of close-in planet pairs quite substantially depending on
the efficiency of star-planet tidal interactions \citep{Papa11, LW12,
  Baty12}. Tidal resonant repulsion could help account for the
diversity of period ratios in Kepler's close-in planet pairs.

We have shown in this paper that disk-planets interactions could also
form planet pairs with a variety of period ratios. We have first
focused on the Kepler-46 planetary system, in which the two
Saturn-mass planets Kepler-46b and Kepler-46c have a period ratio
$\approx 1.69$ \citep{Nes12}.  We have presented results of
hydrodynamical simulations modeling the early evolution of these
planets as they were embedded in their parent disk.  For typical disk
temperatures and turbulent viscosities, both planets are expected to
open a partial gap around their orbit.  A rapid convergent migration
causes the planets to merge their gaps and to evolve into a common
gap. Depending on the density's structure inside the common gap, the
planets' convergent migration is found to stall with a variety of
period ratios between 1.5 and 1.7 (see Figure~\ref{fig:finalopr}). The
observed period ratio of Kepler-46b/c can be reproduced without the
planets being in mean-motion resonance.

Our results also highlight that disk-planets interactions may
significantly increase the orbital period ratio of a planet pair from
a near-resonant value, a mechanism that we term {\it disk-driven
  resonant repulsion}. It has two origins: (i) orbital circularization
by the disk, and (ii) the interaction between each planet and the wake
of its companion. As the wake of a planet turns into a shock at some
distance from the planet, it may deposit angular momentum in the
coorbital region of another planet through an effective corotation
torque.  We have shown that this mechanism can dissipate the total
orbital energy of a planet pair without changing its total angular
momentum, a configuration that increases the planets' period ratio and
decreases their eccentricity. Resonant repulsion caused by wake-planet
interactions can be particularly efficient when at least one of the
planets opens a partial gap around its orbit (see
Section~\ref{sec:divergent}). For the disk and planet parameters
adopted in this work, wake-planet interactions are much more efficient
at driving the repulsion of a planet pair than disk
circularization. However, there may be circumstances for which this
balance is different.

The strength of wake-planet interactions increases as the radial
separation between planets decreases, independently of their
eccentricity. If wake-planet interactions are the dominant process for
the divergent evolution of a planet pair, then one could expect the
period ratio to reach a constant value, which could be different from
a resonant ratio. Our results of simulations report, however, many
cases where the period ratio nearly stalls but then increases.  There
could be two reasons for this behavior. First, wake-planet
interactions cause a (non-linear) evolution of the disk's density
profile near the planets, which causes the torques driving convergent
migration and wake-planet interaction to change over
time. Furthermore, independently of the previous consideration,
preliminary results indicate that the torques responsible for
convergent migration and for wake-driven repulsion have different
dependences on the stellar distance as the planets migrate. Both
effects are found to strengthen wake-driven repulsion as compared to
convergent migration as planets migrate inwards. This would explain
why, in our simulations, the period ratio of inwardly migrating planet
pairs increases after reaching a nearly constant value. The above
considerations indicate that wake-planet interactions may naturally
increase the period ratio of planet pairs away from resonant values.

One of our low-mass disk models for Kepler-46, in which the
Saturn-mass planets are captured in the 2:1 MMR, a period ratio of 2.3
is attained in as short as a few thousand orbits. We note that a
significant divergent evolution of sub-Jovian planets was obtained in
some of the hydrodynamical simulations of \citet{rpk10} and
\citet{Rein43}. The similarity between their results of simulations
and ours (convergent migration nearly stalled close to resonance,
followed by divergent evolution of the planets orbits alongside the
damping of their eccentricity) suggests that the divergent evolution
obtained in these simulations is likely to arise from wake-planet
interactions.

Many planets in Kepler's multiple systems are in the super-Earth to
Neptune-mass range, and have orbital periods below 100 days.  At these
short periods the aspect ratio of a disk should be rather small, thus
many Kepler's planet candidates could have opened partial gaps in
their parent disk. In contrast to the Saturn-mass planets in the
Kepler-46 system, super-Earths are not expected to merge their gaps
and to evolve in a common gap, unless a very fast convergent migration
brings them to very short mutual separations. We have argued instead
that close-in super-Earths are good candidates for efficient resonant
repulsion caused by wake-planet interactions.  For instance, one of
our simulations shows that the orbital period ratio of two $\sim$ 15
Earth-mass planets increases from 1.5 to 1.55 in few thousand
orbits. Disk-driven repulsion via wake-planet interactions could
explain why many planet pairs in Kepler's multiple systems have period
ratios slightly greater than resonant. While resonant repulsion driven
by stellar tides should only be relevant to planet pairs with orbital
periods below a few days, disk-driven repulsion may also apply to
planets on longer periods.

The TTV signals of Kepler's short-period planet pairs indicate that
some of these planets have small but not zero free eccentricities
\citep{LXW12}. Free eccentricities get quickly damped by tidal orbital
circularization. However, the presence of a disk of planetesimals or a
non-smooth gas disk dispersal could give planets some free
eccentricity. Close-in planet pairs that seem to have undergone
resonant repulsion, but which have not zero free eccentricities, like
Kepler-23b/c and Kepler-28b/c, could indicate that repulsion was
primarily mediated by disk-planets interactions rather than by
star-planet tidal interactions.

We foresee at least three directions to pursue work on disk-driven
resonant repulsion. Firstly, the effects of disk turbulence, which we
have modeled with a constant viscosity, need to be addressed
carefully. The levels of stochastic fluctuations required to disrupt
the process as planet masses decrease, and to produce some residual
free eccentricity should be explored. Secondly, planet mass growth
should be taken into account as it may also lead to some divergent
evolution \citep{petrovich}. Thirdly, a Monte-Carlo approach with
three-body integrations using simple prescriptions for the effects of
disk-planet interactions should be used to compare the distribution of
orbital elements among Kepler's planet pairs with the synthetic
distributions predicted by disk-driven resonant repulsion.

\acknowledgements
CB is supported by a Herchel Smith Postdoctoral Fellowship of the
University of Cambridge. We thank Gwena\"{e}l Bou{\'e}, Cathie Clarke,
J{\'e}r{\^o}me Guilet, Cristobal Petrovich and Yanqin Wu for useful
discussions.  We also thank the referee for a very helpful
report. Hydrodynamical simulations were performed on the Darwin
Supercomputer of the University of Cambridge High Performance
Computing Service using Strategic Research Infrastructure Funding from
the Higher Education Funding Council for England.

\appendix

\section{A -- Efficiency of tidal resonant repulsion of a planet pair}
\label{sec:tidal}
We show here that star-planet tidal interactions are unlikely to cause
significant divergent evolution of two planets in the super-Earth mass
range when the orbital period of the inner planet exceeds about 10
days. In the constant time lag model, taking tidal dissipation within
the planet to be much larger than within the star, the orbital
circularization timescale, $\tau_{\rm circ}$, is given by equation
(25) of \citet{GS66} and can be expressed as follows:
\begin{equation}
\tau_{\rm circ} = 4.5\times 10^7\,{\rm yr} 
\times \left( \frac{Q_{\rm i}^{'}}{30} \right)
\times \left( \frac{\rho_{\rm i}}{1\,{\rm g\,cm}^{-3}}  \right)
\times \left( \frac{R_{\rm i}}{0.2\,{\rm R}_{\rm Jup}}  \right)^{-2}
\times \left( \frac{M_{\star}}{M_{\odot}}  \right)^{2/3}
\times \left( \frac{T_{\rm i}}{10\,{\rm days}}  \right)^{13/3},
\label{eq_app1}
\end{equation}
where $Q_{\rm i}^{'}$ is the tidal quality factor of the inner planet,
$\rho_{\rm i}$ its mean density, $R_{\rm i}$ its physical radius,
$T_{\rm i}$ its orbital period and $M_{\star}$ is the star's
mass. Divergent evolution of two resonant planets may occur as a
result of tidal dissipation of orbital energy at approximately
constant orbital angular momentum \citep{Papa11}.  The timescale for
divergent evolution, which we denote by $\tau_{\rm div}$, satisfies
\begin{equation}
\tau_{\rm div} \sim \frac{\tau_{\rm circ}}{e_{\rm i}^2} \times \frac{\Delta a}{a_{\rm i}},
\end{equation}
with $a_{\rm i}$ and $e_{\rm i}$ the semi-major axis and forced
eccentricity of the inner planet, and $\Delta a$ its change in
semi-major axis during divergent evolution (it can be taken to be the
planet's distance from resonance).  We have $e_{\rm i} = C (M_{\rm o}
/ M_{\star}) \times (\Delta a / a_{\rm i})^{-1}$, where $M_{\rm o} =
\mu M_{\rm i}$ is the outer planet's mass, $M_{\rm i}$ the inner
planet's mass ($\mu$ denotes the outer-to-inner planet mass ratio),
and $C$ is a constant of order unity that varies with the resonance
\citep[see, e.g., equation 38 of][]{Papa11}. Using Eq.~(\ref{eq_app1})
and taking $C=1$, we obtain
\begin{equation}
\tau_{\rm div} \sim 2\times 10^{12}\,{\rm yr} 
\times \left( \frac{Q_{\rm i}^{'}}{30} \right)
\times \mu^{-2}
\times \left( \frac{\rho_{\rm i}}{1\,{\rm g\,cm}^{-3}}  \right)^{-1}
\times \left( \frac{R_{\rm i}}{0.2\,{\rm R}_{\rm Jup}}  \right)^{-8}
\times \left( \frac{M_{\star}}{M_{\odot}}  \right)^{8/3}
\times \left( \frac{T_{\rm i}}{10\,{\rm days}}  \right)^{13/3}
\times \left( \frac{\Delta a / a_{\rm i}}{10^{-2}} \right)^3.
\label{eq_app2}
\end{equation}
This order of magnitude estimate shows that planet pairs orbiting
Sun-like stars should experience negligible tidally-driven resonant
repulsion over gigayear timescales when the inner planet is a
super-Earth, with $R_{\rm i} < 2R_{\oplus},$ $\rho_{\rm i} < 5{\rm
  g\,cm^{-3}},$ and orbital period greater or equal to 10 days, even
if $Q_{\rm i}^{'}=1$ and $\mu \gtrsim 1$. Moreover, if the inner
planet is a Jupiter-like planet, then assuming $Q_{\rm i}^{'} \sim
10^5$ and $\rho_{\rm i} < \sim 5\,$ g\,cm$^{-3}$, Eq.~(\ref{eq_app2})
shows that divergent evolution of more than a few percent away from
resonance is expected over a few gigayears as long as the inner
planet's orbital period does not exceed 10 days.

\section{B -- Hydrodynamical simulations of the Kepler-46 system: robustness of results}
\label{sec:robust}
In the course of our numerical experiments, we have tested the
robustness of our results by varying the assumptions made in the
physical model and by varying the grid resolution.
\begin{figure*}
  \centering
  \resizebox{0.8\hsize}{!}
  {
    \includegraphics{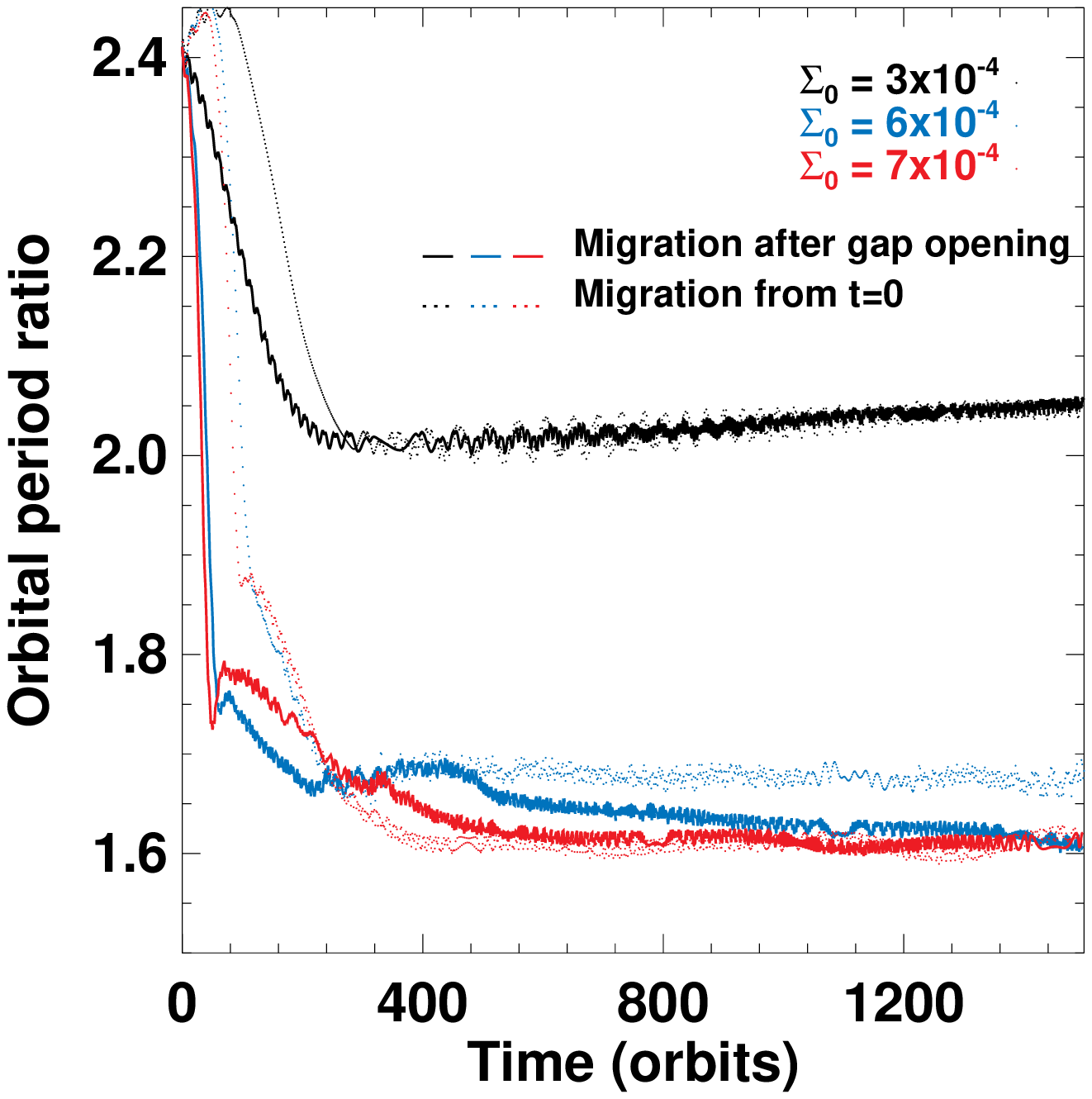}
     \includegraphics{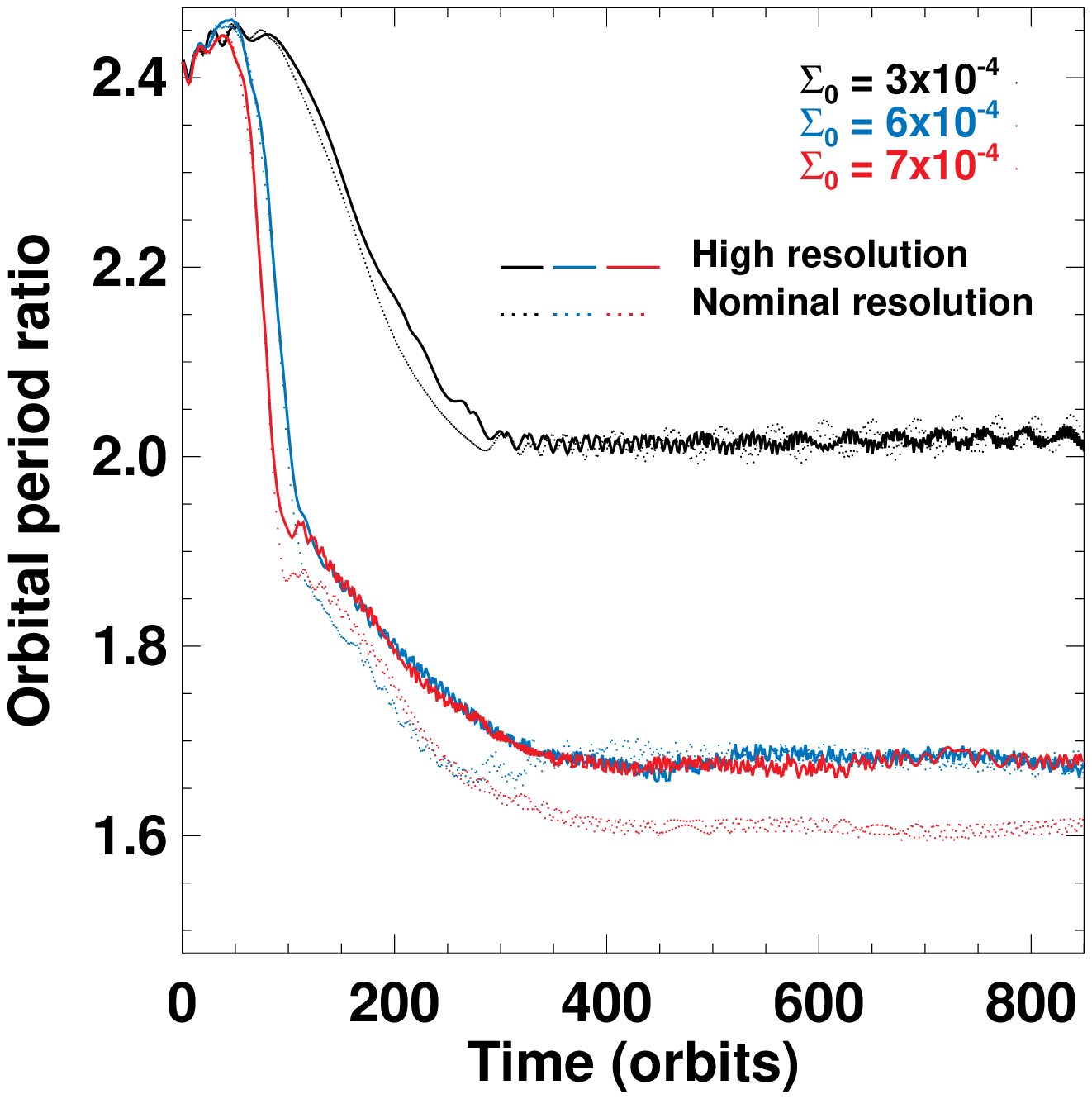}
  }
  \caption{\label{fig:gapres}Ratio of orbital periods for $M_{\rm
      inner}=0.6 M_{\rm J}$ and different values of the unperturbed
    surface density parameter $\Sigma_0$. The left panel compares our
    fiducial case where planets migrate from the beginning of the
    simulations (dotted curves) with the case where planets are first
    held on fixed circular orbits for 500 orbits before migrating
    (solid curves). The right panel shows the overall good convergence
    of our results when increasing the number of grid cells from
    $400\times 800$ (nominal resolution) to $800\times 1600$ (high
    resolution).}
\end{figure*}
\\
\par\noindent\emph{Physical model---} 
We performed several simulations reducing the disk's aspect ratio to
4\% and 3\% while keeping the same viscous alpha parameter. Although
not illustrated here, these simulations show again a variety of
outcomes, but none reproducing the observed period ratio between
Kepler-46c and Kepler-46b. Decreasing the disk's aspect ratio makes
the gaps carved by the planets deeper. When convergent migration is
rapid enough for the planets to cross their 2:1 MMR and to evolve in a
common gap, we find that the planets generally lock themselves into
the 3:2, 4:3 or even 7:5 MMR with very little departure from exact
commensurability. In these simulations, the density in the common gap
is too low for wake-planet interactions to cause significant divergent
evolution (see Section~\ref{sec:divergent}). These simulations thus
suggest that partial gap opening is required to obtain a period ratio
consistent with the observed value. Furthermore, as the structure of
the gaps plays a prominent role in the final evolution of the planets'
period ratio, we carried out additional simulations where the planets
were first held on fixed circular orbits for 500 orbits before
migrating. This preliminary stage allowed enough time for each planet
to build up a gap with a stationary density structure. The results of
the simulations are shown in the left panel of Figure~\ref{fig:gapres}
for $M_{\rm inner}=0.6 M_{\rm J}$ and three values of
$\Sigma_0$. Starting with pre-evolved gaps changes the initial rate of
convergent migration. Notwithstanding quite different initial
evolutions, two out of these three models end up following the same
evolution with or without pre-evolved gaps, and reach the same final
period ratio at the end of the simulations. The model with $\Sigma_0 =
6\times 10^{-4}$ reaches a slightly smaller period ratio (decreased
from 1.65 to 1.6 when starting with pre-evolved gaps).
\\
\par\noindent\emph{Grid resolution---} 
The effect of resolution was checked by doubling the number of grid
cells along each direction for a few simulations. The results of three
high-resolution runs ($800\times1600$) are compared to the
nominal-resolution runs ($400\times800$) in the right panel of
Figure~\ref{fig:gapres}. Doubling the grid resolution slightly changes
the initial rate of convergent migration, leading to slightly
different initial evolution. As for the above series of simulations
with pre-evolved gaps, the model with capture into 2:1 MMR followed by
resonant repulsion is not affected by doubling the resolution. The two
models with common gap evolution seem again more susceptible to a
variation in the initial convergent migration, but lead to small
differences in the final outcome. Interestingly, both high-resolution
models with $\Sigma_0 = 6$ and $7\times 10^{-4}$ yield a period ratio
$\approx 1.68$, in very good agreement with the observed value.

\end{document}